\documentclass[oneside,english]{elsart}
\usepackage[T1]{fontenc}
\usepackage[latin1]{inputenc}
\usepackage{subfigure}
\usepackage{amsmath}
\usepackage{graphicx}
\usepackage{amssymb}
\usepackage[authoryear]{natbib}

\makeatletter
\usepackage{babel}
\makeatother
\begin{document}
\begin{frontmatter}

\title{Bacterial gene regulation in diauxic and nondiauxic growth}

\author{Atul Narang}

\address{Department of Chemical Engineering, University of Florida, Gainesville,
FL~32611-6005.}

\ead{narang@che.ufl.edu}

\ead[url]{http://narang.che.ufl.edu}

\thanks{Corresponding author. Tel: + 1-352-392-0028; fax: + 1-352-392-9513}

\author{Sergei S. Pilyugin}

\address{Department of Mathematics, University of Florida, Gainesville, FL~32611-8105.}

\begin{abstract}
When bacteria are grown in a batch culture containing a mixture of
two growth-limiting substrates, they exhibit a rich spectrum of
substrate consumption patterns including diauxic growth,
simultaneous consumption, and bistable growth. In previous work, we
showed that a minimal model accounting only for enzyme induction and
dilution captures all the substrate consumption patterns (Narang,
Biotech.~Bioeng., 59, 116, 1998; Narang, J. Theoret. Biol.,
accepted, 2006). In this work, we construct the bifurcation diagram
of the minimal model, which shows the substrate consumption pattern
at any given set of parameter values. The bifurcation diagram
explains several general properties of mixed-substrate growth.
(1)~In almost all the cases of diauxic growth, the {}``preferred''
substrate is the one that, by itself, supports a higher specific
growth rate. In the literature, this property is often attributed to
the optimality of regulatory mechanisms. Here, we show that the
minimal model, which contains no regulation, displays the property
under fairly general conditions. This suggests that the higher
growth rate of the preferred substrate is an intrinsic property of
the induction and dilution kinetics. (2)~The model explains the
phenotypes of various mutants containing lesions in the regions
encoding for the operator, repressor, and peripheral enzymes. A
particularly striking phenotype is the {}``reversal of the diauxie''
in which the wild-type and mutant strains consume the very same two
substrates in opposite order. This phenotype is difficult to explain
in terms of molecular mechanisms, such as inducer exclusion or CAP
activation, but it turns out to be a natural consequence of the
model. We show furthermore that the model is robust. The key
property of the model, namely, the competitive dynamics of the
enzymes, is preserved even if the model is modified to account for
various regulatory mechanisms. Finally, the model has important
implications for the problem of size regulation in development. It
suggests that protein dilution is one mechanism for coupling
patterning and growth.
\end{abstract}
\end{frontmatter}

\section{Introduction}

When microbial cells are grown in a batch culture containing a mixture
of two carbon sources, they often exhibit \emph{diauxic} growth~\citep{monod47}.
This phenomenon is characterized by the appearance of two exponential
growth phases separated by a lag phase called \emph{diauxic lag}.
The most well-known example of the diauxie is the growth of \emph{Escherichia
coli} on a mixture of glucose and lactose. Early studies by Monod
showed that in this case, the two exponential growth phases reflect
the sequential consumption of glucose and lactose~\citep{monod1}.
Moreover, only glucose is consumed in the first exponential growth
phase because the synthesis of the \emph{peripheral} enzymes for lactose
is somehow abolished in the presence of glucose. These enzymes include
lactose permease (which catalyzes the transport of lactose into the
cell), $\beta$-galactosidase (which hydrolyzes the intracellular
lactose into products that feed into the glycolytic pathway) and lactose
transacetylase (which is believed to metabolize toxic thiogalactosides
transported by lactose permease). During the period of preferential
growth on glucose, the peripheral enzymes for lactose are diluted
to very small levels. The diauxic lag reflects the time required to
build up these enzymes to sufficiently high levels. After the diauxic
lag, one observes the second exponential phase corresponding to consumption
of lactose.

\begin{figure}
\begin{centering}\subfigure[]{\includegraphics[width=10cm,height=3cm]{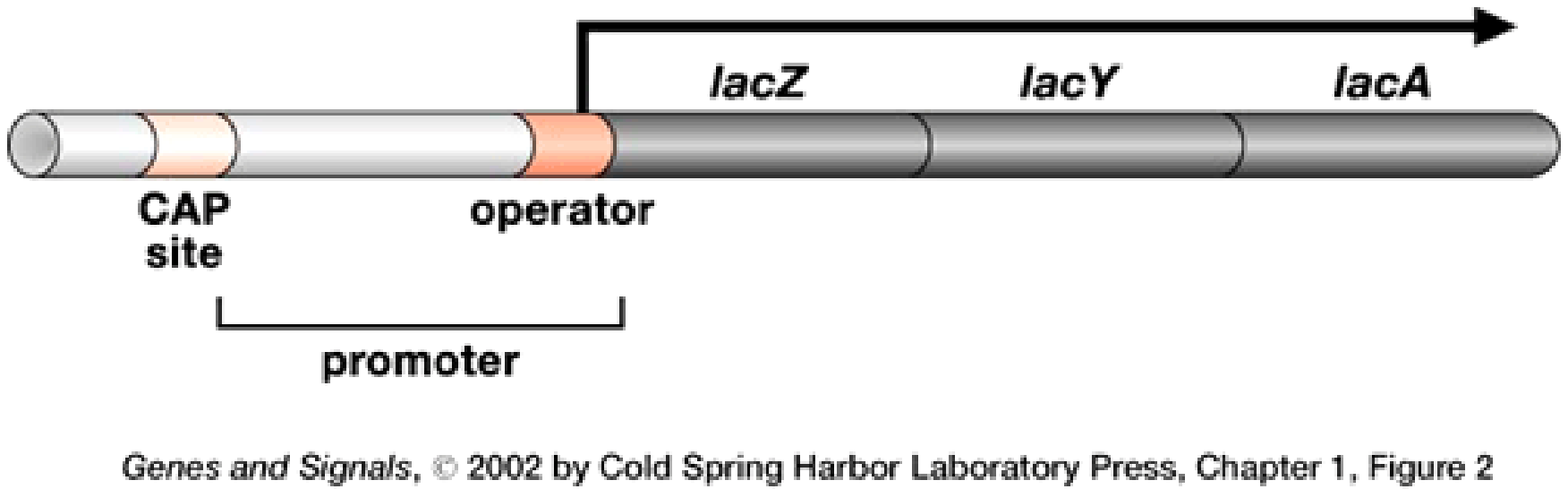}}\par\end{centering}

\begin{centering}\subfigure[]{\includegraphics[width=10cm,height=12cm]{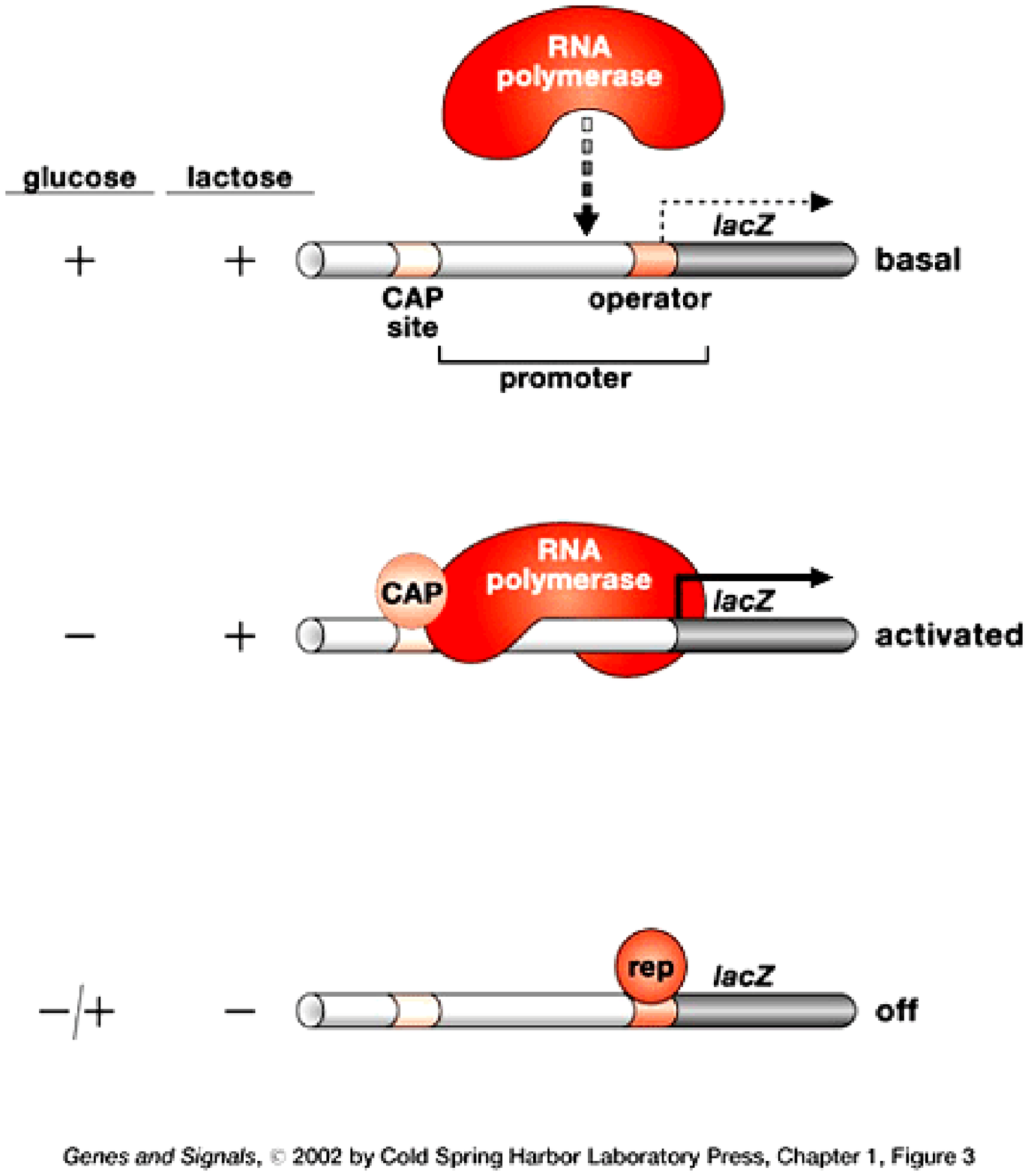}}\par\end{centering}

\caption{\label{f:LacOperon}Molecular mechanism for induction and repression
of the \emph{lac} operon in \emph{E.~coli} ~\citep{Ptashne2}: (a)~Structure
of the \emph{lac} operon. The \emph{lacZ}, \emph{lacY}, and \emph{lacA}
genes code for $\beta$-galactosidase, lactose permease, and lactose
transacetylase, respectively. The \emph{operator, promoter,} and \emph{CAP
site} denote the DNA sequences which bind the repressor, RNA polymerase,
and CAP-cAMP, respectively. (b)~The states of the \emph{lac} operon
in the presence of glucose or/and lactose. The repressor and CAP-cAMP
complex are denoted \emph{rep} and \emph{CAP}, respectively.}
\end{figure}

It turns out that the peripheral enzymes for lactose are synthesized
only if lactose is present in the environment. The mechanism for the
synthesis or \emph{induction} of these enzymes in the presence of
lactose and absence of glucose was discovered by Monod and coworkers~\citep{jacob61}.
It was shown that the genes corresponding to these enzymes are contiguous
on the DNA and transcribed in tandem, an arrangement referred to as
the \emph{lac} operon (Fig.~\ref{f:LacOperon}a). In the absence
of lactose, the \emph{lac} operon is not transcribed because a molecule
called the \emph{lac} repressor is bound to a specific site on the
\emph{lac} operon called the \emph{operator} (Fig.~\ref{f:LacOperon}b,
bottom). This prevents RNA polymerase from attaching to the operon
and initiating transcription. In the presence of lactose, transcription
of \emph{lac} is triggered because allolactose, a product of $\beta$-galactosidase,
binds to the repressor, and renders it incapable of binding to the
operator~(Fig.~\ref{f:LacOperon}b, middle).%
\footnote{A similar mechanism serves to induce the genes for glucose transport~\citep[Fig.~4]{plumbridge03}.%
}

The occurrence of the glucose-lactose diauxie suggests that transcription
of \emph{lac} is somehow repressed in the presence of glucose. Two
molecular mechanisms have been proposed to explain this repression.

\begin{enumerate}
\item \emph{Inducer exclusion}~\citep{Postma1993}: In the presence of
glucose, enzyme IIA$^{{\rm glc}}$, a peripheral enzyme for glucose,
is dephosphorylated. The dephosphorylated~IIA$^{{\rm glc}}$ inhibits
lactose uptake by binding to lactose permease. This reduces the intracellular
concentration of allolactose, and hence, the transcription rate of
the \emph{lac} operon.\\
Genetic evidence suggests that phosphorylated~IIA$^{{\rm glc}}$
activates the enzyme, adenylate cyclase, which catalyzes the synthesis
of cyclic AMP (cAMP). Since the total concentration of IIA$^{{\rm glc}}$
remains constant on the rapid time scale of its dephosphorylation,
exposure of the cells to glucose causes a decrease in the level of
phosphorylated IIA$^{{\rm glc}}$, and hence, cAMP. This reduction
of the cAMP level forms the basis of yet another mechanism of \emph{lac}
repression.
\item \emph{cAMP activation}~\citep[Chap.~1]{Ptashne2}: It has been observed
that RNA polymerase is not recruited to the \emph{lac} operon unless
a protein called catabolite activator protein (CAP) or cAMP receptor
protein (CRP) is bound to a specific site on the \emph{lac} operon
(denoted {}``CAP site'' in Fig.~\ref{f:LacOperon}). Furthermore,
CAP, by itself, has a low affinity for the CAP site, but when bound
to cAMP, its affinity for the CAP site increases dramatically. The
inhibition of \emph{lac} transcription by glucose is then explained
as follows.\\
In the presence of lactose alone (i.e., no glucose), the cAMP level
is high. Hence, CAP becomes cAMP-bound, attaches to the CAP site,
and promotes transcription by recruiting RNA polymerase (Fig.~\ref{f:LacOperon}b,
middle). When glucose is added to the culture, the cAMP level decreases
by the mechanism described above. Consequently, CAP, being cAMP-free,
fails to bind to the CAP site, and \emph{lac} transcription is abolished
(Fig.~\ref{f:LacOperon}b, top).
\end{enumerate}
We show below that neither one of these two mechanisms can fully explain
the glucose-mediated repression of \emph{lac} transcription.

\begin{figure}
\begin{centering}\subfigure[]{\includegraphics[width=7cm,height=5cm]{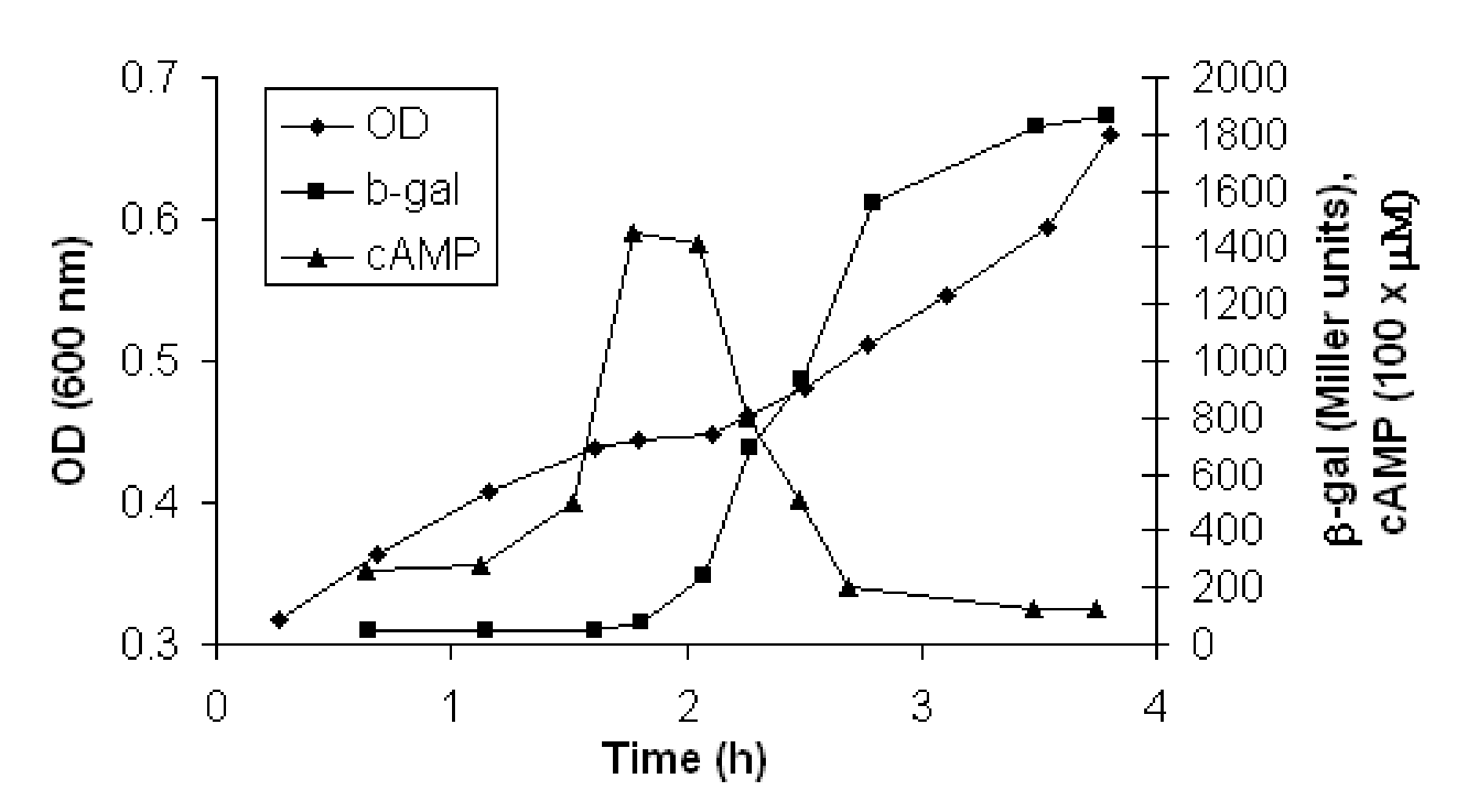}}\subfigure[]{\includegraphics[width=7cm,height=5cm]{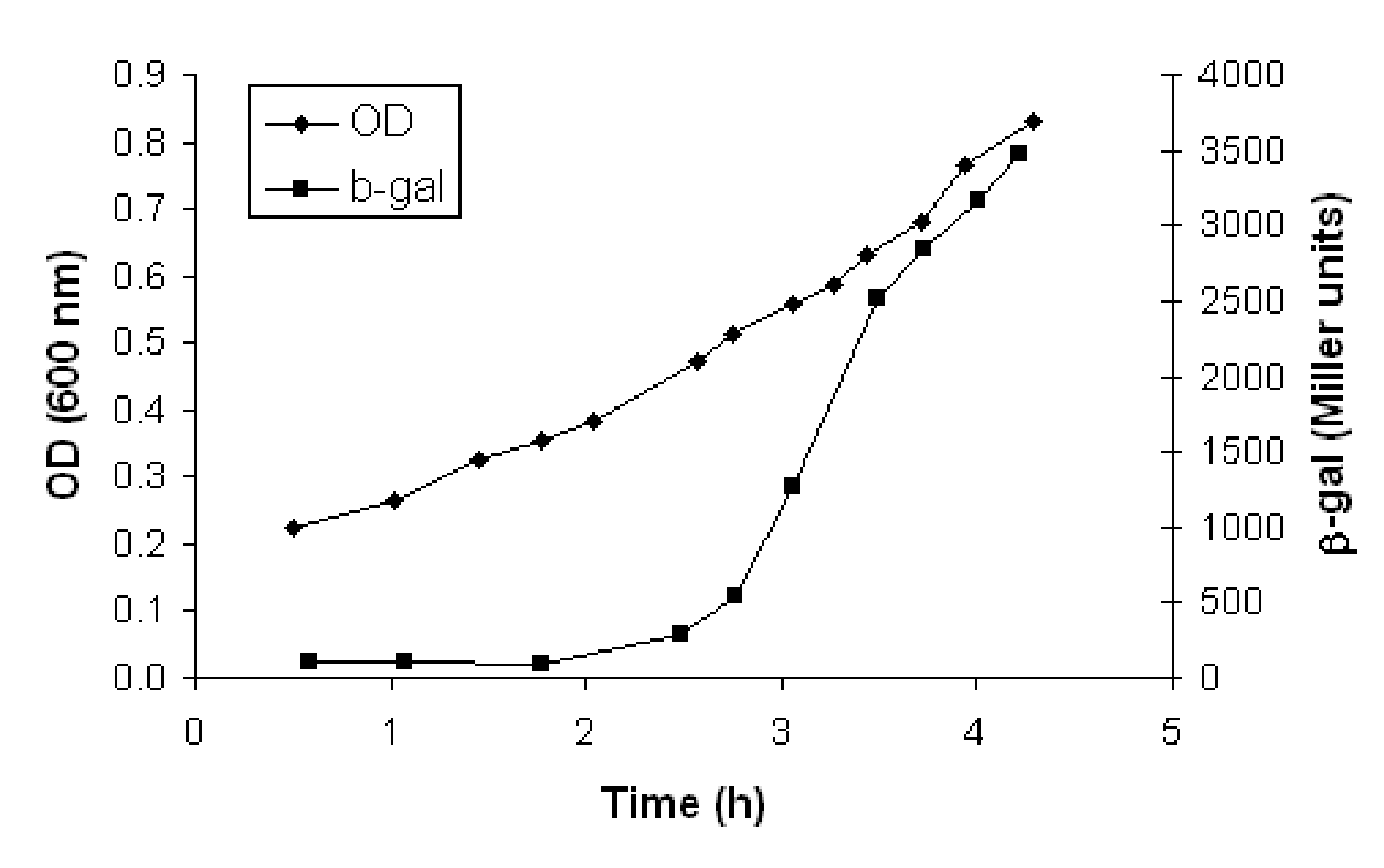}}\par\end{centering}

\begin{centering}\subfigure[]{\includegraphics[width=7cm,height=5cm]{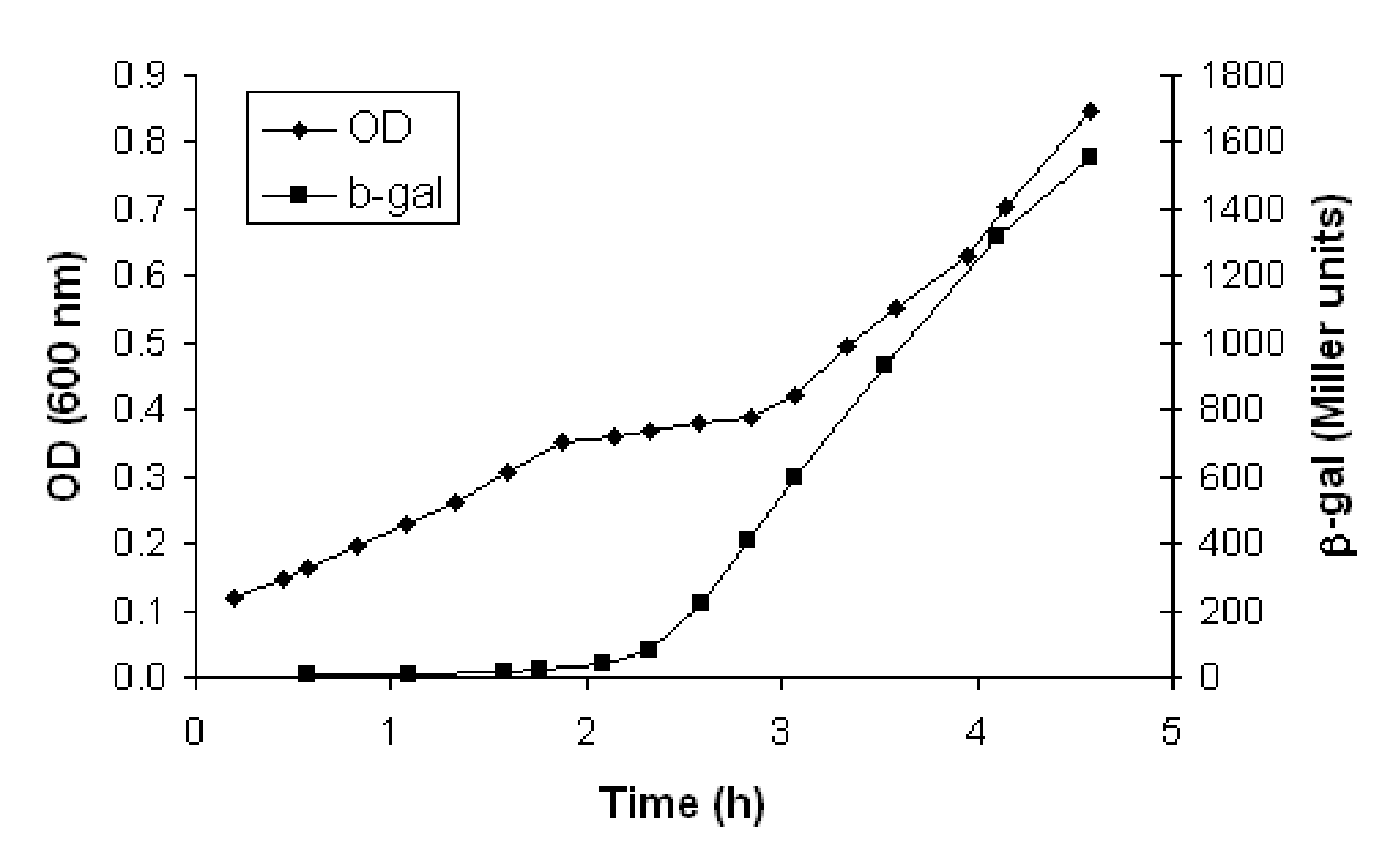}}\subfigure[]{\includegraphics[width=7cm,height=5cm]{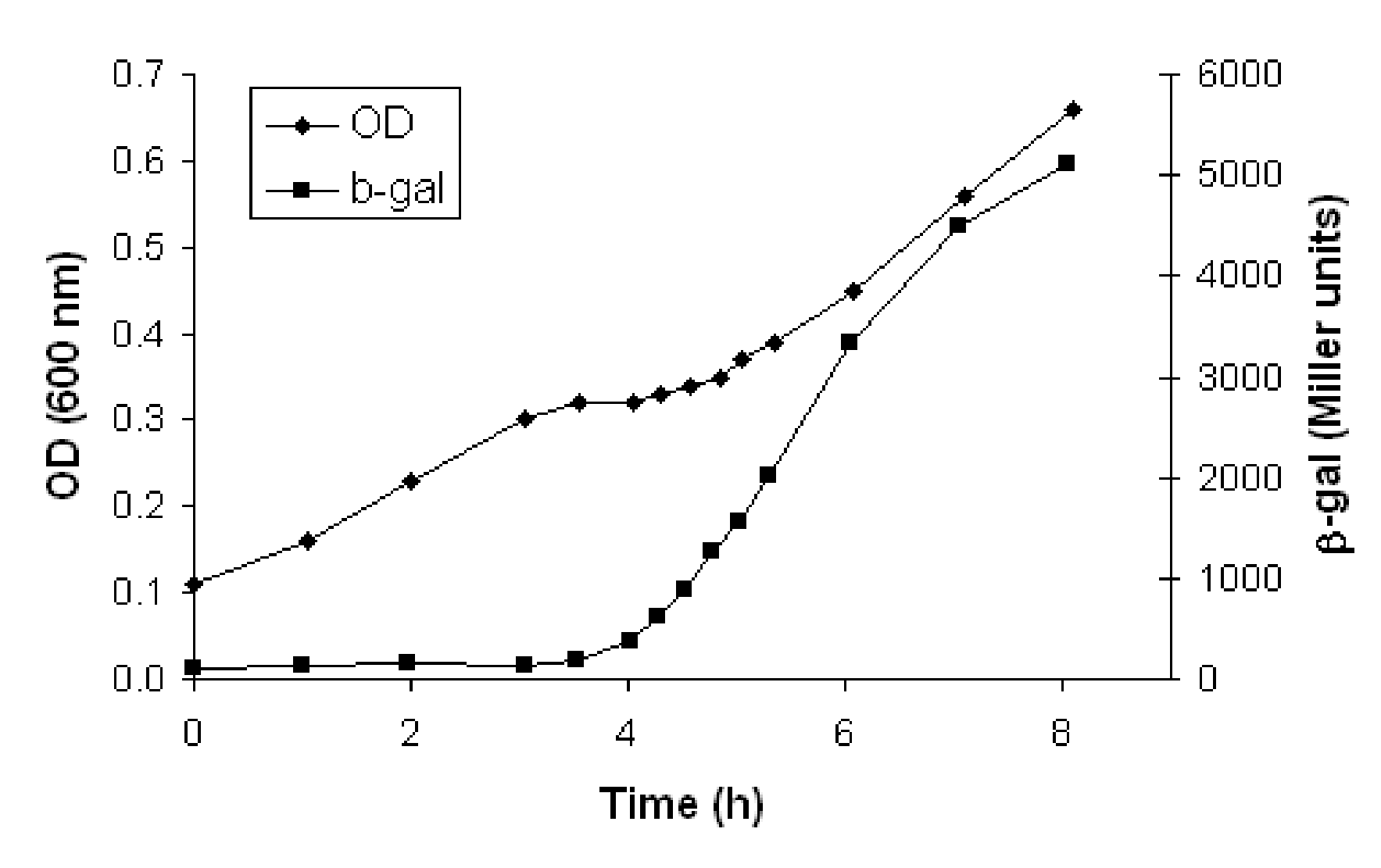}}\par\end{centering}

\caption{\label{f:cAMPeffect}Repression of \emph{lac} transcription in the
presence of glucose is not due to reduced cAMP levels~\citep{inada96,kimata97}.
OD denotes optical density, and Miller units are a measure of $\beta$-galactosidase
activity. (a)~Growth of the wild-type strain, \emph{E.~coli} W3110,
on glucose~+ lactose. The intracellular cAMP levels are comparable
during the two exponential growth phases. (b)~Growth of \emph{E.~coli}
W3110 on glucose~+ lactose in the presence of 5~mM cAMP. Despite
the high cAMP concentration, $\beta$-galactosidase synthesis is repressed
during the first exponential growth phase. (c,~d)~Growth of \emph{E.~coli}
$\triangle$\emph{cya}~\emph{crp}$^{*}$ and PR166 on glucose~+
lactose. The \emph{lac} transcription rate in these strains is independent
of the cAMP level, but $\beta$-galactosidase synthesis is repressed
during the first exponential growth phase.}
\end{figure}

The following three observations contradict the cAMP activation model.

\begin{enumerate}
\item The intracellular cAMP levels during the first exponential growth
phase ($\sim$2.5~$\mu$M) are comparable, if not higher, than those
observed during the second exponential growth phase ($\sim$1.25--2~$\mu$M)
(Fig.~\ref{f:cAMPeffect}a).\\
It follows that the repression of \emph{lac} transcription in the
presence of glucose is not due to lower cAMP levels.%
\footnote{Excess cAMP fails to relieve the repression of transcription during
growth of \emph{E.~coli} on other pairs of substrates, such as glucose~+
melibiose~\citep[Fig.~4]{Okada1981} and glucose~+ galactose (see
Fig.~\ref{f:DataCase1}a of this work).%
}
\item When the culture is exposed to large concentrations (5~mM) of exogenous
cAMP, the diauxic lag vanishes, but the \emph{lac} operon still fails
to be transcribed during the first exponential growth phase (Fig.~\ref{f:cAMPeffect}b).\\
The disappearance of the diauxic lag implies that an elevated level
of intracellular cAMP does stimulate the \emph{lac} transcription
rate. However, it fails to relieve the repression of \emph{lac} transcription
in the presence of glucose.
\item Diauxic growth persists in cells which transcribe the \emph{lac} operon
at a rate that is independent of cAMP levels. This has been demonstrated
with two types of cells (Fig.~\ref{f:cAMPeffect}c,d). In \emph{E.~coli}
$\triangle$\emph{cya}~\emph{crp}$^{*}$ mutants, \emph{crp}, the
gene coding for CAP, is mutated such that CAP binds to the CAP site
even in the absence of cAMP. In \emph{E.~coli} PR166, the \emph{lac}
promoter is mutated such that RNA polymerase binds to the promoter
even if there is no CAP-cAMP at the CAP site. In both cases, transcription
of \emph{lac} is independent of cAMP levels. Yet, $\beta$-galactosidase
synthesis is still repressed during the first exponential growth phase.
\end{enumerate}
These results show that higher cAMP levels do stimulate the \emph{lac}
transcription rate. Indeed, the 5-fold increase in cAMP levels at
the end of the first exponential growth phase in Fig.~\ref{f:cAMPeffect}a
is characteristic of cells exposed to low concentrations (0.3~mM)
of glucose~\citep{McRobb1997}, and it is likely that this serves
to reduce the length of the diauxic lag. However, \emph{lac} transcription
is repressed in the presence of glucose even if the ability of cAMP
to influence \emph{lac} transcription is abolished.

The persistence of the glucose-lactose diauxie in cAMP-independent
cells has led to the hypothesis that inducer exclusion alone is responsible
for inhibiting \emph{lac} transcription~\citep{inada96,kimata97}.
However, inducer exclusion exerts a relatively mild effect on lactose
uptake. In \emph{E.~coli} ML30, the activity of lactose permease
is inhibited no more than $\sim$40\% at saturating concentrations
of glucose~\citep[Table 2]{cohn59a}. This \emph{partial} inhibition
by inducer exclusion cannot explain the \emph{almost complete} inhibition
of \emph{lac} transcription.

Thus, despite several decades of research, no molecular mechanism
has been found to fully explain the glucose-lactose diauxie in \emph{E.
coli}.

In the meantime, microbial physiologists have accumulated a vast body
of work showing that diauxic growth is ubiquitous. It has been observed
in diverse microbial species on many pairs of \emph{substitutable}
substrates (i.e., substrates that satisfy the same nutrient requirements)
including pairs of carbon sources~\citep{egli95,harder82,kovarova98},
nitrogen sources~\citep{Neidhardt1957}, phosphorus sources~\citep{Daughton1979},
and electron acceptors~\citep{Liu1998}. These studies show that
there is no correlation between the chemical identity of a compound
and its ability to act as the preferred substrate. For instance, during
growth on a mixture of glucose and an organic acid, enteric bacteria,
such as \emph{E.~coli}, prefer glucose, whereas soil bacteria\emph{,}
such as \emph{Pseudomonas} and \emph{Arthrobacter}, prefer the organic
acid~\citep{harder76,harder82}. However, there is a correlation
between the maximum specific growth rate on a compound and its ability
to act as a preferred substrate.

\begin{quote}
In most cases, although not invariably, the presence of a substrate
permitting a higher growth rate prevents the utilization of a second,
`poorer', substrate in batch culture~\citep[p.~461]{harder82}.\label{q:CorrelationHD}
\end{quote}
This remarkable correlation, which is reminiscent of anthropomorphic
choice, is often rationalized by appealing to teleological (design-oriented)
arguments. Harder \& Dijkhuizen assert, for instance, that consumption
of lactose is abolished in the presence of glucose because this prevents
{}``unnecessary synthesis of catabolic enzymes in cells that already
have available a carbon and energy source that allows fast growth''~\citep[p.~463]{harder82}.
However, there is no mechanistic explanation for this correlation.

\begin{figure}
\begin{centering}\subfigure[]{\includegraphics[width=7cm,height=5cm]{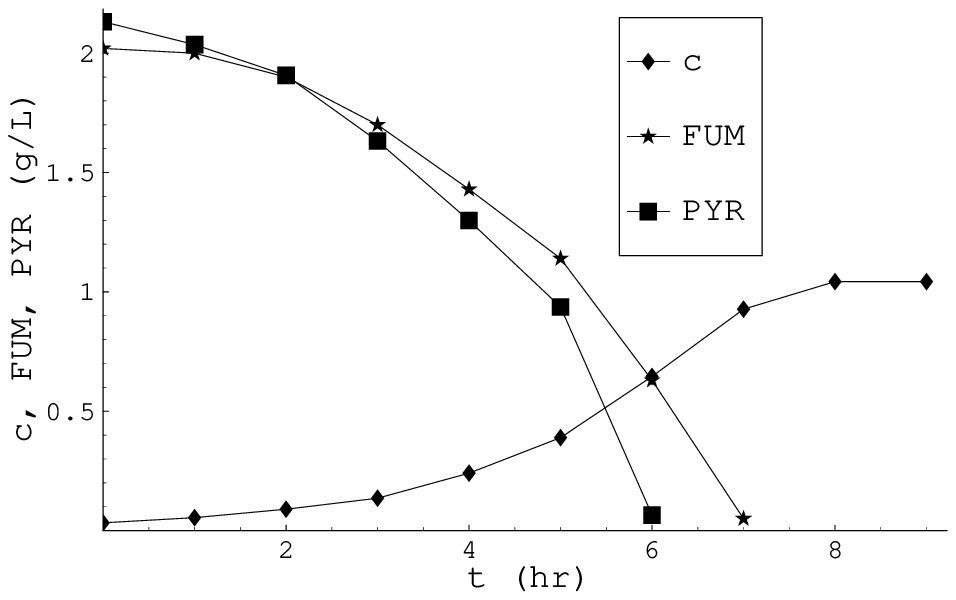}}\subfigure[]{\includegraphics[width=7cm,height=5cm]{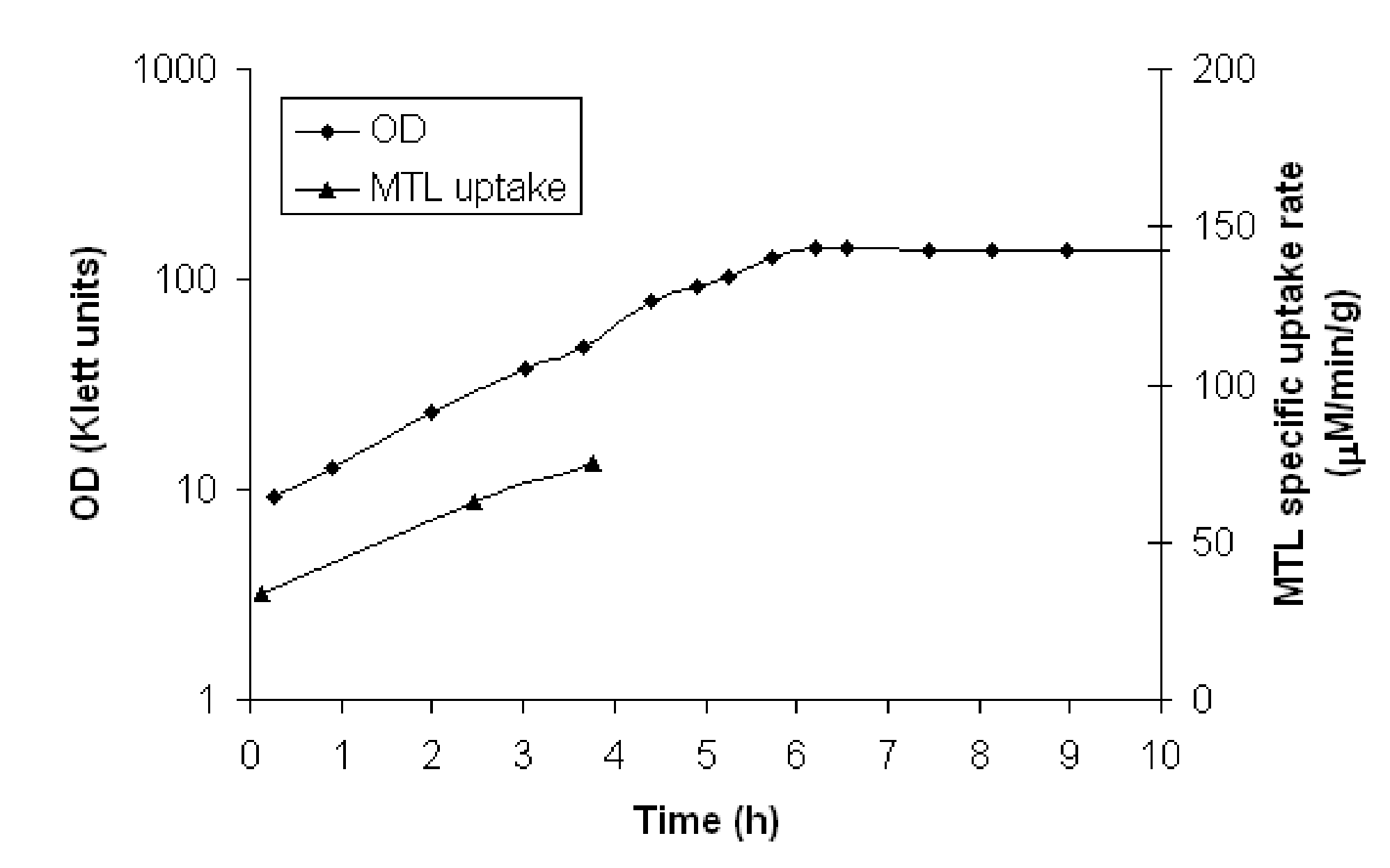}}\par\end{centering}

\caption{\label{f:NondiauxicGrowth}Simultaneous consumption of substrates
in batch cultures: (a)~Simultaneous consumption of fumarate (FUM)
and pyruvate (PYR) during batch growth of \emph{E. coli} K12 ($c$
denotes the cell density in gms dry weight per liter). The single-substrate
maximum specific growth rates on fumarate and pyruvate are 0.41~h$^{-1}$
and 0.28~h$^{-1}$, respectively. This growth pattern is observed
with several pairs of organic acids~\citep{narang97a}. (b)~Simultaneous
consumption of glucose and mannitol (MTL) during batch growth of \emph{E.
coli} strain~158~\citep{Lengeler1972}. There is significant uptake
of mannitol during the first 4~hours even though the cells are precultured
on glucose.}
\end{figure}

Although the diauxie dominates the literature on mixed-substrate growth,
there is ample evidence of nondiauxic growth. In \emph{E. coli} K12,
several pairs of organic acids are consumed simultaneously~\citep{narang97a},
one example of which is shown in Fig.~\ref{f:NondiauxicGrowth}a.
The maximum specific growth rates on these organic acids are in the
range 0.28--0.44~h$^{-1}$, which are low compared to the largest
maximum specific growth rate sustained in a minimal (synthetic) medium
(0.74~h$^{-1}$ on glucose). Similar behavior has been observed in
other species, leading Egli to conclude that

\begin{quote}
Especially combinations of substrates that support medium or low maximum
specific growth rates are utilized simultaneously~\citep[p.~325]{egli95}.
\end{quote}
However, a closer look at data suggests that low or medium growth
rates are not necessary for simultaneous consumption. This is evident
from Monod's early studies with the so-called {}``A-sugars,'' namely,
glucose, fructose, mannitol, mannose, and sucrose~\citep{monod1,monod47}.%
\footnote{It was found later that all the A-sugars are transported by the phosphotransferase
system (PTS)~\citep{Roseman1990}.%
} He found that in \emph{E. coli} and \emph{B. subtilis}, these sugars
supported comparable maximum specific growth rates, but there was
no diauxic lag during growth on a mixture of glucose and any one of
the other A-sugars. Subsequent studies have confirmed that in some
of these cases, both the substrates are consumed simultaneously (Fig.~\ref{f:NondiauxicGrowth}b).
Now, in all the cases of simultaneous consumption described above,
the single-substrate growth rates were comparable. Thus, it is conceivable
that simultaneous consumption occurs whenever the \emph{ratio} of
the single-substrate growth rates is close to 1. It turns out that
this condition may be necessary, but it is certainly not sufficient.
Although the growth rates of \emph{Propionibacterium shermanii} on
glucose and lactate are identical (0.141 and 0.142~h$^{-1}$, respectively),
lactate is consumed preferentially~\citep{lee74}. Similarly, the
growth rates of \emph{E.~coli}~ML308 on glucose and fructose are
comparable (0.91 and 0.73~h$^{-1}$, respectively), but glucose is
consumed preferentially~\citep{clark76}.%
\footnote{The absence of the diauxic lag, observed in Monod's earlier studies
with glucose-fructose mixtures, is due to rapid \emph{de novo} synthesis
of the PTS enzymes for fructose~\citep[Figs.~4--5]{clark76}. Thus,
preferential consumption without a lag does not imply the existence
of new molecular mechanisms --- it can be a consequence of rapid induction
kinetics.%
} Thus, current evidence suggests that the existence of comparable
single-substrate growth rates is, perhaps, necessary, but not sufficient,
for simultaneous consumption.\label{q:CorrelationEgli} It seems desirable
to understand the mechanistic basis of this observation.

In addition to simultaneous substrate utilization, there is some evidence
that the substrate utilization pattern can depend on the history of
the preculture. Hamilton \& Dawes were among the first to observe
such behavior during the growth of \emph{Pseudomonas aeruginosa} on
a mixture of citrate and glucose~\citep{hamilton59,hamilton60,hamilton61}.
Cells precultured on citrate showed diauxic growth with citrate as
the preferred substrate, whereas cells precultured on glucose consumed
both citrate and glucose. We observed a similar substrate consumption
pattern during growth of \emph{E. coli} K12 on glucose and pyruvate~\citep{narang97a}.
An entirely different preculture-dependent pattern was obtained during
the growth of a pseudomonad on glucose and phenol~\citep[Chap.~3, p.~181]{panikov}.
When the cells were precultured on glucose, there was preferential
consumption of glucose. Immediately after the exhaustion of phenol,
when the cells were fully adapted to phenol, the medium was supplemented
with additional glucose and phenol. Once again, there was diauxic
growth, but phenol, rather than glucose, was the preferred substrate.
In earlier work, we have argued that preculture-dependent growth patterns
may be quite common --- the lack of such data reflects the fact that
the effect of preculturing was not investigated in most studies~\citep{narang97a}.
In order to facilitate their identification, it seems appropriate
to determine the feasible preculture-dependent growth patterns.

The goal of this work is to seek mechanistic answers for the following
questions

\begin{enumerate}
\item In diauxic growth, why is the maximum specific growth rate on the
preferred substrate higher than that on the less preferred substrate?
\item Under what conditions are the substrates consumed simultaneously?
\item What types of preculture-dependent growth patterns are feasible?
\end{enumerate}
There are numerous mechanistic models of mixed-substrate growth. Many
of them are based on detailed mechanisms uniquely associated with
the glucose-lactose diauxie in \emph{E. coli}~\citep{Kremling2001,Santillan2004,vandedem73,Wong1997}.
These models cannot address the above questions, which are concerned
with the \emph{general} properties of mixed-substrate growth. Thus,
one led to consider the more general models accounting for only those
processes that are common to most systems of mixed-substrate growth~\citep{Brandt2003,narang97c,narang98b,Thattai2003}.
Recently, we have shown that these general models are similar inasmuch
as the enzymes follow competitive dynamics in all the cases~\citep{Narang06b}.
However, the model in Brandt et al cannot capture nondiauxic growth,
and the model in Thattai \& Shraiman treats the specific growth rate
as a fixed (constant) parameter, an assumption that is not appropriate
for describing the growth of batch cultures.

In this work, we address the questions posed above by appealing to
the minimal model in~\citet{narang98b}. This model accounts for
only enzyme induction and dilution, the two processes that occur in
almost all systems of mixed-substrate growth. Yet, it captures all
the batch growth patterns described above, and its extension to continuous
cultures predicts all the steady states observed in chemostats~\citep{narang98a}.
Here, we show that the minimal model also provides mechanistic explanations
for the foregoing questions. Specifically, we find that

\begin{enumerate}
\item If the induction kinetics are hyperbolic, the maximum specific growth
rate on the preferred substrate is always higher than that on the
less preferred substrate. The manifestation of this correlation in
a minimal model containing no regulation suggests that its existence
is not due to goal-oriented regulatory mechanisms, an assumption that
lies at the heart of models based on optimality principles~\citep{Mahadevan2002,kompala86,ramakrishna96}.
It is an intrinsic property resulting from the kinetics of enzyme
induction and dilution. We also find that the correlation can be violated
when the induction kinetics are sigmoidal, and that the dynamics of
these offending cases are consistent with the data in the literature.
\item The existence of comparable single-substrate growth rates is not sufficient
for simultaneous consumption. This agrees with the data described
above. However, we find that this condition is not necessary either.
This is because the occurrence of simultaneous consumption depends
not only on the relative growth rates, but also on the saturation
constants for induction. If these saturation constants are small,
there is simultaneous consumption, regardless of the relative growth
rates.
\end{enumerate}
We show, furthermore, that the classification of the substrate consumption
patterns predicted by the model explains the phenotypes of several
mutants. The most striking phenotype is the \emph{reversal of the
diauxie}, wherein both the wild-type and the mutant strains display
diauxic growth, but consume the substrates in opposite order. This
phenotype cannot be explained in terms of the standard molecular mechanisms,
but turns out to be a natural consequence of the minimal model.

\section{The model}

\begin{figure}
\begin{centering}\includegraphics[width=8cm,height=6cm]{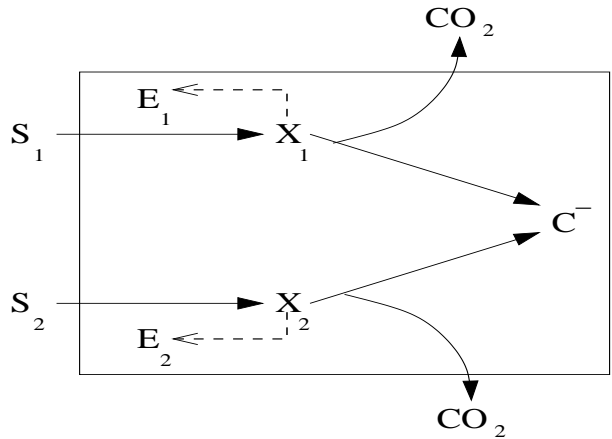}\par\end{centering}

\caption{\label{f:Scheme}Kinetic scheme of the minimal model~\citep{narang98b}. }
\end{figure}

Fig.~\ref{f:Scheme} shows the kinetic scheme of the minimal model.
Here, $S_{i}$ denotes the $i^{{\rm th}}$ exogenous substrate, $E_{i}$
denotes the transport enzyme for $S_{i}$, $X_{i}$ denotes internalized
$S_{i}$, and $C^{-}$ denotes \emph{all} intracellular components
except $E_{i}$ and $X_{i}$ (thus, it includes precursors, free amino
acids, and macromolecules).

In this work, attention will be confined to growth in batch cultures.
We assume that

\begin{enumerate}
\item The concentrations of the intracellular components, denoted $e_{i}$,
$x_{i}$, and $c^{-}$, are based on the dry weight of the cells (g
per g dry weight of cells, i.e., g~gdw$^{-1}$). The concentrations
of the exogenous substrate and cells, denoted $s_{i}$ and $c$, are
based on the volume of the reactor (g/L and gdw/L, respectively).
The rates of all the processes are based on the dry weight of the
cells (g~gdw$^{-1}$~h$^{-1}$). We shall use the term \emph{specific
rate} to emphasize this point.\\
The choice of these units implies that if the concentration of any
intracellular component, $Z$, is $z$ g~gdw$^{-1}$, then the evolution
of $z$ in batch cultures is given by\[
\frac{dz}{dt}=r_{z}^{+}-r_{z}^{-}-\left(\frac{1}{c}\frac{dc}{dt}\right)z\]
where $r_{z}^{+}$ and $r_{z}^{-}$ denote the specific rates of synthesis
and degradation of $Z$ in g~gdw$^{-1}$~h$^{-1}$.
\item The transport and peripheral catabolism of $S_{i}$ is catalyzed by
the {}``lumped'' system of peripheral enzymes, $E_{i}$. The specific
uptake rate of $S_{i}$, denoted $r_{s,i}$, follows the modified
Michaelis-Menten kinetics, $r_{s,i}\equiv V_{s,i}e_{i}s_{i}/(K_{s,i}+s_{i})$.
\item Part of the internalized substrate, denoted $X_{i}$, is converted
to $C^{-}$. The remainder is oxidized to ${\rm CO_{2}}$ in order
to generate energy.

\begin{enumerate}
\item The conversion of $X_{i}$ to $C^{-}$ and ${\rm CO_{2}}$ follows
first-order kinetics, i.e., $r_{x,i}\equiv k_{x,i}x_{i}$.
\item The fraction of $X_{i}$ converted to $C^{-}$ is a constant (parameter),
denoted $Y_{i}$. Thus, the specific rate of synthesis of $C^{-}$
from $X_{i}$ is $Y_{i}r_{x,i}$.%
\footnote{The so-called \emph{conservative} substrates, such as nitrogen and
phosphorus sources, are completely assimilated (as opposed to carbon
sources, which are partially oxidized to generate energy). During
growth on mixtures of such substrates, $Y_{i}=1$ for both the substrates.%
}
\end{enumerate}
\item The internalized substrate also induces the synthesis of $E_{i}$.

\begin{enumerate}
\item The specific synthesis rate of $E_{i}$ follows Hill kinetics, i.e.,
$r_{e,i}\equiv V_{e,i}x_{i}^{n_{i}}/(K_{e,i}^{n_{i}}+x_{i}^{n_{i}})$,
where $n_{i}=1\textnormal{ or }2$. Kinetic analysis of the data shows
that enzyme induction can be hyperbolic ($n_{i}=1$) or sigmoidal
($n_{i}=2$)~\citep{yagil71}.\\
By appealing to a molecular model of induction, we can express $n_{i}$,
$V_{e,i}$, and $K_{e,i}$ in terms of the parameters associated with
repressor-operator and repressor-inducer binding. It is shown in Appendix~\ref{a:MolecularParameters}
that the Yagil \& Yagil model of induction implies that $n_{i}$ is
the number of inducer molecules that bind to 1 repressor molecule.
Furthermore, if the enzyme is \emph{inducible},\begin{equation}
V_{e,i}=\nu_{e,i}o_{t,i},\; K_{e,i}^{n_{i}}=\frac{K_{x,i}}{K_{o,i}}r_{t,i},\label{eq:MolecularParameters}\end{equation}
where $\nu_{e,i}$ is the enzyme synthesis rate per unit mass of operator;
$o_{t,i},r_{t,i}$ are the total concentrations of the operator and
repressor (g gdw$^{-1}$), respectively; and $K_{x,i},K_{o,i}$ are
the dissociation constants for repressor-inducer and repressor-operator
binding, respectively.
\item The synthesis of the enzymes occurs at the expense of the biosynthetic
constituents, $C^{-}$.
\item Enzyme degradation is negligibly small.
\end{enumerate}
\end{enumerate}
Given these assumptions, the mass balances yield the equations\begin{align}
\frac{ds_{i}}{dt} & =-r_{s,i}c,\; r_{s,i}\equiv V_{s,i}e_{i}\frac{s_{i}}{K_{s,i}+s_{i}},\label{eq:NsO}\\
\frac{dx_{i}}{dt} & =r_{s,i}-r_{x,i}-\left(\frac{1}{c}\frac{dc}{dt}\right)x_{i},r_{x,i}\equiv k_{x,i}x_{i},\label{eq:NxO}\\
\frac{de_{i}}{dt} & =r_{e,i}-\left(\frac{1}{c}\frac{dc}{dt}\right)e_{i},\; r_{e,i}\equiv V_{e,i}\frac{x^{n_{i}}}{K_{e,i}^{n_{i}}+x^{n_{i}}},\label{eq:NeO}\\
\frac{dc^{-}}{dt} & =(Y_{1}r_{x,1}+Y_{2}r_{x,2})-(r_{e,1}+r_{e,2})-\left(\frac{1}{c}\frac{dc}{dt}\right)c^{-}.\label{eq:NcMO}\end{align}
It is shown in Appendix~\ref{a:NmodelEquations} that since $x_{1}+x_{2}+e_{1}+e_{2}+c^{-}=1$,
Eqs.~(\ref{eq:NxO})--(\ref{eq:NcMO}) implicitly define the specific
growth rate, denoted $r_{g}$, and the evolution of the cell density
via the relations\begin{equation}
\frac{dc}{dt}=r_{g}c,\; r_{g}\equiv\sum_{i=1}^{2}r_{s,i}-\sum_{i=1}^{2}(1-Y_{i})r_{x,i}.\label{eq:NrG}\end{equation}
Furthermore, since $x_{i}$ is small, it attains quasisteady state
on a time scale of seconds, thus resulting in the simplified equations\begin{align}
\frac{ds_{i}}{dt} & =-r_{s,i}c,\label{eq:Ns}\\
x_{i} & \approx\frac{V_{s,i}e_{i}s_{i}/(K_{s,i}+s_{i})}{k_{x,i}},\label{eq:Nx}\\
\frac{de_{i}}{dt} & =r_{e,i}-r_{g}e_{i},\; r_{e,i}\approx V_{e,i}\frac{[e_{i}s_{i}/(K_{s,i}+s_{i})]^{n_{i}}}{\bar{K}_{e,i}^{n_{i}}+[e_{i}s_{i}/(K_{s,i}+s_{i})]^{n_{i}}},\;\bar{K}_{e,i}\equiv K_{e,i}\frac{k_{x,i}}{V_{s,i}},\label{eq:Ne}\\
\frac{dc}{dt} & =r_{g}c,\; r_{g}\approx Y_{1}r_{s,1}+Y_{2}r_{s,2},\label{eq:Nc}\\
c^{-} & =1-x_{1}-x_{2}-e_{1}-e_{2},\label{eq:NcM}\end{align}
where (\ref{eq:Nx}) is obtained from the quasisteady state relation,
i.e., $0\approx r_{s,i}-r_{x,i}$.

We are particularly interested in the dynamics of the peripheral enzymes
during the first exponential growth phase, since it is these finite-time
dynamics that determine the substrate utilization pattern. If the
peripheral enzymes for one of the substrates vanish during this period,
there is diauxic growth; if the peripheral enzymes for both substrates
persist, there is simultaneous substrate utilization.

It turns out that the motion of the enzymes during the first exponential
growth phase is governed by only two equations. To see this, observe
that during the first exponential growth phase, both substrates are
in excess, i.e., $s_{i}\gg K_{s,i}$. Hence, even though the exogenous
substrate concentrations are decreasing, the transport enzymes see
a quasiconstant environment ($s_{i}/(K_{s,i}+s_{i})\approx$1), and
approach the quasisteady state levels corresponding to exponential
(balanced) growth. This motion is approximated by the equations \begin{align}
\frac{de_{1}}{dt} & =V_{e,1}\frac{e_{1}^{n_{1}}}{\bar{K}_{e,1}^{n_{1}}+e_{1}^{n_{1}}}-\left(Y_{1}V_{s,1}e_{1}+Y_{2}V_{s,2}e_{2}\right)e_{1},\label{eq:E1}\\
\frac{de_{2}}{dt} & =V_{e,2}\frac{e_{2}^{n_{2}}}{\bar{K}_{e,2}^{n_{2}}+e_{2}^{n_{2}}}-\left(Y_{1}V_{s,1}e_{1}+Y_{2}V_{s,2}e_{2}\right)e_{2},\label{eq:E2}\end{align}
 obtained from~(\ref{eq:Ne}) by replacing $s_{i}/(K_{s,i}+s_{i})$
with~1. We shall refer to these as the \emph{reduced} equations.
It should be emphasized that the steady states of the reduced equations
are \emph{quasisteady states} of the full system of equations (see~\citealp{narang97c}
for a rigorous derivation of the reduced equations).

The reduced equations are formally similar to the equations of the
standard Lotka-Volterra model for two competing species, namely,\begin{align}
\frac{dN_{1}}{dt} & =r_{1}N_{1}(1-a_{11}N_{1}-a_{12}N_{2}),\label{eq:LV1}\\
\frac{dN_{2}}{dt} & =r_{2}N_{2}(1-a_{21}N_{1}-a_{22}N_{2}),\label{eq:LV2}\end{align}
where $N_{i}$ is the population density of the $i^{{\rm th}}$ species,
$r_{i}$ is the (unrestricted) specific growth rate of the $i^{{\rm th}}$
species in the absence of any competition, and $a_{i1},a_{i2}$ are
parameters that quantify the reduction of the unrestricted specific
growth rate due to intra- and inter-specific competition~\citep{murray}.
Thus, enzyme induction is the correlate of unrestricted growth, and
the two dilution terms are the correlates of intra- and inter-specific
competition. In what follows, we shall constantly appeal to this dynamical
analogy.

\begin{figure}
\begin{centering}\includegraphics[width=7cm,height=7.5cm]{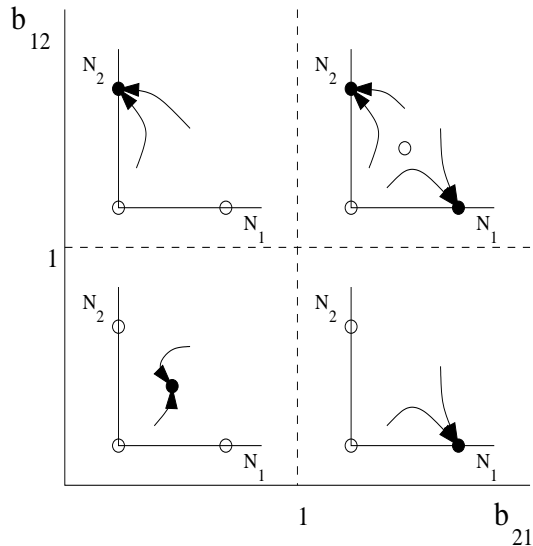}\par\end{centering}

\caption{\label{f:GlobalDynamicsLV}Classification of the global dynamics
for the standard Lotka-Volterra model. The full and open circles show
stable and unstable steady states, respectively. }
\end{figure}

The dynamics of the standard Lotka-Volterra model are well understood.
Indeed, the bifurcation diagram of the model is completely determined
by the two dimensionless parameters, $b_{21}\equiv a_{21}/a_{11}$
and $b_{12}\equiv a_{12}/a_{22}$~(Fig.~\ref{f:GlobalDynamicsLV}).
These parameters characterize the extent to which each species inhibits
the other species relative to the extent to which it inhibits itself.
Both species coexist precisely when they inhibit themselves more than
they inhibit the other species, i.e., $b_{21},b_{12}<1$. Under all
other conditions, coexistence is impossible. If the interaction between
the species is asymmetric ($b_{21}<1,b_{12}>1$ or $b_{21}>1,b_{12}<1$),
one of them is rendered extinct (species~1 and~2, respectively).
If both species inhibit the other species more than they inhibit themselves,
i.e. $b_{21},b_{12}>1$, the outcome depends on the initial population
densities.

Given the formal similarity of the reduced equations to the Lotka-Volterra
model, we expect them to display {}``extinction'' and {}``coexistence''
dynamics. Importantly, these dynamics have simple biological interpretations.
Extinction of one of the enzymes corresponds to diauxic growth, and
coexistence of both enzymes corresponds to simultaneous consumption.
It is therefore clear that the bifurcation diagram for the reduced
equations is a useful analytical tool. It furnishes a classification
of the substrate consumption patterns, which can then be used to systematically
address the questions posed in the Introduction. Our first goal is
to construct this bifurcation diagram.

To minimize the number of parameters in the bifurcation diagram, we
rescale the reduced equations by defining the dimensionless variables\[
\epsilon_{i}\equiv\frac{e_{i}}{\sqrt{V_{e,i}/(Y_{i}V_{s,i})}},\;\tau\equiv t\sqrt{V_{e,1}Y_{1}V_{s,1}}\;.\]
The choice of the reference variables in this scaling is suggested
by the following fact: $\sqrt{V_{e,i}/(Y_{i}V_{s,i})}$ and $\sqrt{V_{e,i}Y_{i}V_{s,i}}$
are upper bounds for the enzyme level and maximum specific growth
rate attained during single-substrate exponential growth on saturating
concentrations of $S_{i}$. Indeed, under these conditions, the mass
balance for $E_{i}$ becomes\[
0=V_{e,i}\frac{e_{i}^{n_{i}}}{\bar{K}_{e,i}^{n_{i}}+e_{i}^{n_{i}}}-Y_{i}V_{s,i}e_{i}^{2}.\]
Hence, $e_{i}<\sqrt{V_{e,i}/(Y_{i}V_{s,i})}$, and the maximum specific
growth rate on $S_{i}$, denoted $r_{g,i}^{{\rm max}}$, satisfies
the relation\[
r_{g,i}^{{\rm max}}\approx Y_{i}V_{s,i}e_{i}<Y_{i}V_{s,i}\sqrt{\frac{V_{e,i}}{Y_{i}V_{s,i}}}=\sqrt{Y_{i}V_{s,i}V_{e,i}}.\]
 The above scaling yields the dimensionless reduced equations\begin{align}
\frac{d\epsilon_{1}}{d\tau} & =\frac{\epsilon_{1}^{n_{1}}}{\kappa_{1}^{n_{1}}+\epsilon_{1}^{n_{1}}}-\left(\epsilon_{1}+\alpha\epsilon_{2}\right)\epsilon_{1},\label{eq:scaledE1}\\
\frac{d\epsilon_{2}}{d\tau} & =\alpha\frac{\epsilon_{2}^{n_{2}}}{\kappa_{2}^{n_{2}}+\epsilon_{2}^{n_{2}}}-\left(\epsilon_{1}+\alpha\epsilon_{2}\right)\epsilon_{2},\label{eq:scaledE2}\end{align}
with dimensionless parameters\begin{eqnarray}
\kappa_{i} & \equiv & \frac{\bar{K}_{e,i}}{\sqrt{V_{e,i}/(Y_{i}V_{s,i})}}=K_{e,i}k_{x,i}\sqrt{\frac{Y_{i}}{V_{s,i}V_{e,i}}}=\frac{K_{x,i}r_{t,i}k_{x,i}}{K_{o,i}}\sqrt{\frac{Y_{i}}{V_{s,i}V_{e,i}}},\label{eq:KappaI}\\
\alpha & \equiv & \;\frac{\sqrt{V_{e,2}Y_{2}V_{s,2}}}{\sqrt{V_{e,1}Y_{1}V_{s,1}}}.\label{eq:Alpha}\end{eqnarray}
These dimensionless parameters have simple biological interpretations.
We can view $\kappa_{i}$ as a dimensionless saturation constant for
induction, and $\alpha$ as a measure of the maximum specific growth
rate on $S_{2}$ relative to that on $S_{1}$.

\section{Results}

We wish to construct the bifurcation diagram for Eqs.~(\ref{eq:scaledE1})--(\ref{eq:scaledE2}).
Since limit cycles are impossible in Lotka-Volterra models for competing
species~\citep{hirschEcol}, it suffices to determine the steady
states and their stability.

Eqs.~(\ref{eq:scaledE1})--(\ref{eq:scaledE2}) admit at most four
types of steady states: The trivial steady state ($\epsilon_{1}=\epsilon_{2}=0$),
the semitrivial steady states ($\epsilon_{1}>0,\epsilon_{2}=0$ and
$\epsilon_{1}=0,\epsilon_{2}>0$), and the nontrivial steady state,
$\epsilon_{1},\epsilon_{2}>0$. We denote these steady states by $E_{00}$,
$E_{10}$, $E_{01}$, and $E_{11}$, respectively.

We shall consider two cases: $n_{1}=n_{2}=1$ and $n_{1}=2$, $n_{2}=1$.
The second case will serve to show the qualitative changes engendered
by sigmoidal induction kinetics.

\subsection{Case 1 ($n_{1}=n_{2}=1$)}

In this case, the scaled equations are\begin{align*}
\frac{d\epsilon_{1}}{dt} & =\frac{\epsilon_{1}}{\kappa_{1}+\epsilon_{1}}-\left(\epsilon_{1}+\alpha\epsilon_{2}\right)\epsilon_{1},\\
\frac{d\epsilon_{2}}{dt} & =\alpha\frac{\epsilon_{2}}{\kappa_{2}+\epsilon_{2}}-\left(\epsilon_{1}+\alpha\epsilon_{2}\right)\epsilon_{2}.\end{align*}
The bifurcation diagrams for these equations are shown in Fig.~\ref{f:BDcase1}.
They were inferred from the following facts derived in Appendix~\ref{a:StabilityAnalysis1}.

\begin{figure}[t]
\begin{centering}\subfigure[]{\includegraphics[width=6cm,height=6cm]{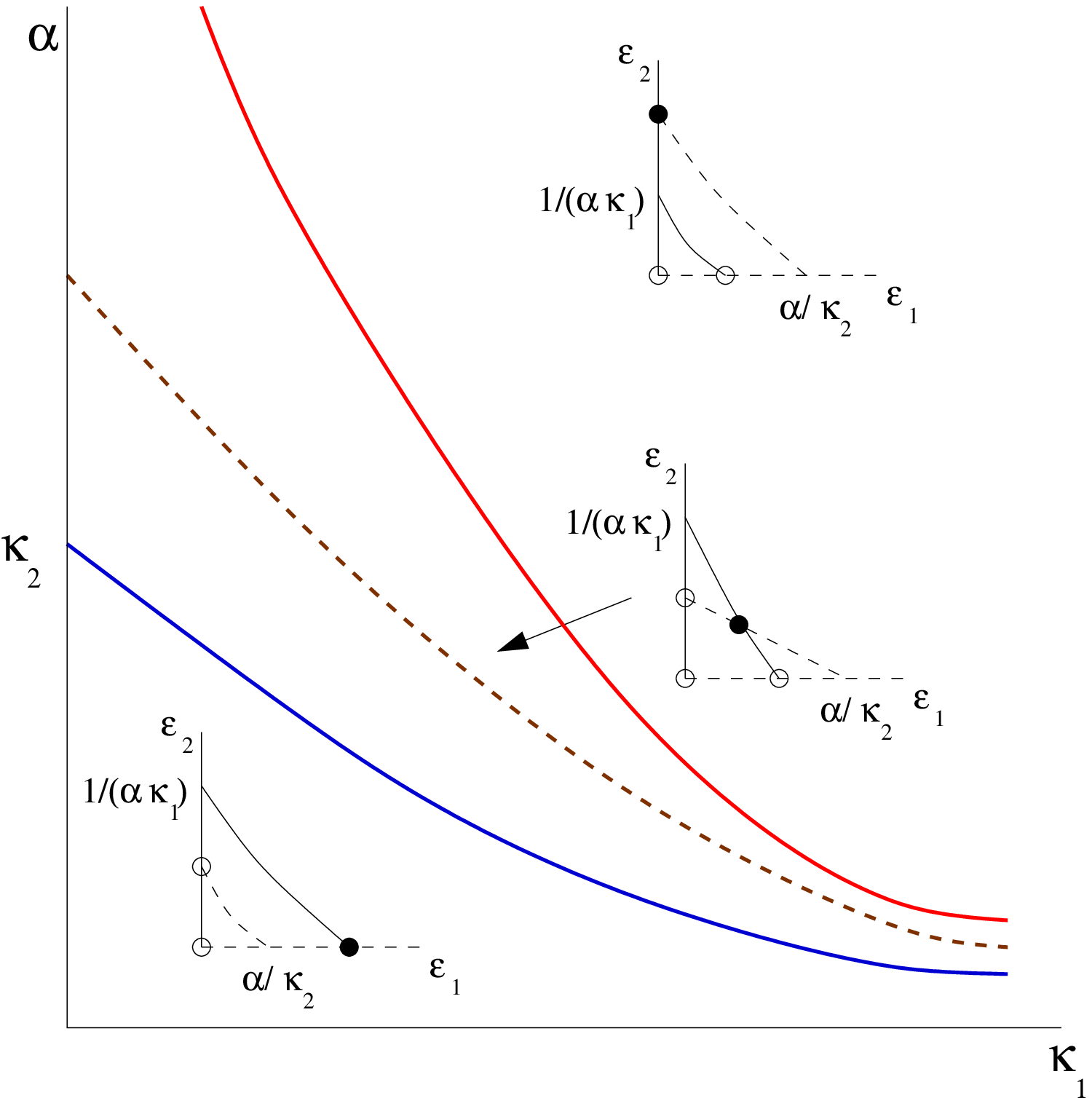}}\hspace*{0.5cm}\subfigure[]{\includegraphics[width=6cm,height=6cm]{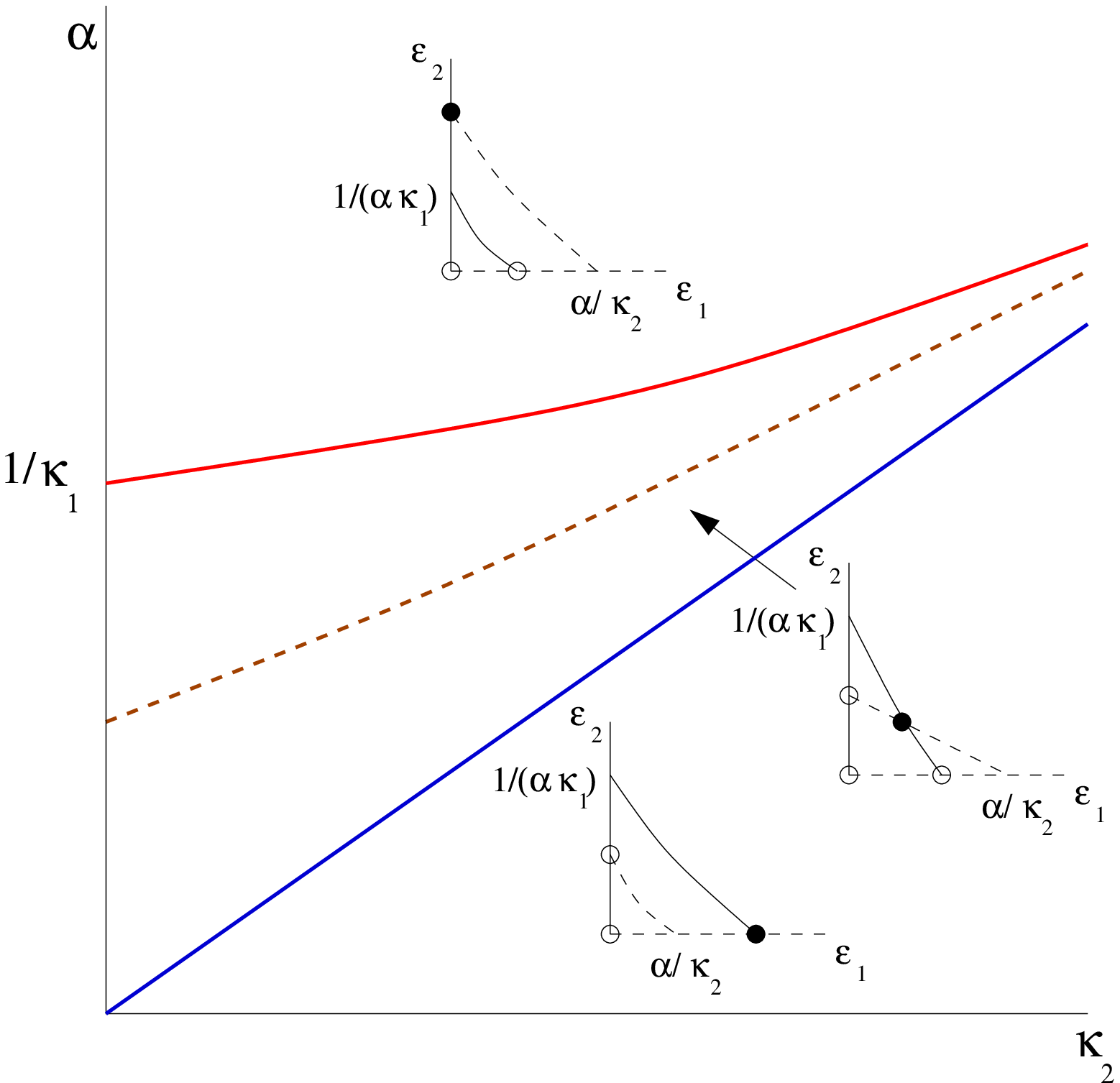}}\par\end{centering}

\caption{\label{f:BDcase1}Bifurcation diagrams for the case $n_{1}=n_{2}=1$
at (a)~fixed $\kappa_{2}>0$ and (b)~fixed $\kappa_{1}>0$. In the
phase portraits, the nullclines for $\epsilon_{1}$ and $\epsilon_{2}$
are represented by full and dashed lines, respectively; stable and
unstable steady states are represented by full and open circles, respectively.
The graphs of $\alpha_{*}$, $\alpha^{*}$, and $\alpha_{g}$ are
represented by blue, red, and dashed brown curves.}
\end{figure}

\begin{enumerate}
\item The trivial steady, $E_{00}$, always exists (for all $\alpha,\kappa_{1},\kappa_{2}>0$),
but it is always unstable (as a node).
\item The semitrivial steady state, $E_{10}$, always exists. It is (uniquely)
given by\[
\epsilon_{1}=\frac{-\kappa_{1}+\sqrt{\kappa_{1}^{2}+4}}{2},\;\epsilon_{2}=0,\]
and is stable (as a node) precisely if $\left.\epsilon_{1}\right|_{E_{10}}$
exceeds $\alpha/\kappa_{2}$, the $\epsilon_{1}$-intercept of the
nontrivial nullcline for $\epsilon_{2}$.%
\footnote{The nullclines for $\epsilon_{i}$ refer to the locus of points on
the $\epsilon_{1}\epsilon_{2}$-plane at which $d\epsilon_{i}/dt=0$.
In the case of the reduced equations, the nullclines for $\epsilon_{i}$
consist of two curves. One of these curves is the \emph{trivial} nullcline,
$\epsilon_{i}=0$; the other curve is called the \emph{nontrivial}
nullcline.%
} That is\begin{equation}
\left.\epsilon_{1}\right|_{E_{10}}>\frac{\alpha}{\kappa_{2}}\Leftrightarrow\alpha<\alpha_{*}(\kappa_{1},\kappa_{2})\equiv\frac{\kappa_{2}\left(-\kappa_{1}+\sqrt{\kappa_{1}^{2}+4}\right)}{2}.\label{eq:E10StabilityCase1}\end{equation}

\item The steady state, $E_{01}$, always exists. It is given by \[
\epsilon_{1}=0,\;\epsilon_{2}=\frac{-\kappa_{2}+\sqrt{\kappa_{2}^{2}+4}}{2},\]
and it is stable (as a node) precisely if $\left.\epsilon_{2}\right|_{E_{01}}$
exceeds $1/(\alpha\kappa_{1})$, the $\epsilon_{2}$-intercept of
the nontrivial nullcline for $\epsilon_{1}$, i.e.,\begin{equation}
\left.\epsilon_{2}\right|_{E_{01}}>\frac{1}{\alpha\kappa_{1}}\Leftrightarrow\alpha>\alpha^{*}(\kappa_{1},\kappa_{2})\equiv\frac{2}{\kappa_{1}\left(-\kappa_{2}+\sqrt{\kappa_{2}^{2}+4}\right)}.\label{eq:E01StabilityCase1}\end{equation}

\item The surface of $\alpha_{*}(\kappa_{1},\kappa_{2})$ lies below the
surface of $\alpha^{*}(\kappa_{1},\kappa_{2})$, i.e., \begin{equation}
\alpha_{*}(\kappa_{1},\kappa_{2})<\alpha^{*}(\kappa_{1},\kappa_{2})\label{eq:SurfaceDispositionCase1}\end{equation}
for all $\kappa_{1},\kappa_{2}>0$. The notation was chosen to reflect
this fact: The functions, $\alpha_{*}(\kappa_{1},\kappa_{2})$ and
$\alpha^{*}(\kappa_{1},\kappa_{2})$, represent the lower and upper
surfaces of the bifurcation diagram.
\item The steady state, $E_{11}$, exists if and only if both $E_{10}$
and $E_{01}$ are unstable, i.e., \begin{equation}
\alpha_{*}(\kappa_{1},\kappa_{2})<\alpha<\alpha^{*}(\kappa_{1},\kappa_{2}).\label{eq:E11StabilityCase1}\end{equation}
It is unique and stable whenever it exists.
\end{enumerate}
The bifurcation diagrams imply the following classification of the
substrate utilization patterns.

\begin{enumerate}
\item If $\alpha<\alpha_{*}(\kappa_{1},\kappa_{2})$, only $E_{10}$ is
stable, which corresponds to preferential consumption of $S_{1}$.
\item If $\alpha_{*}(\kappa_{1},\kappa_{2})<\alpha<\alpha^{*}(\kappa_{1},\kappa_{2})$,
only $E_{11}$ is stable, and there is simultaneous consumption of
$S_{1}$ and $S_{2}$.
\item If $\alpha>\alpha^{*}(\kappa_{1},\kappa_{2})$, only $E_{01}$ is
stable, which corresponds to preferential consumption of $S_{2}$.
\end{enumerate}
Thus, the surfaces of $\alpha_{*}(\kappa_{1},\kappa_{2})$ and $\alpha^{*}(\kappa_{1},\kappa_{2})$
delineate the boundaries of the substrate consumption patterns.%
\footnote{An analogous classification is also obtained when the model is extended
to continuous cultures~\citep[Fig.~10]{narang98a}. However, the
control parameters consist of the dilution rate and feed concentrations
(rather than the physiological parameters, $\alpha$, $\kappa_{1}$,
and $\kappa_{2}$).%
}

\begin{figure}[t]
\begin{centering}\subfigure[]{\includegraphics[width=7cm,height=5cm]{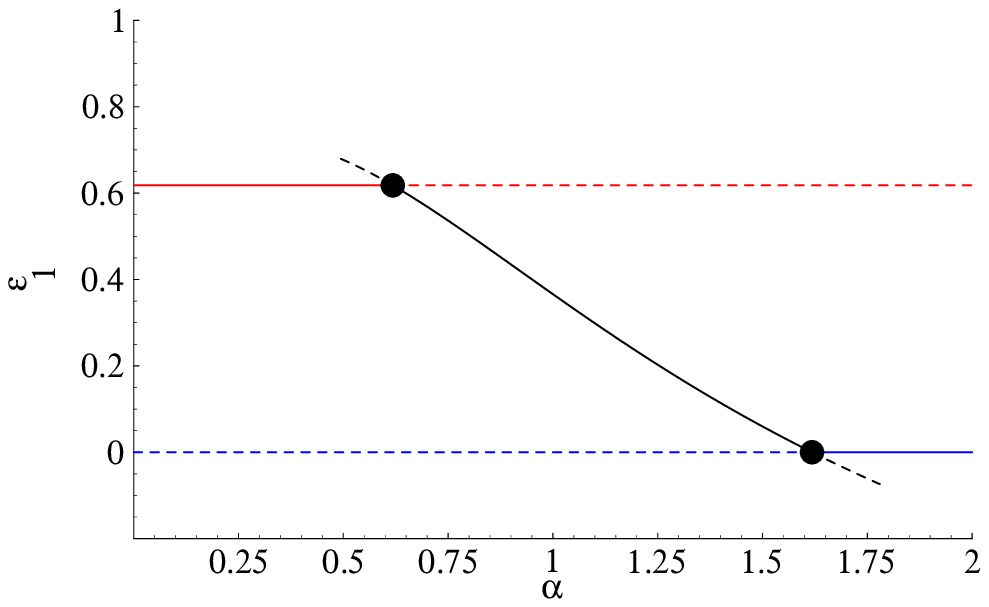}}\subfigure[]{\includegraphics[width=7cm,height=5cm]{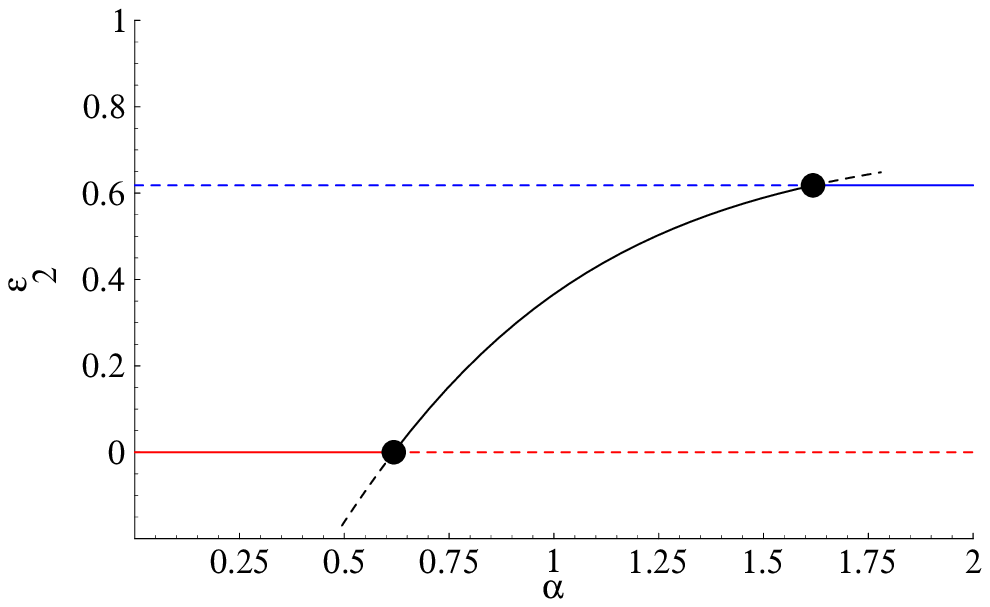}}\par\end{centering}

\caption{\label{f:CDcase1}The bifurcation diagram obtained when $\alpha$
is increased at $\kappa_{1}=\kappa_{2}=1$. The red, blue, and black
curves represent the locus of $E_{10}$, $E_{01}$, and $E_{11}$,
respectively. The curves are full (resp., dashed) if the steady state
is stable (resp., unstable). The bifurcation points at $\alpha=\alpha_{*}(1,1)=0.62$
and $\alpha=\alpha^{*}(1,1)=1.62$ are represented by full circles.}
\end{figure}

If the point, $(\kappa_{1},\kappa_{2},\alpha)$, crosses either one
of these boundaries, there is an abrupt transition in the substrate
consumption pattern due to transcritical bifurcations. This becomes
evident if $\alpha$ is increased at any fixed $\kappa_{1},\kappa_{2}>0$
(Fig.~\ref{f:CDcase1}). At $\alpha=\alpha_{*}(\kappa_{1},\kappa_{2})$,
the substrate consumption pattern switches from preferential consumption
of $S_{1}$ to simultaneous consumption of $S_{1}$ and $S_{2}$ through
a transcritical bifurcation in which $E_{10}$ (red curve) yields
its stability to $E_{11}$ (black curve). As $\alpha$ is increased
further, there is another transition at $\alpha=\alpha^{*}(\kappa_{1},\kappa_{2})$
wherein simultaneous consumption switches to preferential consumption
of $S_{2}$ via a transcritical bifurcation involving the transfer
of stability from $E_{11}$ (black curve) to $E_{01}$ (blue curve).

We gain intuitive insight into the bifurcation diagram by considering
two limiting cases. Fig.~\ref{f:BDcase1} shows that if $\kappa_{1}$
or $\kappa_{2}$ are large, the curves for $\alpha_{*}$ and $\alpha^{*}$
converge, and simultaneous consumption is virtually impossible. In
contrast, if both $\kappa_{1}$ and $\kappa_{2}$ are small, there
is simultaneous consumption for almost all $\alpha$. To understand
these limiting cases, observe that when $\kappa_{1},\kappa_{2}$ are
large, Eqs.~(\ref{eq:scaledE1})--(\ref{eq:scaledE2}) are approximated
by the equations\begin{align*}
\frac{d\epsilon_{1}}{dt} & \approx\frac{1}{\kappa_{1}}\epsilon_{1}\left(1-\kappa_{1}\epsilon_{1}-\alpha\kappa_{1}\epsilon_{2}\right)\epsilon_{1},\\
\frac{d\epsilon_{2}}{dt} & \approx\frac{\alpha}{\kappa_{2}}\epsilon_{2}\left(1-\frac{\kappa_{2}}{\alpha}\epsilon_{1}-\kappa_{2}\epsilon_{2}\right)\epsilon_{2},\end{align*}
which are formally identical to the standard Lotka-Volterra model
with $a_{11}=\kappa_{1}$, $a_{12}=\alpha\kappa_{1}$, $a_{21}=\kappa_{2}/\alpha$,
and $a_{22}=\kappa_{2}$. However, there is an important difference.
The parameters, $b_{21}\equiv a_{21}/a_{11},\; b_{12}\equiv a_{12}/a_{22}$,
are not independent since $b_{21}=\kappa_{2}/(\alpha\kappa_{1})=1/b_{12}$.
But if $b_{12}$ and $b_{21}$ are restricted to the curve $b_{21}b_{12}=1$,
Fig.~\ref{f:GlobalDynamicsLV} implies that coexistence (i.e., simultaneous
consumption) is impossible: $E_{1}$ becomes extinct if $b_{21}<1$,
and $E_{2}$ becomes extinct if $b_{21}>1$. On the other hand, if
$\kappa_{1},\kappa_{2}$ are small, the enzyme synthesis rate is essentially
constant (quasi-constitutive). The enzymes therefore resist extinction,
and coexist for almost all $\alpha$.

\subsubsection{Dependence of substrate consumption pattern on genotype}

In the experimental literature, the influence of the physiological
parameters is often studied by altering the genetic make-up (genotype)
of the cells, and observing the resultant change in the substrate
consumption pattern (phenotype) of the cells. We show below that the
bifurcation diagrams are consistent with the phenotypic changes observed
in response to various genotypic alterations.

Before doing so, however, it is useful to note that in all the experiments
described below, the phenotype of the wild-type strain is preferential
consumption of a substrate (glucose, in most cases). Since Eqs.~(\ref{eq:scaledE1})
and (\ref{eq:scaledE2}) are formally the same, there is no loss of
generality in assuming that the preferred substrate is $S_{2}$, and
the parameters, $\kappa_{1},\kappa_{2},\alpha$, for the wild-type
strain lie in the region, $\alpha>\alpha^{*}$ (above the red curve
in Fig.~\ref{f:BDcase1}).

We begin by considering the cases in which the genetic perturbation
transforms the substrate consumption pattern from preferential to
simultaneous consumption.

\begin{figure}
\begin{centering}\subfigure[]{\includegraphics[width=7cm,height=5cm]{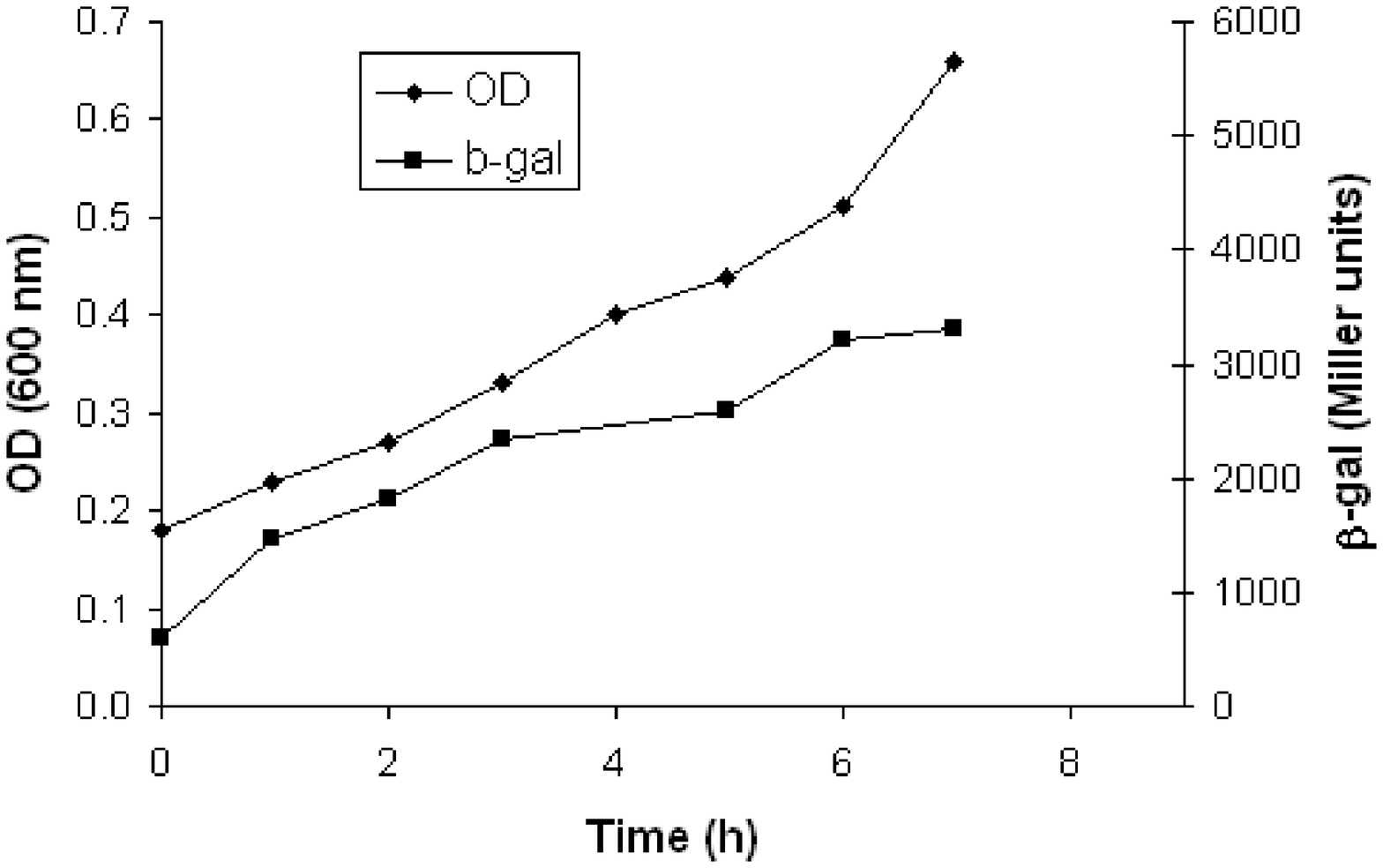}}\subfigure[]{\includegraphics[width=7cm,height=5cm]{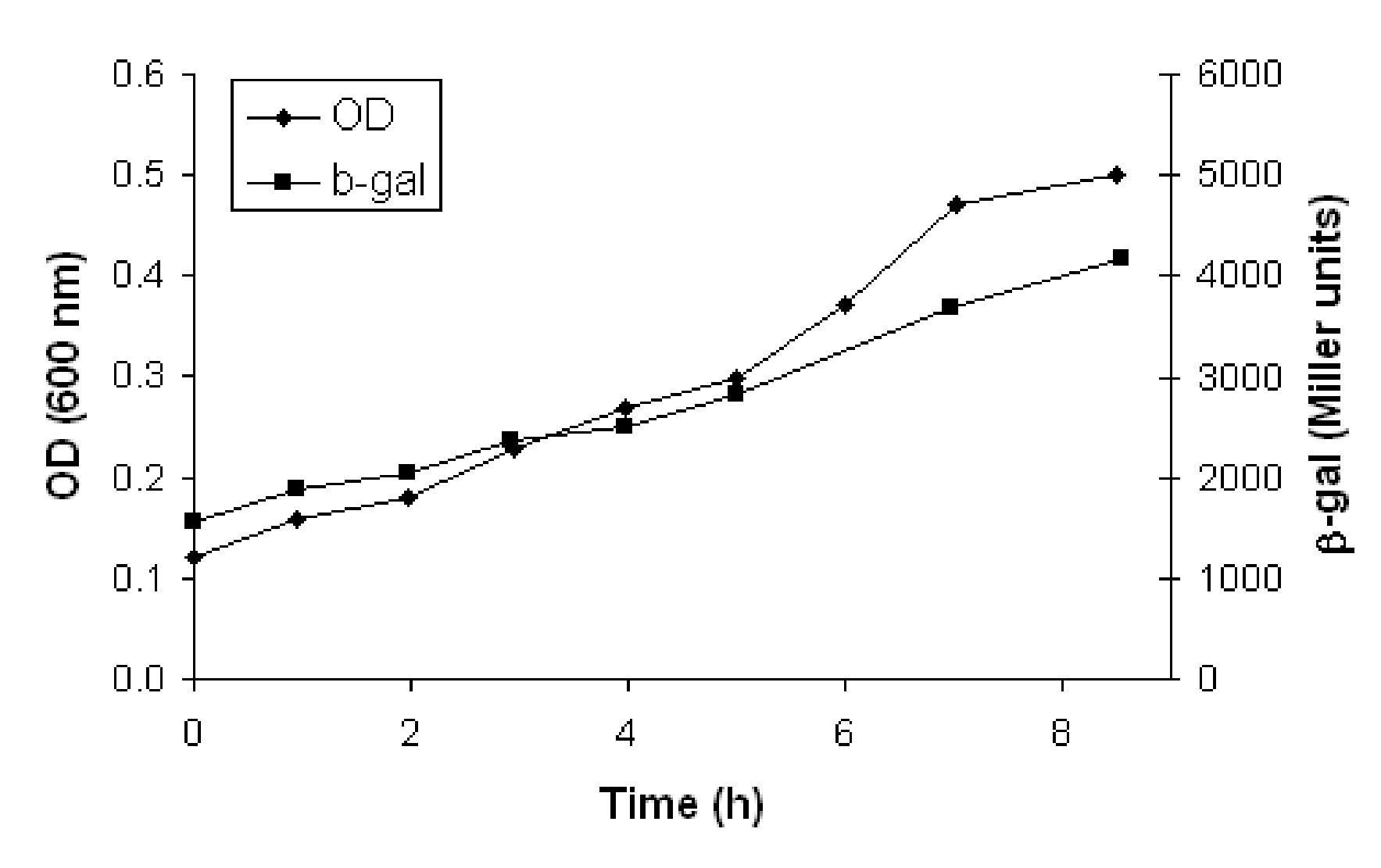}}\par\end{centering}

\caption{\label{f:RemovalOfDiauxie}In \emph{E. coli} PR166, synthesis of
$\beta$-galactosidase persists in the presence of glucose if (a)\emph{~lacY}
is overexpressed or (b)\emph{~crr}, the gene for enzyme IIA$^{\textnormal{glc}}$,
is deleted~\citep{kimata97}. The mutant cells in (b) grow on glucose
despite the absence of IIA$^{\textnormal{glc}}$ because of slow uptake
of glucose by the PTS enzymes for mannose.}
\end{figure}

In wild-type \emph{E. coli}, transcription of \emph{lac} is abolished
in the presence of glucose. However, mutants with lesions in the \emph{lac}
operator synthesize $\beta$-galactosidase even in the presence of
glucose~\citep{jacob61}. Thus, the mutation transforms the substrate
consumption pattern from preferential consumption of glucose to simultaneous
consumption of glucose and lactose. The very same phenotypic change
is also observed in mutants with a defective \emph{lacI}, the gene
encoding the \emph{lac} repressor~\citep{jacob61}. To explain these
phenotypic changes in terms of the model, observe that mutations in
the \emph{lac} operator or \emph{lacI} impair the \emph{lac} repressor-operator
binding, i.e., they increase the dissociation constant, $K_{o,1}$.
It follows from Eqs.~(\ref{eq:KappaI})--(\ref{eq:Alpha}) that $\kappa_{1}$
decreases at fixed $\kappa_{2}$ and $\alpha$. Inspection of Fig.~\ref{f:BDcase1}a
shows that such a change can shift the substrate consumption pattern
from preferential consumption of~$S_{2}$ to simultaneous consumption.

If \emph{lacY}, the gene encoding lactose permease, is overexpressed
in \emph{E.~coli} PR166, synthesis of $\beta$-galactosidase persists
in the presence of glucose~(Fig.~\ref{f:RemovalOfDiauxie}a). Now,
in the model, overexpression of \emph{lacY} corresponds to higher
$V_{e,1}$. It follows from Eqs.~(\ref{eq:KappaI})--(\ref{eq:Alpha})
that $\kappa_{1},\alpha$ decrease at fixed $\kappa_{2}$, and Fig.~\ref{f:BDcase1}a
implies that the observed phenotype is indeed feasible.

In \emph{E. coli} PR166, $\beta$-galactosidase is synthesized despite
the presence of glucose if \emph{crr}, the gene for enzyme IIA$^{\textnormal{glc}}$,
is deleted~(Fig.~\ref{f:RemovalOfDiauxie}a). Similarly, in the
wild-type strain, \emph{E.~coli} K12 W3110, glucose is consumed before
galactose. However, mutants with lesions in a gene encoding a transport
enzyme for glucose consume the two substrates simultaneously~\citep{Kamogawa1967}.
In these cases, the effect of the mutation is to decrease $V_{s,2}$,
so that $\kappa_{2}$ and $\alpha$ decrease at fixed $\kappa_{1}$.
It follows from Fig.~\ref{f:BDcase1}b that such a change could lead
to simultaneous consumption of the substrates.

\begin{figure}[t]
\begin{centering}\subfigure[]{\includegraphics[width=7cm,height=5cm]{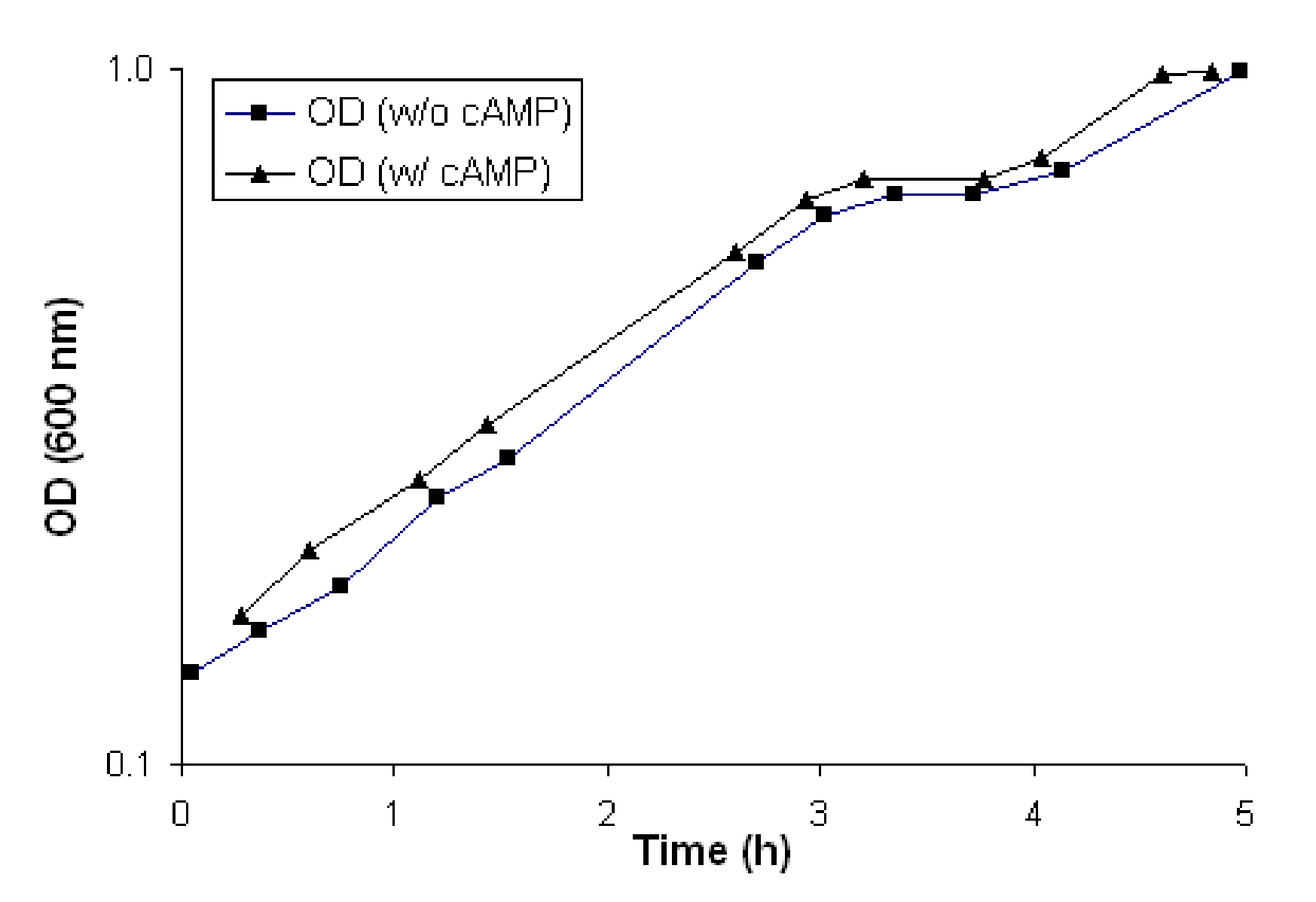}}\subfigure[]{\includegraphics[width=7cm,height=5cm]{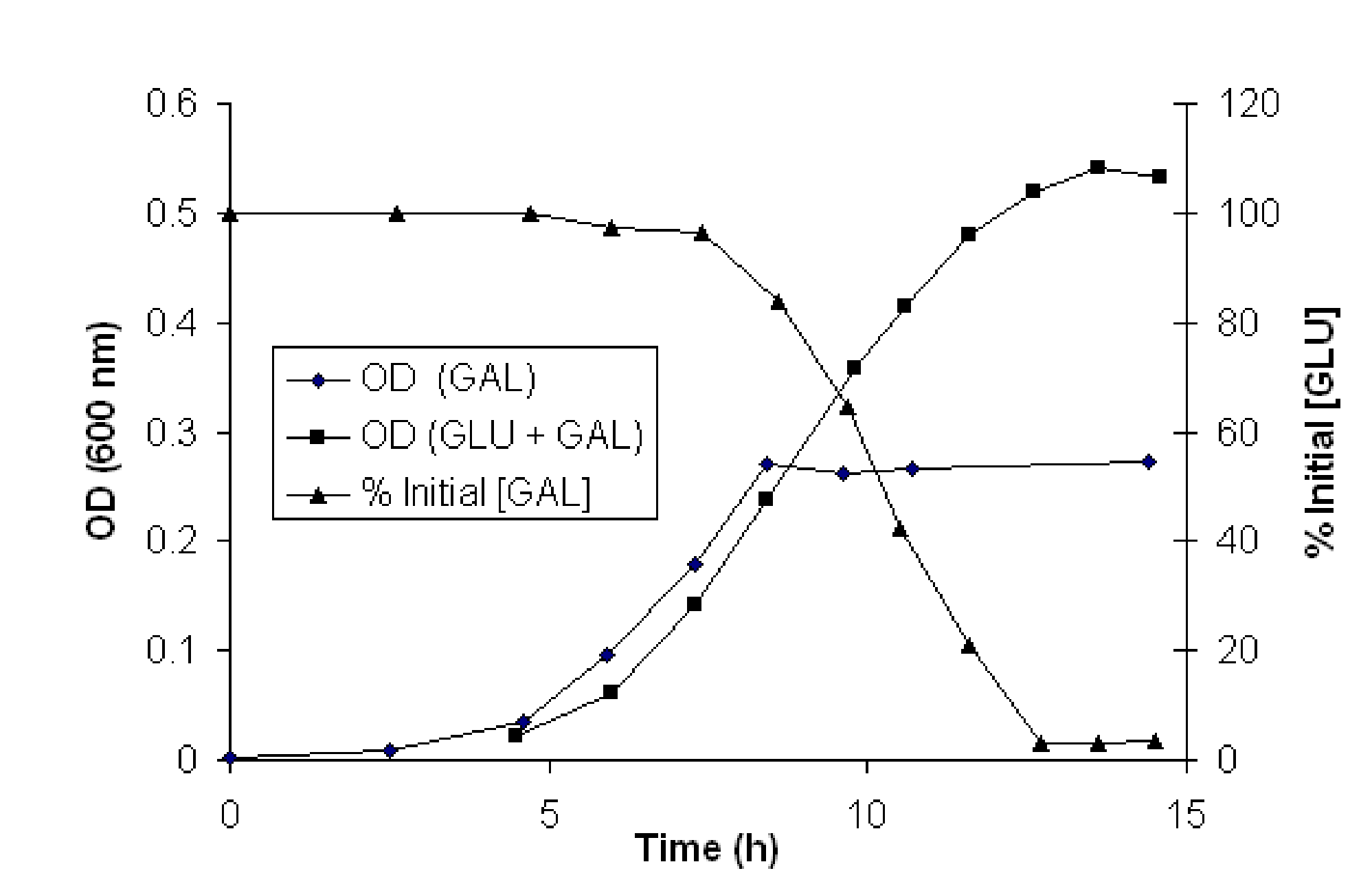}}\par\end{centering}

\begin{centering}\subfigure[]{\includegraphics[bb=0bp 0bp 577bp 354bp,width=7cm,height=5cm]{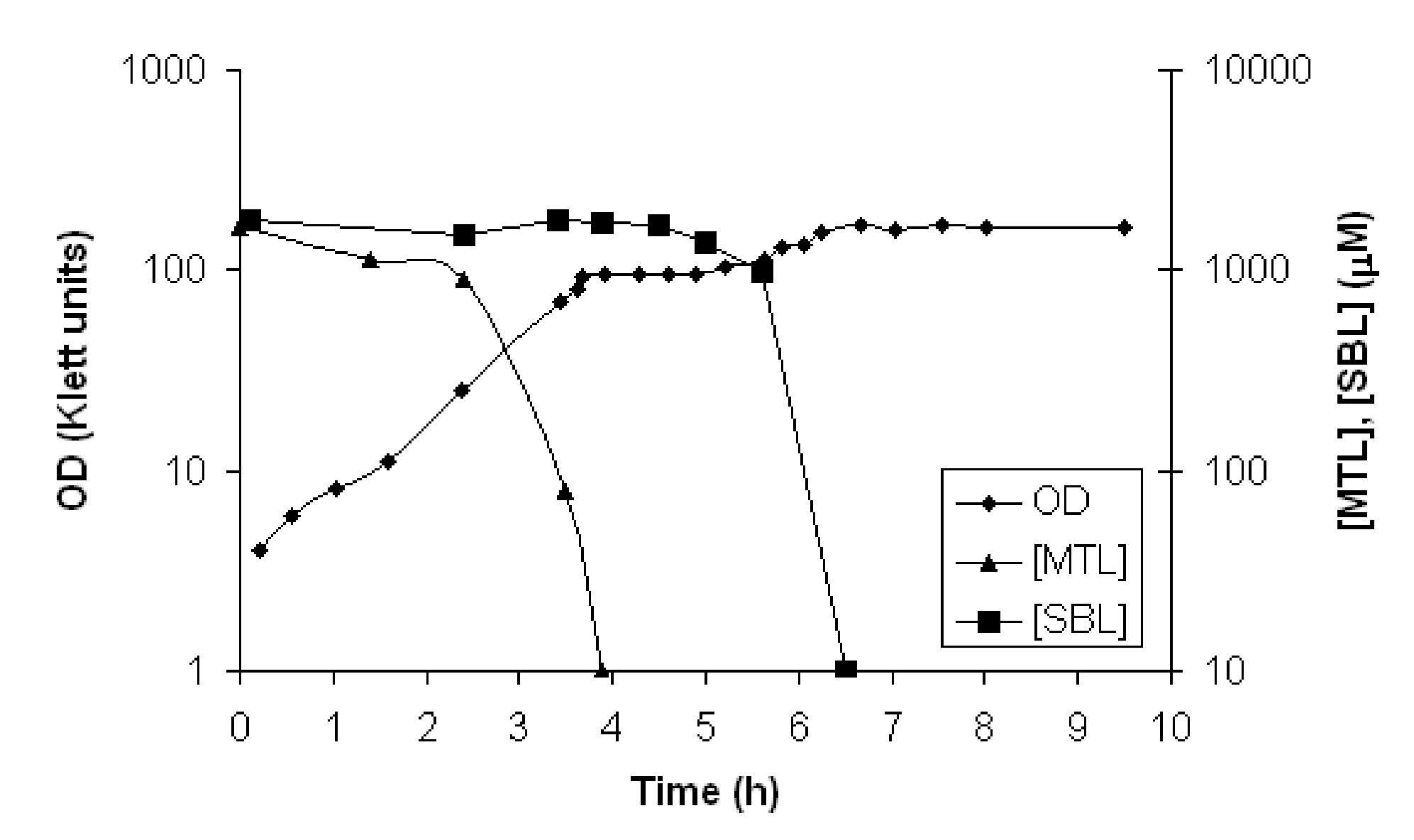}}\subfigure[]{\includegraphics[width=7cm,height=5cm]{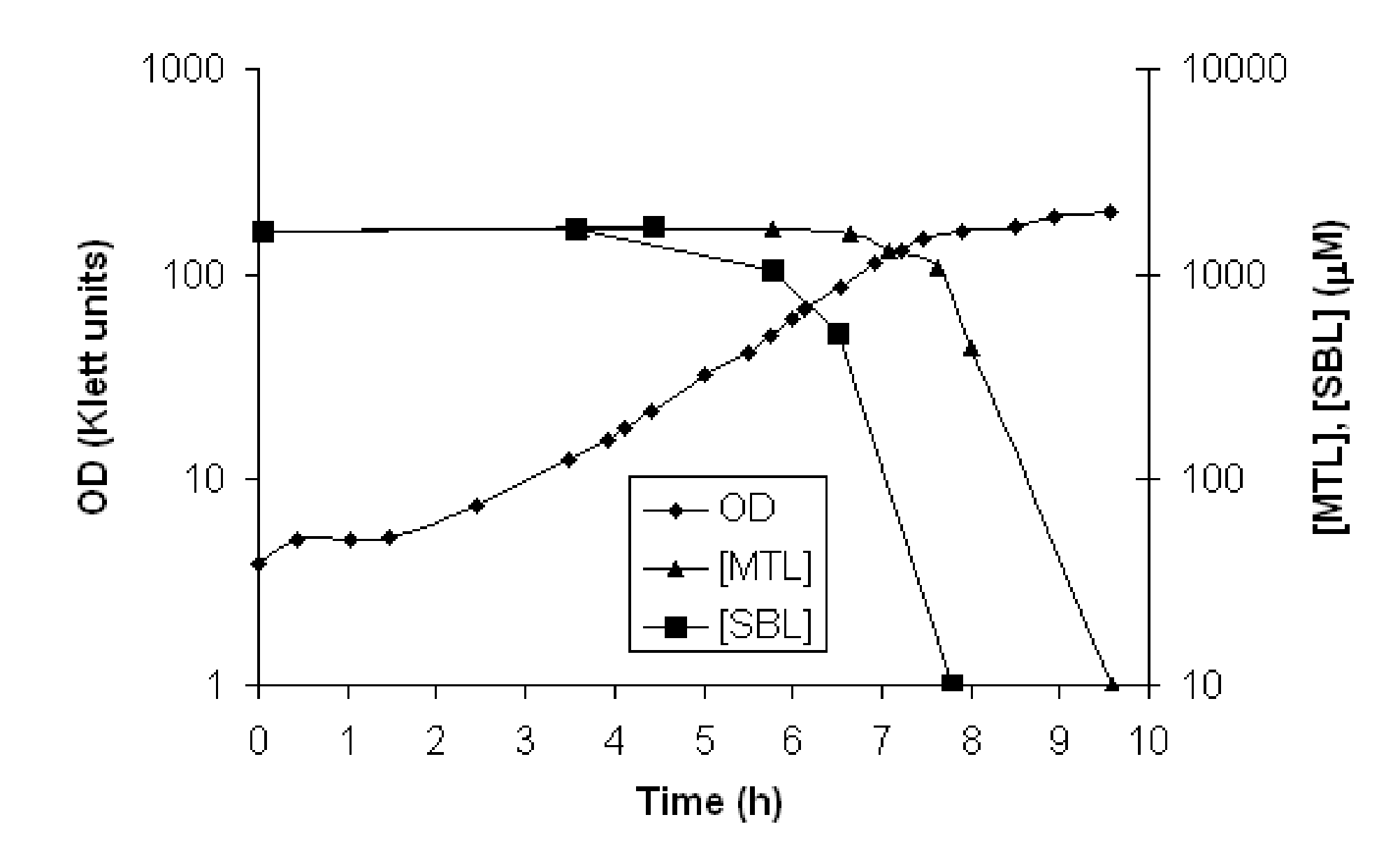}}\par\end{centering}

\caption{\label{f:DataCase1}Reversal of the diauxie in mutants of \emph{E.
coli}: \textbf{Upper panel:} (a) In strain Hfr3000, glucose is consumed
before galactose~\citep{Joseph1981}. The optical density (OD) shows
a pronounced diauxic lag, regardless of the presence of cAMP in the
culture. (b)~In the corresponding PTS-deficient strain, galactose
is consumed before glucose~\citep{ASENSIO1963}. Note that the evolution
of the OD during the first 8~h is the same during growth on galactose
{[}OD (GAL)] and glucose + galactose {[}OD (GLU+GAL)]. Furthermore,
there is no consumption of glucose during this period. \textbf{Lower
panel:} (c)~In strain 159, mannitol (MTL) is consumed before sorbitol
(SBL). (d) In the corresponding enzyme~II$^{{\rm mtl}}$-deficient
strain, sorbitol is consumed before mannitol~\citep{Lengeler1972}.}
\end{figure}

Now, all the mutant phenotypes discussed above can be explained just
as well by alternative hypotheses appealing only to the molecular
mechanisms. Indeed, the first case is obviously due to impaired repressor-operator
binding, and one can argue that the remaining two cases are due to
diminished inducer exclusion. However, the next two examples, which
involve the \emph{reversal of the diauxie}, are difficult to explain
from the molecular point of view.

Fig.~\ref{f:DataCase1}a shows that in \emph{E. coli} Hfr3000, glucose
is consumed before galactose. However, the mutant strain MM6, which
contains a lesion in the PTS enzyme~I~\citep{Tanaka1967}, consumes
galactose before glucose (Fig.~\ref{f:DataCase1}b). Likewise, \emph{E.~coli}
strain~159 consumes mannitol before sorbitol (Fig.~\ref{f:DataCase1}c),
but the corresponding mutant strain~157, which contains a lesion
in the PTS enzyme II$^{{\rm mtl}}$, consumes sorbitol before mannitol
(Fig.~\ref{f:DataCase1}d). These phenotypic changes fall within
the scope of the minimal model. In both mutants, the transport enzyme
for the preferred substrate is impaired, i.e., $V_{s,2}$ decreases,
so that $\kappa_{1}$ remains unchanged, but $\kappa_{2}$ increases
and $\alpha$ decreases. If the changes in $\kappa_{2}$ and $\alpha$
are sufficiently large, Fig.~\ref{f:BDcase1}b implies that the substrate
consumption pattern will shift from preferential consumption of~$S_{2}$
to preferential consumption of~$S_{1}$.

It should be emphasized that the {}``reversal of the diauxie'' is
a natural consequence of the minimal model. This is because each enzyme
inhibits the other enzyme due to dilution by growth, i.e., the inhibition
is \emph{mutual} or \emph{competitive}. Consequently, suppressing
the uptake (and hence, the growth) on one of the substrates automatically
tilts the balance of power in favor of the other substrate. In contrast,
the {}``reversal of the diauxie'' is difficult to explain in terms
of molecular mechanisms alone. This is because in all the molecular
mechanisms, the inhibition is \emph{unilateral} rather than mutual.
In \emph{E. coli}, for instance, there are numerous mechanisms that
allow PTS sugars, such as glucose and mannitol, to inhibit the synthesis
of the enzymes for non-PTS substrates. But there is no mechanism for
non-PTS substrates to inhibit the synthesis of PTS enzymes. This difficulty
did not escape the attention of Asensio et al, who observed the reversal
of the glucose-galactose diauxie (Fig.~\ref{f:DataCase1}, top panel).
Faced with the {}``reversal of the diauxie,'' they were compelled
to conclude that the {}``diauxie is, at least in part, due to competitive
effects at the permease level.''

\subsubsection{Dependence of substrate consumption pattern on relative growth rates}

In order to consider the relationship between the substrate consumption
pattern and the ratio of the single-substrate maximum specific growth
rates, define\[
\rho\equiv\frac{r_{g,2}^{{\rm max}}}{r_{g,1}^{{\rm max}}},\]
where $r_{g,i}^{{\rm max}}$ denotes maximum specific growth rate
during single-substrate growth on saturating concentrations of $S_{i}$.
Now, the model implies that\begin{align*}
r_{g,1}^{{\rm max}} & =Y_{1}V_{s,1}\left.e_{1}\right|_{E_{10}}=\sqrt{V_{e,1}Y_{1}V_{s,1}}\left.\epsilon_{1}\right|_{E_{10}},\\
r_{g,2}^{{\rm max}} & =Y_{2}V_{s,2}\left.e_{2}\right|_{E_{01}}=\sqrt{V_{e,2}Y_{2}V_{s,2}}\left.\epsilon_{2}\right|_{E_{01}},\end{align*}
 so that\[
\rho=\alpha\frac{\left.\epsilon_{2}\right|_{E_{01}}}{\left.\epsilon_{1}\right|_{E_{10}}}=\frac{\alpha}{\alpha_{g}(\kappa_{1},\kappa_{2})},\;\alpha_{g}(\kappa_{1},\kappa_{2})\equiv\frac{\left.\epsilon_{1}\right|_{E_{10}}}{\left.\epsilon_{2}\right|_{E_{01}}}=\frac{-\kappa_{1}+\sqrt{\kappa_{1}^{2}+4}}{-\kappa_{2}+\sqrt{\kappa_{2}^{2}+4}}.\]
It follows that

\begin{figure}
\begin{centering}\includegraphics[width=10cm,height=6cm]{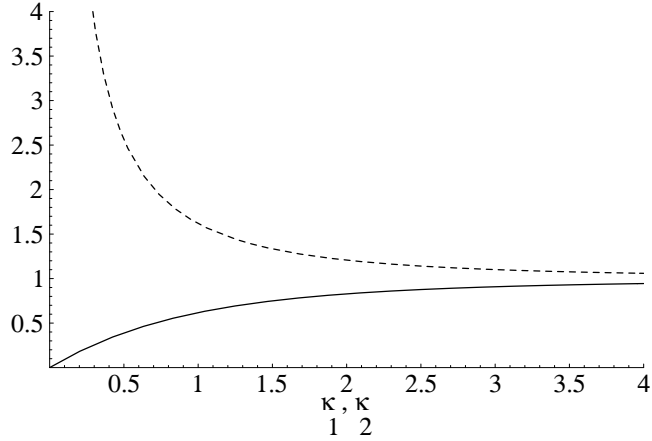}\par\end{centering}

\caption{\label{f:BDCase1growthRates}Classification of substrate consumption
patterns in terms of $\rho$, the ratio of the single-substrate maximum
specific growth rates. The full and dashed lines show the graphs of
$\rho_{*}(\kappa_{2})$ and $\rho^{*}(\kappa_{1})$, respectively.}
\end{figure}

\begin{enumerate}
\item The surface of $\alpha_{g}(\kappa_{1},\kappa_{2})$ separates the
parameter space into two distinct regions: Above the surface, $\rho>1$,
i.e., $r_{g,2}^{{\rm max}}>r_{g,1}^{{\rm max}}$, and below the surface,
$\rho<1$, i.e., $r_{g,2}^{{\rm max}}<r_{g,1}^{{\rm max}}$.
\item The surface of $\alpha_{g}(\kappa_{1},\kappa_{2})$ lies between the
surfaces of $\alpha_{*}(\kappa_{1},\kappa_{2})$ and $\alpha^{*}(\kappa_{1},\kappa_{2})$,
i.e.,\begin{equation}
\alpha_{*}(\kappa_{1},\kappa_{2})<\alpha_{g}(\kappa_{1},\kappa_{2})<\alpha^{*}(\kappa_{1},\kappa_{2})\label{eq:alphaRelations}\end{equation}
for all $\kappa_{1},\kappa_{2}>0$ (see Appendix~\ref{a:StabilityAnalysis1}).
Thus, the graph of $\alpha_{g}$, denoted by the dashed brown line
in Fig.~\ref{f:BDcase1}, lies between the graphs of $\alpha_{*}$
(blue curve) and $\alpha^{*}$ (red curve).
\end{enumerate}
Given these results, we can recast the classification of the substrate
consumption patterns in terms of $\rho$. To this end, define\begin{align*}
\rho_{*}(\kappa_{2}) & \equiv\frac{\alpha_{*}(\kappa_{1},\kappa_{2})}{\alpha_{g}(\kappa_{1},\kappa_{2})}=\kappa_{2}\frac{-\kappa_{2}+\sqrt{\kappa_{2}^{2}+4}}{2},\\
\rho^{*}(\kappa_{1}) & \equiv\frac{\alpha^{*}(\kappa_{1},\kappa_{2})}{\alpha_{g}(\kappa_{1},\kappa_{2})}=\frac{1}{\kappa_{1}}\frac{2}{-\kappa_{1}+\sqrt{\kappa_{1}^{2}+4}}.\end{align*}
Then, there is preferential consumption of $S_{1}$ (resp., $S_{2}$)
precisely when $\rho<\rho_{*}(\kappa_{2})$ (resp., $\rho>\rho^{*}(\kappa_{1})$),
and simultaneous consumption if and only if $\rho_{*}(\kappa_{2})<\rho<\rho^{*}(\kappa_{1})$.
Thus, $\rho_{*}$ and $\rho^{*}$ define the limits of $\rho$ at
which there is simultaneous consumption. It turns out that $\rho_{*}(\kappa_{2})$
increases from 0 to 1 as $\kappa_{2}$ goes from 0 to $\infty$, and
$\rho^{*}(\kappa_{1})$ decreases from $\infty$ to $0$ as $\kappa_{1}$
goes from $0$ to $\infty$ (Fig.~\ref{f:BDCase1growthRates}). We
are now ready to discuss the relationship between the substrate consumption
patterns and the ratio of the single-substrate maximum specific growth
rates.

The Harder \& Dijkhuizen correlation states that when growth is diauxic,
the preferred substrate is the one that, by itself, supports a higher
maximum specific growth rate~(p.~\pageref{q:CorrelationHD}). The
model predictions are consistent with this correlation. This is already
evident from Fig.~\ref{f:BDcase1}: $\alpha<\alpha_{g}$, i.e., $\rho<1$
in the region, $\alpha<\alpha_{*}$, corresponding to preferential
consumption of $S_{1}$, and $\alpha>\alpha_{g}$, i.e., $\rho>1$
in the region $\alpha>\alpha^{*}$ corresponding to preferential consumption
of $S_{2}$. The same property is also manifested in Fig.~\ref{f:BDCase1growthRates},
e.g., in the region, $\rho<\rho_{*}(\kappa_{2})$, corresponding to
preferential consumption of $S_{1}$, $\rho<1$ because the graph
of $\rho_{*}(\kappa_{2})$ is always below 1. The manifestation of
the Harder-Dijkhuizen correlation in this minimal model suggests that
is an intrinsic property of the induction and dilution kinetics. It
can be explained without invoking goal-oriented regulatory mechanisms,
which form the basis of models based on optimality principles~\citep{kompala86,Mahadevan2002,ramakrishna96}.

Current experimental evidence suggests that the existence of comparable
single-substrate maximum specific growth rates is, perhaps, necessary
but not sufficient for simultaneous consumption~(p.~\pageref{q:CorrelationEgli}).
However, Fig.~\ref{f:BDCase1growthRates} shows that this condition
($\rho\approx1$) is neither necessary nor sufficient for simultaneous
consumption. It is not necessary because when $\kappa_{1},\kappa_{2}\ll1$,
there is simultaneous consumption for almost all $\rho$. It is not
sufficient for simultaneous consumption because when $\kappa_{1},\kappa_{2}\gg1$,
simultaneous consumption is virtually impossible --- it cannot be
obtained unless $\rho$ lies in a vanishingly small neighborhood of
1. These results can be understood in terms of the limiting cases
discussed above. If $\kappa_{1},\kappa_{2}$ are small, the enzymes
are quasi-constitutive, and they resist extinction, regardless of
the maximum specific growth rates. As $\kappa_{1}$ and $\kappa_{2}$
increase, the enzymes become progressively more vulnerable to extinction,
and in the limit of large $\kappa_{1},\kappa_{2}$, they cannot coexist.

We note finally that unlike the standard Lotka-Volterra model for
competing species, there are no parameter values that yield bistable
enzyme dynamics (compare Figs.~\ref{f:GlobalDynamicsLV} and~\ref{f:BDcase1}).
We show below that bistability becomes feasible when the induction
kinetics are sigmoidal.

\subsection{Case 2 ($n_{1}=2,\; n_{2}=1$)}

In this case, the scaled equations are\begin{align*}
\frac{d\epsilon_{1}}{dt} & =\frac{\epsilon_{1}^{2}}{\kappa_{1}^{2}+\epsilon_{1}^{2}}-\left(\epsilon_{1}+\alpha\epsilon_{2}\right)\epsilon_{1}\\
\frac{d\epsilon_{2}}{dt} & =\alpha\frac{\epsilon_{2}}{\kappa_{2}+\epsilon_{2}}-\left(\epsilon_{1}+\alpha\epsilon_{2}\right)\epsilon_{2}\end{align*}
The key results, which are shown in detail in Appendix~\ref{a:StabilityAnalysis2},
are as follows

\begin{enumerate}
\item The trivial steady, $E_{00}$, always exists, regardless of the parameter
values. It is always unstable.
\item The semitrivial steady state, $E_{10}$, exists if and only if $\kappa_{1}<1$,
in which case it is unique, and given by\[
\epsilon_{1}=\sqrt{1-\kappa_{1}^{2}},\;\epsilon_{2}=0.\]
It is stable (as a node) if and only if $\left.\epsilon_{1}\right|_{E_{10}}$
exceeds the $\epsilon_{1}$-intercept of the nontrivial nullcline
for $\epsilon_{2}$, i.e.,\begin{equation}
\left.\epsilon_{1}\right|_{E_{10}}>\frac{\alpha}{\kappa_{2}}\Leftrightarrow\alpha<\kappa_{2}\sqrt{1-\kappa_{1}^{2}}\equiv\alpha_{*}(\kappa_{1},\kappa_{2}).\label{eq:E10StabilityCase2}\end{equation}

\item The semitrivial steady state, $E_{01}$, always exists, and is given
by \begin{equation}
\epsilon_{1}=0,\;\epsilon_{2}=\frac{-\kappa_{2}+\sqrt{\kappa_{2}^{2}+4}}{2}.\label{eq:E01StabilityCase2}\end{equation}
 It is always stable (as a node).
\item Nontrivial steady states exist only if $\kappa_{1}<1$. Under these
conditions, there are at most two nontrivial steady states. There
is a unique nontrivial steady state if and only if \[
0<\alpha<\alpha_{*}(\kappa_{1},\kappa_{2}),\]
and it is unstable whenever it exists. There are two nontrivial steady
states if and only if \[
0<\kappa_{1}<\beta\equiv\sqrt{\frac{2+\kappa_{2}^{2}}{2(1+\kappa_{2}^{2})}},\;\alpha_{*}(\kappa_{1},\kappa_{2})<\alpha<\alpha^{*}(\kappa_{1},\kappa_{2}),\]
where $\alpha^{*}(\kappa_{1},\kappa_{2})$ is the value of $\alpha$
at which the nontrivial nullclines for $\epsilon_{1}$ and $\epsilon_{2}$
touch. One of these steady states is stable and the other is unstable.
\item The surface of $\alpha_{*}(\kappa_{1},\kappa_{2})$ lies below the
surface of $\alpha^{*}(\kappa_{1},\kappa_{2})$ for all $0<\kappa_{1}<\beta$
and $\kappa_{2}>0$.
\end{enumerate}
\begin{figure}[t]
\begin{centering}\includegraphics[width=6cm,height=6cm]{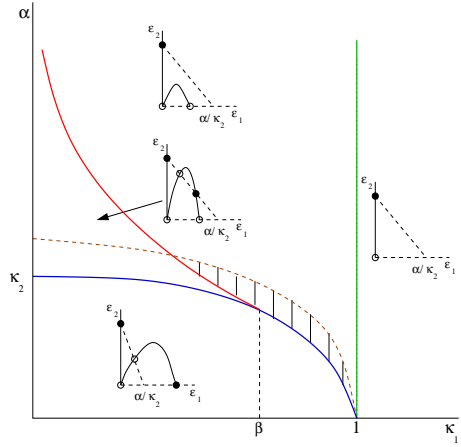}\par\end{centering}

\caption{\label{f:BDcase2}The bifurcation diagram for the case $n_{1}=2$,
$n_{2}=1$ at any fixed $\kappa_{2}>0$. In the phase portraits, the
nullclines for $\epsilon_{1}$ and $\epsilon_{2}$ are represented
by full and dashed lines, respectively; stable and unstable steady
states are represented by full and open circles, respectively. The
graphs of $\alpha_{*}$, $\alpha^{*}$, $\kappa_{1}=1$, and $\alpha_{g}$
are represented by blue, red, green, and dashed brown curves, respectively.
In the hatched region, $S_{2}$ is the preferred substrate for all
preculturing conditions, even though it supports a maximum specific
growth rate lower than that on $S_{1}$.}
\end{figure}

The bifurcation diagram shown in Fig.~\ref{f:BDcase2} implies the
following classification of the substrate utilization patterns.

\begin{figure}[t]
\begin{centering}\subfigure[]{\includegraphics[width=7cm,height=5cm]{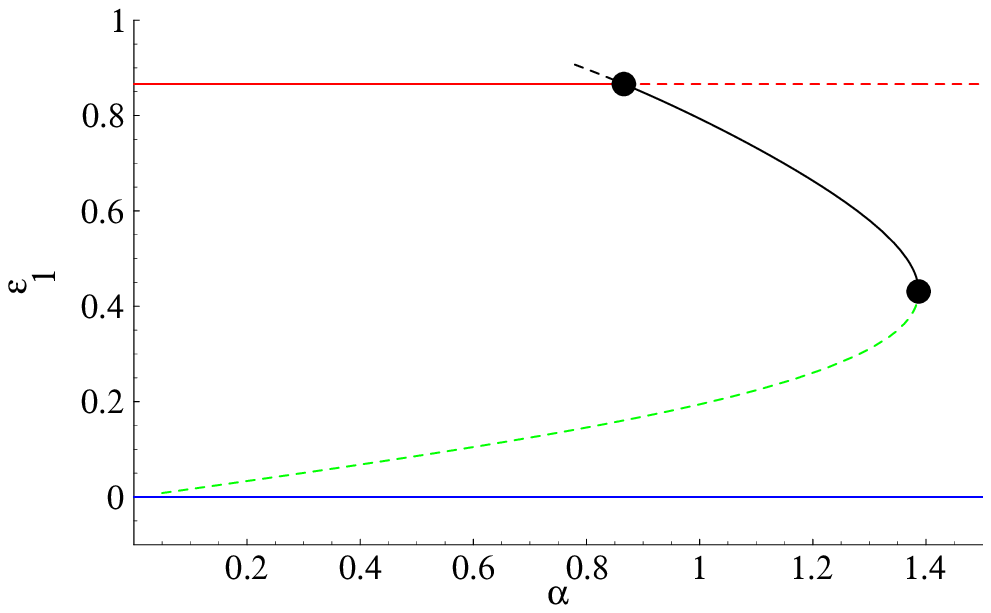}}\subfigure[]{\includegraphics[width=7cm,height=5cm]{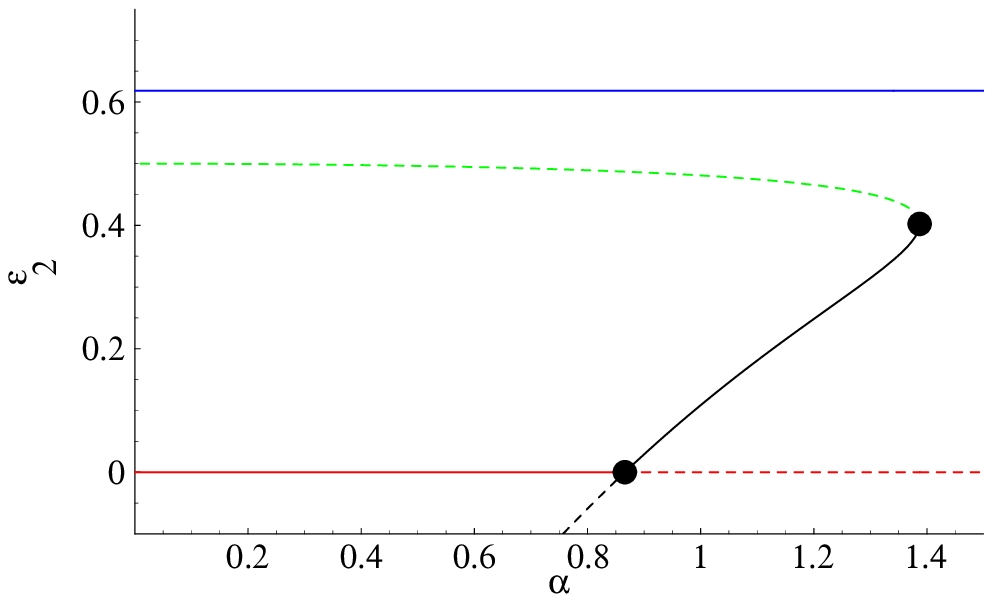}}\par\end{centering}

\caption{\label{f:CDcase2}The bifurcation diagram obtained when $\alpha$
is increased at $\kappa_{1}=0.5$ and $\kappa_{2}=1$. The red and
blue curves show the semitrivial steady states, $E_{10}$, $E_{01}$,
respectively. The green and black curves show the two nontrivial states
($E_{11}$). The curves are full (resp., dashed) if the corresponding
steady state is stable (resp., full). The bifurcation points are represented
by full circles.}
\end{figure}

\begin{enumerate}
\item If $\alpha<\alpha_{*}$, $E_{10}$ and $E_{01}$ are stable, i.e.,
there is preferential consumption of $S_{1}$ or $S_{2}$, depending
on the initial conditions.
\item If $0<\kappa_{1}<\beta$ and $\alpha_{*}<\alpha<\alpha^{*}$, $E_{01}$
and $E_{11}$ are stable, i.e., there is preferential consumption
of $S_{2}$ or simultaneous consumption of $S_{1}$ and $S_{2}$,
depending on the initial conditions.
\item If $\beta<\kappa_{1}<1,\alpha>\alpha_{*}$ or $\kappa_{1}>1$, there
is preferential consumption of $S_{2}$, regardless of the initial
conditions.
\end{enumerate}
The surfaces of $\alpha_{*}$ and $\alpha^{*}$ define the locus of
transcritical and fold (saddle-node) bifurcations, respectively (Fig.~\ref{f:CDcase2}).
If $\alpha$ is increased at any fixed $0<\kappa_{1}<\beta$ and $\kappa_{2}>0$,
the substrate consumption pattern changes at $\alpha=\alpha_{*}$
from bistable dynamics involving preferential consumption of $S_{1}$
or $S_{2}$ to bistable dynamics involving preferential consumption
of $S_{2}$ or simultaneous consumption. This transition occurs via
a transcritical bifurcation. At $\alpha=\alpha^{*}$, the substrate
consumption pattern switches to preferential consumption of $S_{2}$
via a fold bifurcation.

Comparison of Fig.~\ref{f:BDcase2} with Fig.~\ref{f:BDcase1} shows
that certain features are preserved. Specifically, preferential consumption
of $S_{1}$ is feasible only at low $\alpha$, and simultaneous consumption
occurs only if $\alpha$ has intermediate values and $\kappa_{1},\kappa_{2}$
are not too large. However, a unique property emerges in Fig.~\ref{f:BDcase2},
namely, bistability. This is due to the sigmoidal induction kinetics
for $E_{1}$, which ensure that preferential consumption of $S_{2}$
is feasible at all parameter values.

It is also worth examining the relationship between the classification
predicted by the model and the empirical classification based on the
single-substrate maximum specific growth rates. In this case\[
\rho=\alpha\frac{\left.\epsilon_{2}\right|_{E_{01}}}{\left.\epsilon_{1}\right|_{E_{10}}}=\frac{\alpha}{\alpha_{g}(\kappa_{1},\kappa_{2})},\;\alpha_{g}\equiv\frac{2\sqrt{1-\kappa_{1}^{2}}}{-\kappa_{2}+\sqrt{\kappa_{2}^{2}+4}}.\]
 Now, $\alpha_{g}>\alpha_{*}$ because $-\kappa_{2}+\sqrt{\kappa_{2}^{2}+4}<2/\kappa_{2}$
(see Appendix~\ref{a:StabilityAnalysis1}). Furthermore, $\alpha_{g}$
is zero at $\kappa_{1}=1$. Thus, the graph of $\alpha_{g}$ lies
above the graph of $\alpha_{*}$ (dashed brown line in Fig.~\ref{f:BDcase2}).
This implies that a substrate can be consumed preferentially even
if it supports a lower maximum specific growth rate. Indeed, if the
parameters lie in the region, $\alpha<\alpha_{g}$, then $S_{2}$
supports a lower maximum specific growth rate than $S_{1}$, and yet,
cells precultured on $S_{2}$ consume this substrate preferentially.
If the parameter values lie in the hatched region of Fig.~\ref{f:BDcase2},
$S_{2}$ is the preferred substrate, regardless of the manner in which
the cells are precultured.

\begin{figure}[t]
\begin{centering}\subfigure[]{\includegraphics[bb=0bp 0bp 576bp 360bp,width=7cm,height=5cm]{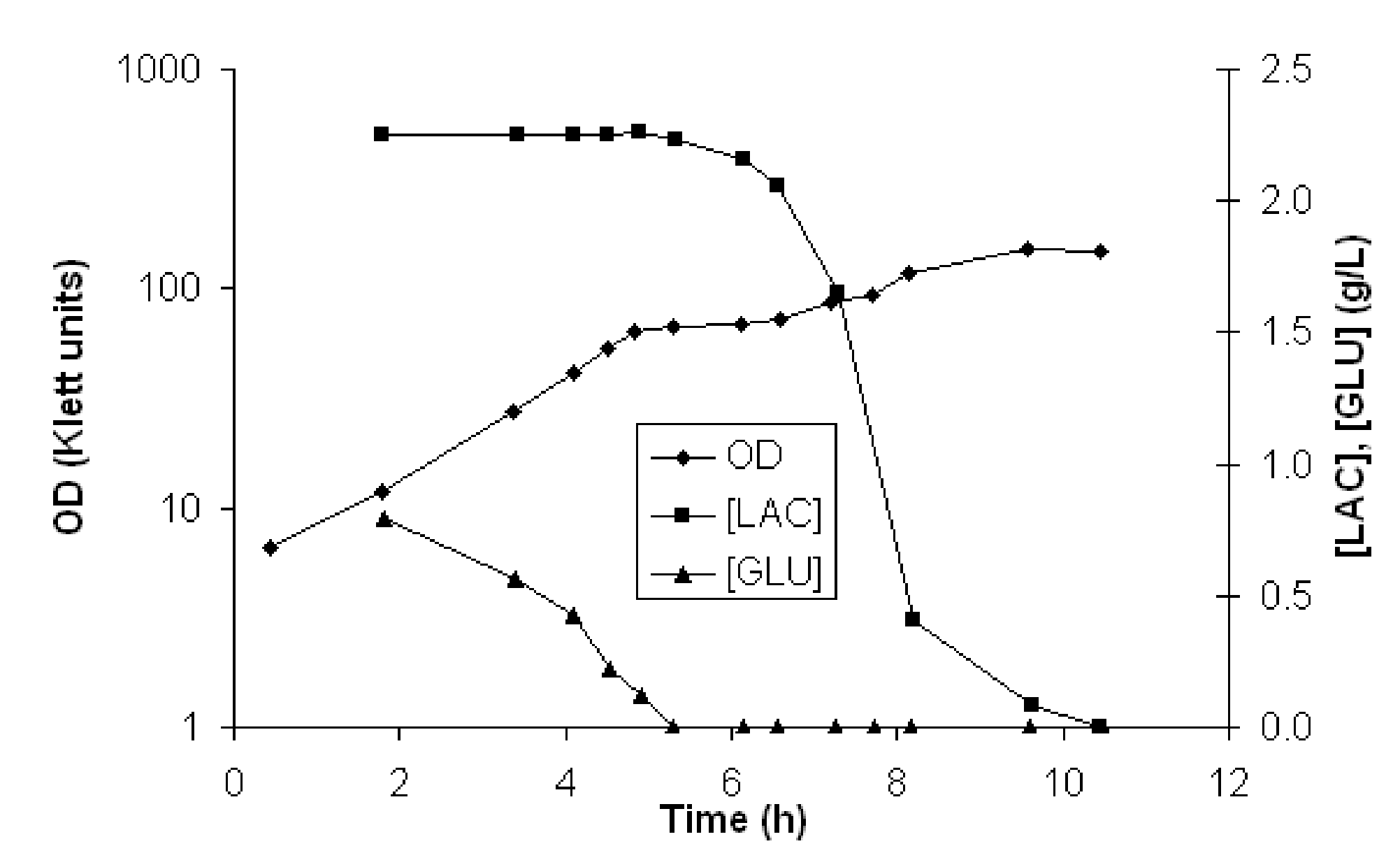}}\subfigure[]{\includegraphics[bb=0bp 0bp 576bp 360bp,width=7cm,height=5cm]{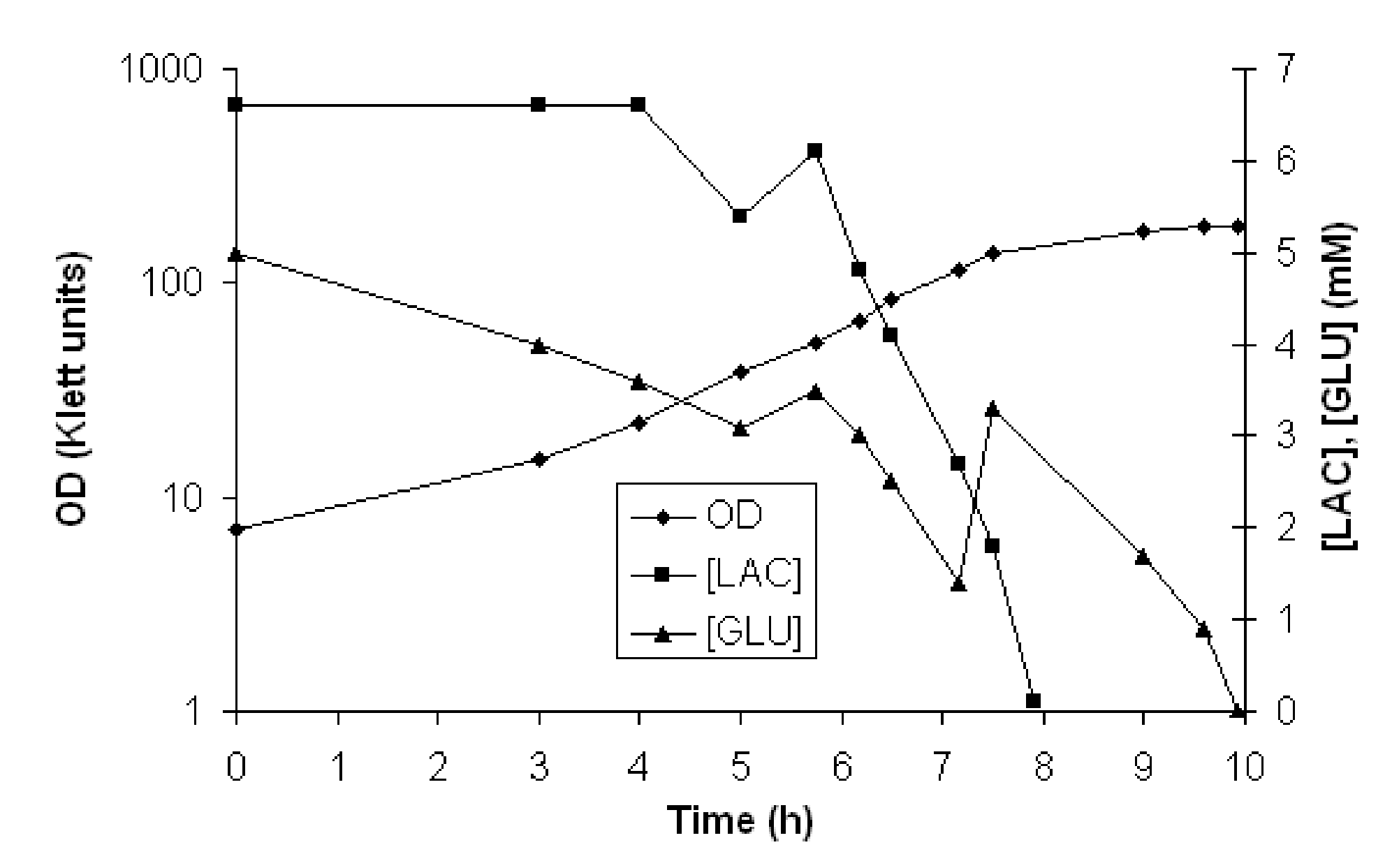}}\par\end{centering}

\begin{centering}\subfigure[]{\includegraphics[width=7cm,height=5cm]{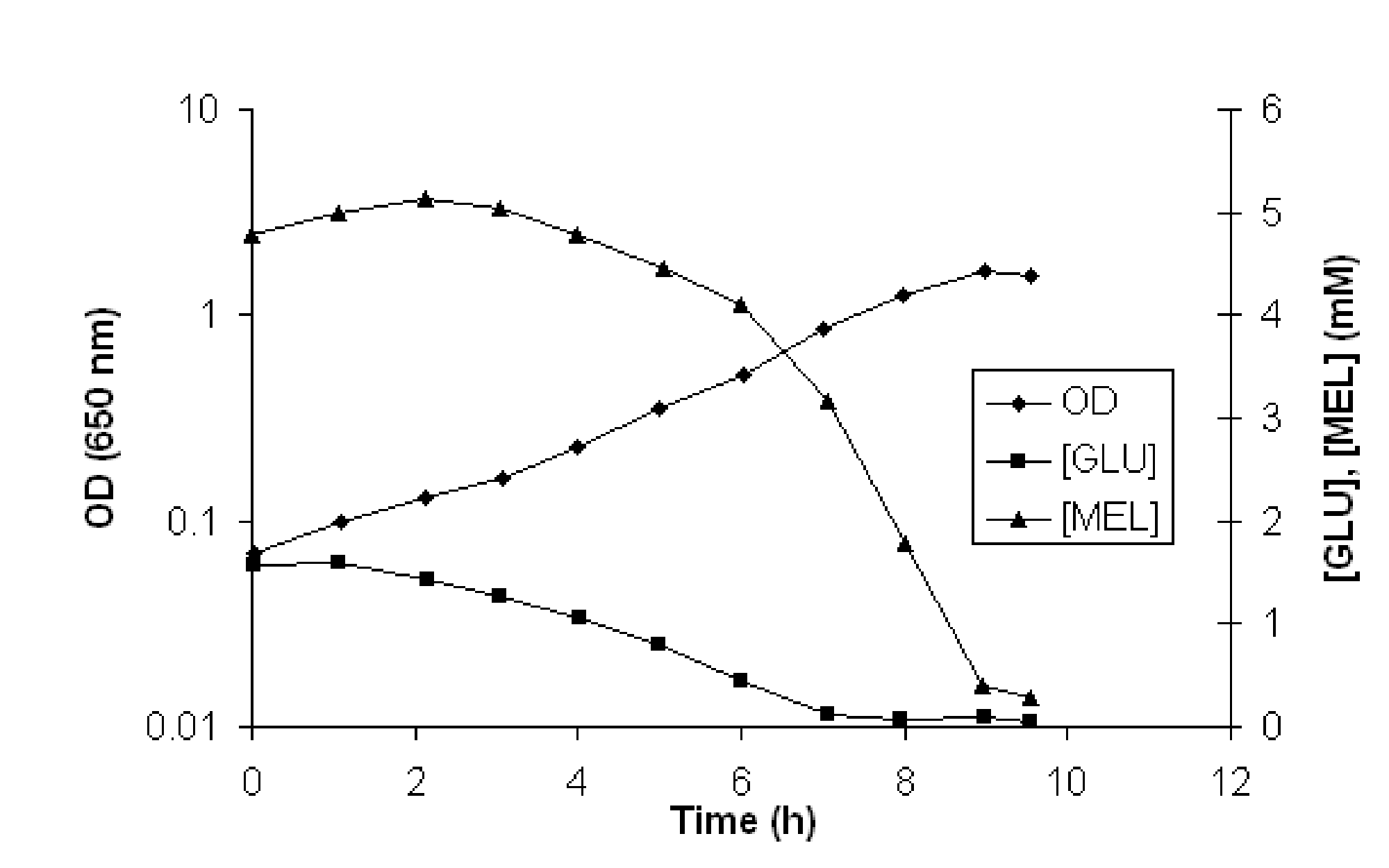}}\subfigure[]{\includegraphics[bb=0bp 0bp 576bp 360bp,width=7cm,height=5cm]{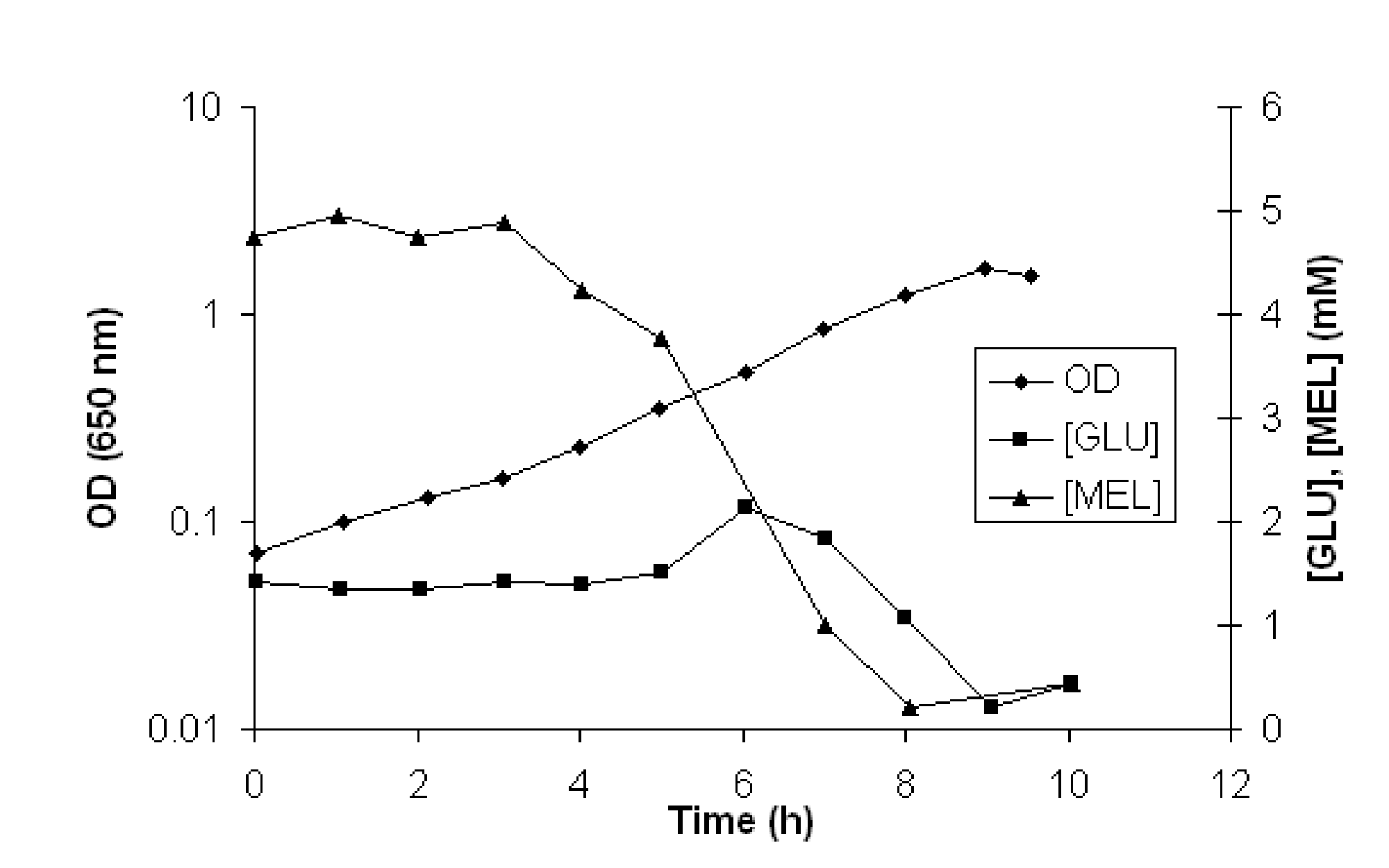}}\par\end{centering}

\caption{\label{f:DataCase2}Bistability in mixed-substrate growth: \textbf{Upper
panel:} Growth of \emph{Streptococcus mutans} GS5 on a mixture of
lactose (LAC) and glucose (GLU)~\citep{Liberman1984b}. (a)~Glucose
is consumed preferentially if the cells are precultured on glucose.
(b)~ Glucose and lactose are consumed simultaneously if the cells
are precultured on lactose. \textbf{Lower panel:} Growth of \emph{Salmonella
typhimurium} SB1476 on a mixture on a mixture of glucose (GLU) and
melibiose (MEL)~\citep{Kuroda1992}. (c)~Glucose-precultured cells
consume glucose before melibiose. (d)~Melibiose-precultured cells
consume melibiose before glucose. In (b) and (d), the concentration
of glucose increases at $t\approx8$ and $t\approx6$~h, respectively.
It is believed that is due to expulsion of the glucose produced from
intracellular hydrolysis of lactose and melibiose, respectively.}
\end{figure}

\subsubsection{Evidence of bistable substrate consumption patterns}

The bistable dynamics predicted by Fig.~\ref{f:BDcase2} have been
observed in experiments.

The bistable dynamics in the region, $\alpha_{*}<\alpha<\alpha^{*}$,
correspond to preferential consumption of $S_{2}$ if the preculture
is grown on $S_{2}$, and simultaneous consumption if the preculture
is grown on $S_{1}$. Two examples of this substrate consumption pattern
were described in the Introduction, namely, growth of \emph{P. aeruginosa}
on glucose plus citrate~\citep{hamilton59,hamilton60,hamilton61}
and growth of \emph{E. coli} K12 on a mixture of glucose and pyruvate~\citep{narang97a}.
Fig.~\ref{f:DataCase2} shows another example of this substrate consumption
pattern. When \emph{Streptococcus mutans} GS5 is grown on a mixture
of glucose and lactose, glucose-precultured cells consume glucose
before lactose (Fig.~\ref{f:DataCase2}a), whereas lactose-precultured
cells consume both glucose and lactose (Fig.~\ref{f:DataCase2}b).

The bistable dynamics in the region, $\alpha<\alpha_{*}$, correspond
to preferential consumption of $S_{1}$ if the preculture is grown
on $S_{1}$, and preferential consumption of $S_{2}$ if the preculture
is grown on $S_{2}$. Furthermore, the maximum specific growth on
$S_{2}$ is lower than that on $S_{1}$. There is evidence suggesting
the existence of this substrate consumption pattern. Tsuchiya and
coworkers studied the growth of \emph{Salmonella typhimurium} on a
mixture of glucose and melibiose~\citep{Kuroda1992,Okada1981}. They
found that the wild-type strain LT2 consumed glucose before melibiose.
However, the PTS enzyme I mutant, SB1476, yielded the bistable substrate
consumption pattern corresponding to the region, $\alpha<\alpha_{*}$.
Cells precultured on glucose consumed glucose preferentially (Fig.~\ref{f:DataCase2}c),
and cells precultured on melibiose consumed melibiose preferentially
(Fig.~\ref{f:DataCase2}d). Moreover, the maximum specific growth
rate on glucose (0.24~h$^{-1}$) is significantly lower than that
on melibiose (0.41~h$^{-1}$). It should be noted that these experiments
were done in the presence of 5~mM cAMP in the culture. However, at
least in the case of glucose-precultured cells, the same phenotype
was observed even in the absence of cAMP.

\section{Discussion}

We have shown that a minimal model accounting for only enzyme induction
and dilution captures and explains all the substrate consumption patterns
observed in the experimental literature. In what follows, we discuss
the robustness of the model, and its implications for the problem
of size regulation in development.

\subsection{Robustness of the model}

Given the simplicity of the model, it is necessary ask whether the
properties of the model will be preserved if additional metabolic
details and regulatory mechanisms are incorporated in the model. Now,
the defining property of the minimal model is that the enzymes follow
competitive dynamics. We show below that this property is not a consequence
of the particular kinetics assumed in the model. It is the outcome
of two very general characteristics possessed by most systems of mixed-substrate
growth.

To see this, it is useful to consider the generalized Lotka-Volterra
model for competing species~\citep[Chap.~12]{hirschEcol}. This model
postulates that the competitive interactions between two species are
captured by the relations\[
\frac{dN_{i}}{dt}=f_{i}(N_{1},N_{2}),\; f_{1}(0,N_{2})=f_{2}(N_{1},0)=0\textnormal{ and }\;\frac{\partial f_{1}}{\partial N_{2}},\frac{\partial f_{2}}{\partial N_{1}}<0.\]
In other words, the essence of competitive interactions can be distilled
into two properties: \renewcommand{\theenumi}{\alph{enumi}}

\begin{enumerate}
\item The growth of a species is impossible in the absence of that species
($dN_{i}/dt=0$ whenever $N_{i}=0$).
\item Each species inhibits the growth of the other species ($\partial f_{1}/\partial N_{2},\partial f_{2}/\partial N_{1})<0$.
\end{enumerate}
These properties, by themselves, imply the existence of all the dynamics
associated with competitive interactions, namely, the absence of limit
cycles, and the existence of extinction and coexistence steady states.
\renewcommand{\theenumi}{\arabic{enumi}}

Now, properties (a) and (b) will be manifested in most systems of
mixed-substrate growth. Indeed, the evolution of the enzymes during
the first exponential growth phase can be described by the relations\[
\frac{de_{i}}{dt}=g_{i}(e_{1},e_{2})\equiv r_{e,i}(e_{1},e_{2})-r_{g}(e_{1},e_{2})e_{i}.\]
If we assume that

\begin{enumerate}
\item each enzyme is necessary for its own synthesis, i.e., $r_{e,i}=0$
whenever $e_{i}=0$,
\item each enzyme has either no effect or inhibits the synthesis of the
other enzyme, i.e., $\partial r_{e,1}/\partial e_{2}$, $\partial r_{e,2}/\partial e_{1}\le0$,
\item the specific growth rate is an increasing function of $e_{1}$ and
$e_{2}$, i.e., $\partial r_{g}/\partial e_{1}$, $\partial r_{g}/\partial e_{2}>0$,
\end{enumerate}
then the enzymes satisfy both the hypotheses of the generalized model
for competing species: (a)~There is no enzyme synthesis in the absence
of the enzyme ($de_{i}/dt=0$ whenever $e_{i}=0$), and (b)~each
enzyme inhibits the synthesis of the other enzyme ($\partial g_{1}/\partial e_{2}$,
$\partial g_{2}/\partial e_{1}<0$). Consequently, they will display
extinction and coexistence dynamics.

It remains to consider the generality of assumptions 1--3.

\begin{figure}
\begin{centering}\subfigure[]{\includegraphics[width=7cm,height=5cm]{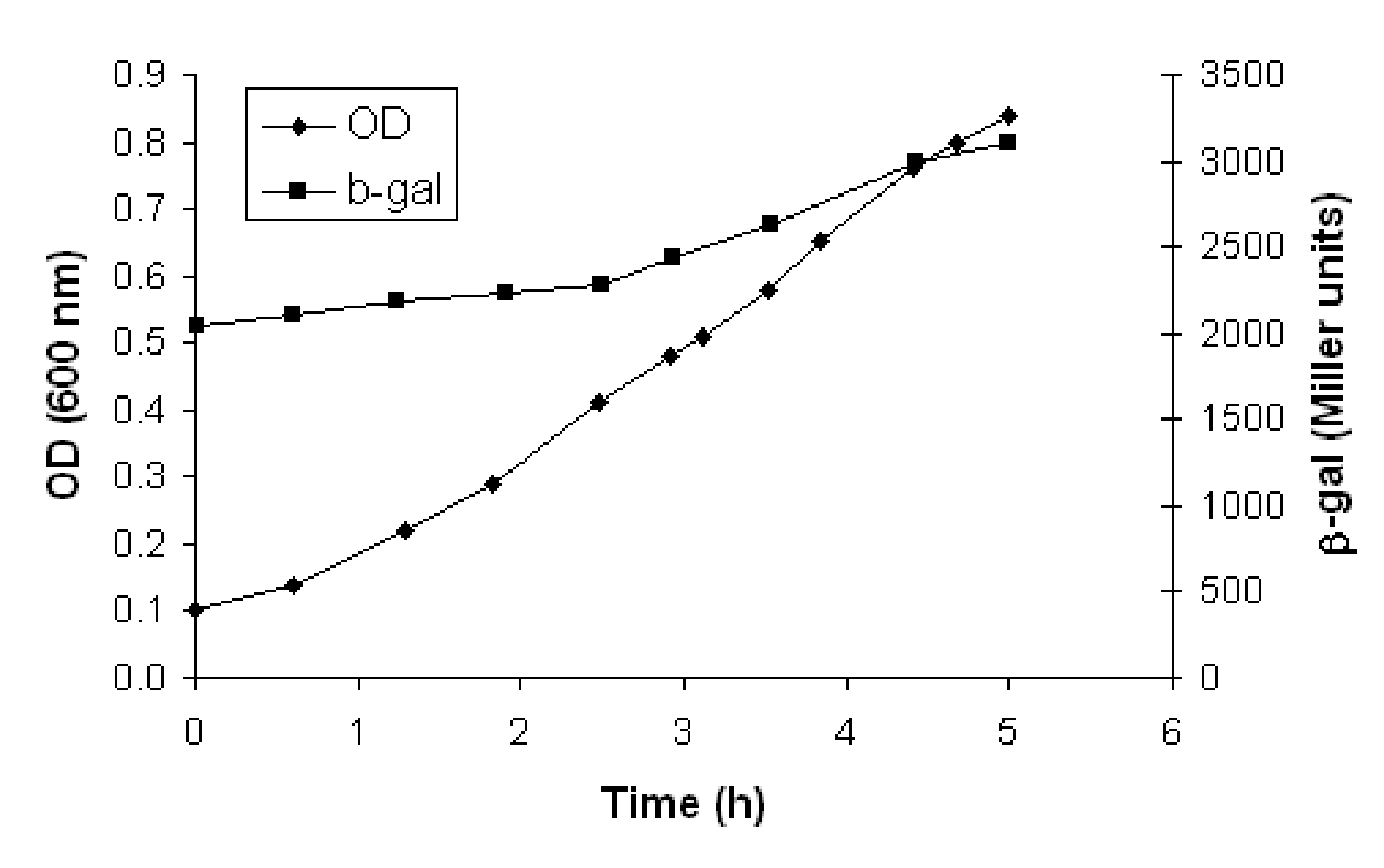}}\subfigure[]{\includegraphics[width=7cm,height=5cm]{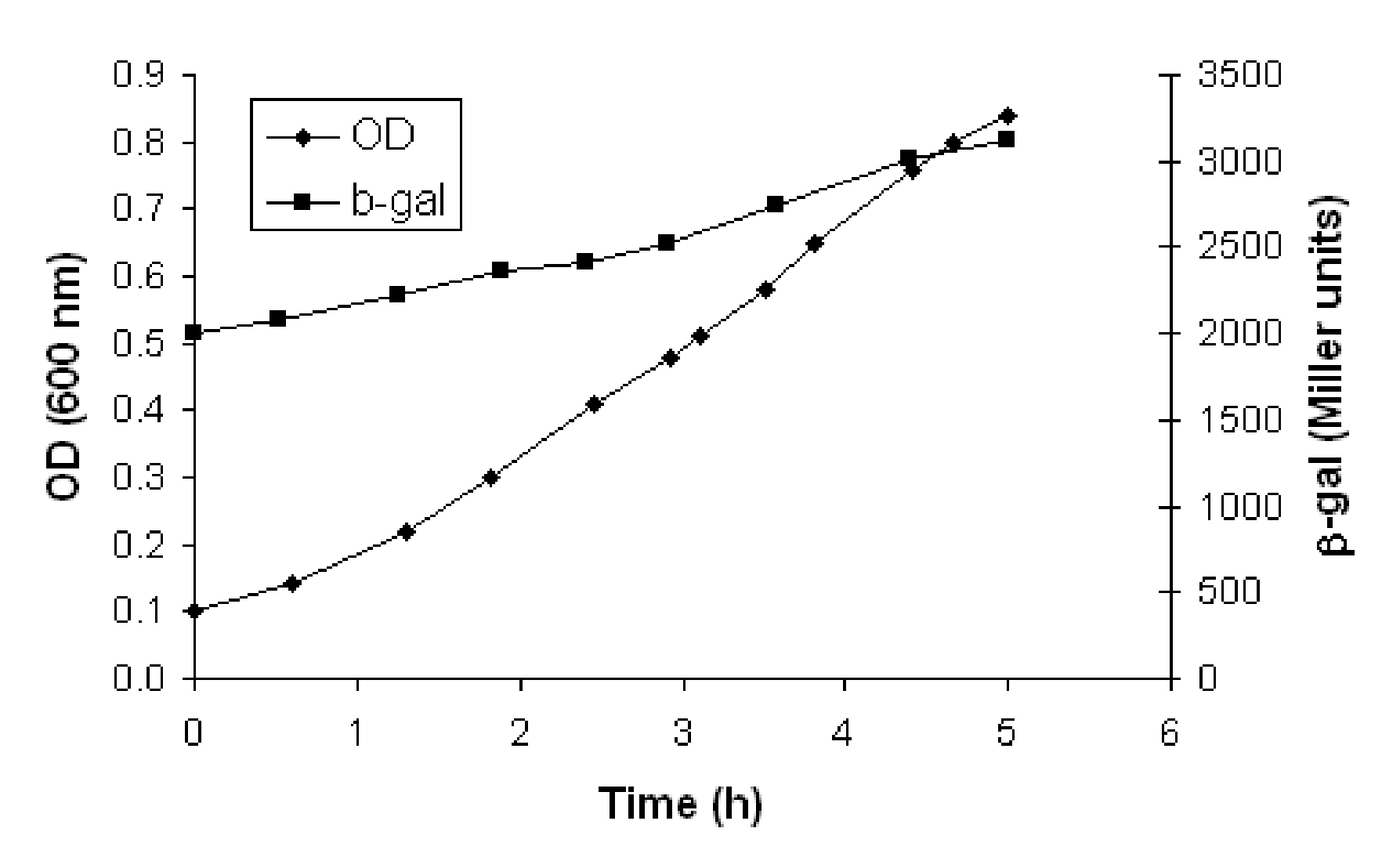}}\par\end{centering}

\caption{\label{f:ViolationOfAssumption1}Disappearance of the
diauxie when the enzymes are not necessary for their own
synthesis~\citep{inada96}. Synthesis of $\beta$-galactosidase
persists in the presence of glucose if (a)\emph{~lacI} is mutated so
that \emph{lac} transcription becomes constitutive or (b)~IPTG is
present in the medium. In the first case, \emph{lac} transcription
is constitutive, i.e., it persists even in the absence of the
inducer. In the second case, \emph{lac} transcription is no longer
dependent on the existence of the permease.}
\end{figure}

Assumption~1 will be satisfied whenever the substrates are transported
by unique inducible enzymes. In these cases, the enzymes are required
for the existence of the inducer ($e_{i}=0\Rightarrow x_{i}=0$),
and the inducers are necessary for the synthesis of the enzymes ($x_{i}=0\Rightarrow r_{e,i}=0$);
hence, $e_{i}=0\Rightarrow r_{e,i}=0$. One can imagine two cases
in which assumption~1 is violated. First, if an enzyme is constitutive,
it is synthesized even in the absence of the inducer ($x_{i}=0\nRightarrow r_{e,i}=0$).
Second, in the presence of a gratuitous inducer, such as IPTG, which
can enter the cell even in the absence of lactose permease, the enzyme
is not required for the existence of the inducer ($e_{i}=0\nRightarrow x_{i}=0$).
In both cases, the {}``extinction'' steady state ceases to exist,
and the substrates will be consumed simultaneously. This is consistent
with experiments (Fig.~\ref{f:ViolationOfAssumption1}).

Assumption~2 will be satisfied provided the enzymes do not activate
each other. But all the known regulatory mechanisms invariably entail
direct or indirect inhibition of one of the enzymes by the other enzyme.
This includes inducer exclusion (dephosphorylated enzyme~II$^{{\rm glc}}$
inhibits lac permease), and cAMP activation (dephosphorylation of
II$^{{\rm glc}}$ causes a reduction of cAMP levels, which in turn
inhibits \emph{lac} transcription).

Assumption~3 will be satisfied if the yield of biomass on a substrate
during single-substrate growth does not change markedly during mixed-substrate
growth. In the model, the yields were assumed to be constant. This
is certainly true for conservative substrates since $Y_{i}=1$. It
is also observed to hold in many mixtures of carbon sources~\citep{egli82a,narang97a}.
However, it is conceivable that there are systems in which the yields
vary with the enzyme levels. In such cases, the specific growth rate
will have the form, $r_{g}(e_{1},e_{2})=Y_{1}(e_{1},e_{2})V_{s,i}e_{1}+Y_{2}(e_{1},e_{2})V_{s,2}e_{2}$.
At present, the data is not sufficient for determining the extent
to which the yields with the enzyme levels.

It is therefore clear that inclusion of various regulatory mechanisms
will enhance the mutual inhibition due from dilution. However, the
qualitative behavior will be preserved, since the enzymes will still
follow Lotka-Volterra dynamics. Thus, the key property of the model,
namely, competitive dynamics of the enzymes, is quite robust insofar
as the perturbations with respect to regulatory mechanisms are concerned.

The notion that diauxic growth is the outcome of competitive interactions
between the enzymes is not new. It can be found in the earliest papers
on diauxic growth. In 1947, Monod noted that~\citep[p.~254]{monod47}

\begin{quote}
{}``it appears that the mechanisms involved in diauxic inhibition
have the character of \emph{competitive interactions} between different
specific enzyme-forming systems.''
\end{quote}
He observed, furthermore, that the~\citep[p.~259]{monod47}.

\begin{quote}
{}``existence of competitive interactions in the synthesis of different
specific enzymes appears to be a fact of fundamental significance
in enzymatic adaptation, and one for which any conception of the phenomenon
should be able to account.''
\end{quote}
However, these conclusions were based on the kinetics of the enzyme
levels during diauxic growth, and had no mechanistic basis.

The above argument, first made in~\citet{narang98b}, shows the mechanistic
basis of the competitive interactions in a mathematically precise
fashion.

\subsection{Implication of the model for development}

Diauxic growth has played a critical role in shaping models of patterning
in development. The first link between genetics and development was
established in the late 40's by appealing to the following argument~\citep{Gilbert2002}.
During diauxic growth, cells possessing identical genes synthesize
different proteins at distinct times (namely, the first and second
exponential growth phases). By analogy, patterning in differentiation
could be viewed as the synthesis of different proteins at distinct
times and locations~\citep{monod47,Spiegelman1948}. From this standpoint,
diauxic growth and developmental patterning can be viewed as {}``temporal''
and {}``spatiotemporal'' differentiation, respectively.

The subsequent discovery of the molecular mechanisms involved in developmental
patterning have confirmed the above hypothesis. It has been found
that developmental patterns are generated by mechanisms similar in
principle, but more complex in detail, than those involved diauxic
growth~\citep[Chap.~3]{Ptashne2}.

Despite remarkable successes in developmental patterning, there are
outstanding questions about \emph{size regulation}, i.e., the mechanisms
by which patterning is coupled to growth~\citep{Day2000,Hafen2003,Serrano1997}.
Examples of such questions include: What determines the size of organs
and organisms, i.e., why does their growth cease at a certain time?
And why is development \emph{scale-invariant}, i.e., why is the size
of the organs is proportional to the size of the organism?

The model presented here may be relevant to the problem of size regulation.
It shows that the {}``temporal'' differentiation in the diauxie
is coupled to growth, and this coupling is mediated by the process
of enzyme dilution. Inasmuch as the diauxie is a paradigm of the mechanisms
controlling cellular differentiation, a similar mechanism may lie
at the heart of the coupling between developmental patterning and
growth. Based on the minimal model, one can speculate, for instance,
that organ growth ceases at a certain time because growth-promoting
enzymes are driven to {}``extinction'' at sufficiently high growth
rates.

The model also has implications for the problem of scale invariance.
In many mathematical models of development, pattern formation occurs
when a homogeneous steady state of a reaction-diffusion system\begin{equation}
\frac{\partial c}{\partial t}=D\nabla^{2}c-r(c,p)\label{eq:MassBalanceWithoutGrowth}\end{equation}
becomes unstable due to the onset of a Turing instability~\citep{murray}.
Here, $c(x,t)$ denotes the vector of morphogen concentrations, $D$
is the matrix of diffusivities, and $r(c,p)$ is the reaction rate
vector expressed as a function of $c$ and a vector of parameters,
$p$. In general, the patterns predicted by these models are not scale-invariant.
However, this problem can be resolved if the system is fed more information
about its size (say, $L$). For instance, perfect scale invariance
is obtained if the diffusivities or rate constants are proportional
to $L^{2}$, and plausible mechanisms for such a dependence have been
proposed~\citep{Othmer1980,Ishihara2006}.

In growing systems, however, information regarding the growth rate
is constantly fed to the mechanism driving pattern formation. Indeed,
in the presence of growth, Eq.~(\ref{eq:MassBalanceWithoutGrowth})
becomes \begin{equation}
\frac{\partial c}{\partial t}+v.\nabla c=D\nabla^{2}c-r(c,p)-c\nabla\cdot v\label{eq:MassBalanceWithGrowth}\end{equation}
where $v(x,t)$ is the velocity vector field, $v\cdot\nabla c$ is
the accumulation of the morphogens due to convection, $\nabla\cdot v$
is the specific growth rate, and $c\nabla\cdot v$ is the dilution
of the morphogens due to growth. Crampin et al have shown that these
equations exhibit a certain degree of scale invariance --- as the
system grows, the number of pattern elements remains the same despite
a doubling of the system size~\citep{Crampin1999,Crampin2002}. Further
analysis of this class of equations offers the promise of deeper insights
into the coupling between patterning and growth.

\section{Conclusions}

\begin{enumerate}
\item We showed that a minimal model accounting for enzyme induction and
dilution, but not cAMP activation and inducer exclusion, captures
and explains all the observed substrate consumption patterns, including
diauxic growth, simultaneous consumption, and bistable growth. This
suggests that the dynamics characteristic of mixed-substrate growth
are already inherent in the minimal structure associated with induction
and dilution. We find that many of the molecular mechanisms, such
as inducer exclusion, serve to amplify these inherent dynamics.
\item We constructed bifurcation diagrams showing the parameter values at
which the various substrate consumption patterns will be observed.
The bifurcation diagrams explain the phenotypic responses to various
genetic perturbations, including lesions in the genes for the repressor,
operator, and the transport enzymes. Importantly, they provide a simple
explanation for the {}``reversal of the diauxie,'' a phenomenon
which is quite difficult to explain in terms of molecular mechanisms.
The bifurcation diagrams also provide deep insights into the mechanisms
underlying the empirically observed correlations between the substrate
consumption patterns and the single-substrate growth rates. We found
that

\begin{enumerate}
\item When the induction kinetics are hyperbolic, the preferred substrate
is always the one that that supports a higher growth rate. This correlation
is, therefore, unlikely to be the outcome of optimal design. It is
a natural consequence of the fact that the enzymatic dynamics are
governed by the rates of induction and dilution.\\
If induction is sigmoidal, it is possible for the preferred substrate
to support a lower growth rate than the less preferred substrate.
We presented experimental data illustrating this case.
\item The existence of comparable growth rates is neither necessary nor
sufficient for simultaneous consumption. When the saturation constants
are small, simultaneous consumption occurs regardless of the maximum
specific growth rates, since induction is quasi-constitutive. If the
saturation constants are large, simultaneous consumption is impossible
even if the growth rates are comparable.
\end{enumerate}
\item The key property of the model, namely, competitive dynamics of the
enzymes, is quite robust with respect to structural perturbations.
\item The model may have implications for the problem of size regulation
in development, since it provides a mechanism for coupling differentiation
and growth, namely, protein dilution.
\end{enumerate}
\begin{ack}
This research was supported in part with funds from the National Science
Foundation under contract NSF DMS-0517954.
\end{ack}
\appendix

\section{\label{a:MolecularParameters}Interpretation of $n_{i}$, $V_{e,i}$,
and $K_{e,i}$ in terms of molecular interactions}

To express $n_{i}$, $V_{e,i}$, and $K_{e,i}$ in terms of molecular
parameters, we appeal to the Yagil \& Yagil model~\citep{yagil71}.
For notational clarity, we shall ignore the subscript $i$ for the
substrate; thus, the operator, inducer, and repressor will be denoted
by $O$, $X$, and $R$, respectively. Furthermore, their concentrations
will be denoted $[O]$, $[X]$, and $[R]$, respectively.

The Yagil \& Yagil model views induction as the outcome of a competition
for the repressor between the operator and the inducer. Induction
occurs when the inducer molecules sequester the repressors away from
the operator. The competitive interactions are represented by two
binding equilibria\begin{eqnarray}
R+O & \rightleftharpoons & R\cdot O,\; K_{o}\equiv\frac{[R][O]}{[R\cdot O]},\label{eq:YY1}\\
R+nX & \rightleftharpoons & R\cdot X_{n}\; K_{x}\equiv\frac{[R][X]}{[R\cdot X_{n}]},\label{eq:YY2}\end{eqnarray}
where $n$ denotes the number of inducer molecules that bind to 1
molecule of repressor; $[R\cdot O]$, $[R\cdot X_{n}]$ denote the
concentrations of the complexes, $R\cdot O$, $R\cdot X_{n}$, respectively;
and $K_{o}$, $K_{x}$ denote the dissociation constants for the two
equilibria.

It is assumed that

\begin{enumerate}
\item Enzyme synthesis is limited by the transcription rate, i.e., translation
is not limiting. Thus, the specific enzyme synthesis rate is proportional
to the specific transcription rate. Furthermore, the specific transcription
rate is proportional to the concentration of the free operator, i.e.,
\begin{equation}
r_{e}=\nu[O],\label{eq:YYrE}\end{equation}
where $\nu$ denotes the enzyme synthesis rate per unit mass of operator.
\item The total concentrations of $O$ and $R$, denoted $[O]_{t}$ and
$[R]_{t}$, respectively, are conserved, i.e.,\begin{align}
[O]_{t} & =[O]+[R\cdot O],\label{eq:YY3}\\
{}[R]_{t} & =[R]+[R\cdot O]+[R\cdot X_{n}].\label{eq:YY4}\end{align}
These two relations, together with Eqs.~(\ref{eq:YY1})--(\ref{eq:YY2}),
constitute 4 equations in 4 unknowns, namely, $[O]$, $[R]$, $[R\cdot$O],
and $[R\cdot X_{n}]$. In principle, these equations can be solved
for $[O]$, and substituted in (\ref{eq:YYrE}) to obtain $r_{e}$.
However, since the solution is cumbersome, it is convenient to make
the following additional assumption.
\item The repressor is bound primarily to the inducer (rather than the operator),
i.e., \[
[R\cdot O]\ll[R\cdot X_{n}].\]
This assumption is valid under most conditions because the operator
concentration ($\sim$2 per cell) is significantly smaller than the
inducer concentration.
\end{enumerate}
These assumptions yield\[
r_{e}=\nu[O]=V_{e}\frac{K_{x}+[X]^{n}}{K_{x}(1+[R]_{t}/K_{o})+[X]^{n}},\]
where $V_{e}\equiv\nu[O_{t}]$.

In the case of \emph{constitutive} enzymes, the repressor has a weak
affinity for the operator, i.e., $K_{o}\gg[R]_{t}$, so that \[
r_{e}\approx V_{e}\]
regardless of the inducer concentration.

In the case of \emph{inducible} enzymes, the repressor has a high
affinity for the operator, i.e., $K_{o}\ll[R]_{t}$, so that \begin{align*}
r_{e} & \approx V_{e}\frac{K_{x}+[X]^{n}}{K_{x}[R]_{t}/K_{o}+[X]^{n}}.\end{align*}
This is a monotonically increasing function of $[X]$ with a small
nonzero intercept. Neglecting this small {}``basal'' enzyme synthesis
rate yields%
\footnote{Recent evidence suggests that in the case of the \emph{lac} operon,
the cooperativity does not arise from the binding of two inducer molecules
to a single repressor molecule. Instead, it might be due to the cooperative
binding of a single repressor molecule to two operators~\citep{Oehler2006}.%
}\[
r_{e}\approx V_{e}\frac{[X]^{n}}{K_{e}^{n}+[X]^{n}},\; K_{e}^{n}\equiv\frac{K_{x}}{K_{o}}[R]_{t}.\]

\section{\label{a:NmodelEquations}Derivation of the equations}

Equations (\ref{eq:NxO}--\ref{eq:NcMO}) implicitly define the specific
growth rate and the evolution of the cell density. To see this, observe
that since all the intracellular concentrations are expressed as mass
fractions (g/gdw), their sum equals~1, i.e., $x_{1}+x_{2}+e_{1}+e_{2}+c^{-}=1$.
Hence, addition of equations (\ref{eq:NxO}--\ref{eq:NcMO}) yields\[
0=\sum_{i=1}^{2}r_{s,i}-\sum_{i=1}^{2}(1-Y_{i})r_{x,i}-\frac{1}{c}\frac{dc}{dt}\]
which can be rewritten in the more familiar form\[
\frac{dc}{dt}=r_{g}c,\; r_{g}\equiv\sum_{i=1}^{2}r_{s,i}-\sum_{i=1}^{2}(1-Y_{i})r_{x,i}\]
where $r_{g}$ denotes the specific growth rate.

We can simplify the model by observing that $x_{i}\sim10^{-3}$~g/gdw~\citep{chung96}
and $r_{s,i},r_{x,i}\sim1$~g gdw$^{-1}$ h$^{-1}$. Thus, $x_{i}$
attains quasisteady state on a time scale of $10^{-3}$~h. Moreover,
the dilution term $r_{g}x_{i}\sim10^{-4}$~g~gdw$^{-1}$~h$^{-1}$
is negligibly small compared to $r_{s,i},r_{x,i}$. Hence, within
a few seconds, (\ref{eq:NxO}) becomes, $0\approx r_{s,i}-r_{x,i}$,
which implies that $r_{g}\approx\sum_{i}Y_{i}r_{s,i}$, i.e., $Y_{i}$
is essentially the yield of biomass on $S_{i}$. Thus, we arrive at
the equations \begin{align*}
\frac{dc}{dt} & =(Y_{1}r_{s,1}+Y_{2}r_{s,2})c,\\
\frac{ds_{i}}{dt} & =-r_{s,i}c\\
\frac{de_{i}}{dt} & =r_{e,i}-(Y_{1}r_{s,1}+Y_{2}r_{s,2})e_{i},\\
x_{i} & \approx\frac{V_{s,i}e_{i}s_{i}/(K_{s,i}+s_{i})}{k_{x,i}}\\
c^{-} & =1-x_{1}-x_{2}-e_{1}-e_{2}\end{align*}
where $x_{i}$ is obtained by solving the quasisteady state relation,
$r_{x,i}\approx r_{s,i}$.

\section{\label{a:StabilityAnalysis1}Stability analysis of case 1 ($n_{1}=n_{2}=1$)}

In this case, the steady states satisfy the equations\begin{align*}
0 & =\left(\frac{1}{\kappa_{1}+\epsilon_{1}}-\epsilon_{1}-\alpha\epsilon_{2}\right)\epsilon_{1},\\
0 & =\left(\alpha\frac{1}{\kappa_{2}+\epsilon_{2}}-\epsilon_{1}-\alpha\epsilon_{2}\right)\epsilon_{2}.\end{align*}
and the Jacobian at any $(\epsilon_{1},\epsilon_{2})$ is\[
J(\epsilon_{1},\epsilon_{2})=\left[\begin{array}{cc}
\frac{\kappa_{1}}{(\kappa_{1}+\epsilon_{1})^{2}}-2\epsilon_{1}-\alpha\epsilon_{2} & -\alpha\epsilon_{1}\\
-\epsilon_{2} & \frac{\alpha\kappa_{2}}{(\kappa_{2}+\epsilon_{2})^{2}}-\epsilon_{1}-2\alpha\epsilon_{2}\end{array}\right].\]

\subsection{Trivial steady state}

It is evident that $E_{00}$ exists, regardless of the parameter values.
It is always an unstable node since\[
J(E_{00})=\left[\begin{array}{cc}
\frac{1}{\kappa_{1}} & 0\\
0 & \frac{1}{\kappa_{2}}\end{array}\right]\]
which implies that both eigenvalues, $1/\kappa_{1}$ and $1/\kappa_{2}$,
are positive.

\subsection{Semitrivial steady states}

The semitrivial steady state, $E_{10}$, always exists. It is unique
and given by\[
\epsilon_{1}=\frac{-\kappa_{1}+\sqrt{\kappa_{1}^{2}+4}}{2},\;\epsilon_{2}=0.\]
Since $1/(\kappa_{1}+\epsilon_{1})=\epsilon_{1}$ at $E_{10}$, the
Jacobian at this steady state is\[
\left[\begin{array}{cc}
\frac{\kappa_{1}\epsilon_{1}}{\kappa_{1}+\epsilon_{1}}-2\epsilon_{1} & -\alpha\epsilon_{1}\\
0 & \frac{\alpha}{\kappa_{2}}-\epsilon_{1}\end{array}\right],\]
and the eigenvalues are\[
\lambda_{1}=\frac{\kappa_{1}\epsilon_{1}}{\kappa_{1}+\epsilon_{1}}-2\epsilon_{1}=-\epsilon_{1}\frac{\kappa_{1}+2\epsilon_{1}}{\kappa_{1}+\epsilon_{1}}<0,\;\lambda_{2}=\frac{\alpha}{\kappa_{2}}-\epsilon_{1}.\]
Hence, $E_{10}$ is stable (as a node) if and only if

\begin{equation}
\lambda_{2}=\frac{\alpha}{\kappa_{2}}-\epsilon_{1}<0\;\Leftrightarrow\alpha<\kappa_{2}\left.\epsilon_{1}\right|_{E_{10}}\Leftrightarrow\frac{\kappa_{2}}{\alpha}-\kappa_{1}>\frac{\alpha}{\kappa_{2}}.\label{eq:E10StabilityCase1App}\end{equation}

A similar analysis of the semitrivial steady state, $E_{01}$, shows
that it always exists. It is unique and given by \[
\epsilon_{1}=0,\;\epsilon_{2}=\frac{-\kappa_{2}+\sqrt{\kappa_{2}^{2}+4}}{2}.\]
It is stable (as a node) if and only if\begin{equation}
\alpha>\frac{1}{\kappa_{1}\left.\epsilon_{2}\right|_{E_{01}}}\Leftrightarrow\alpha(\alpha\kappa_{1}-\kappa_{2})>\frac{1}{\kappa_{1}}.\label{eq:E01StabilityCase1App}\end{equation}

\subsection{Nontrivial steady state(s)}

The nontrivial steady state(s), $E_{11}$, satisfy the equations\begin{align}
0 & =\frac{1}{\kappa_{1}+\epsilon_{1}}-\epsilon_{1}-\alpha\epsilon_{2},\label{eq:E11SS1Case1}\\
0 & =\alpha\frac{1}{\kappa_{2}+\epsilon_{2}}-\epsilon_{1}-\alpha\epsilon_{2}.\label{eq:E11SS2Case1}\end{align}
Eliminating $\epsilon_{2}$ from these equations yields the equation\[
\frac{1}{\kappa_{1}+\epsilon_{1}}=(1+\alpha^{2})\epsilon_{1}+\alpha(\alpha\kappa_{1}-\kappa_{2}),\]
which has at most 1 positive root, and it exists if and only if\[
\alpha(\alpha\kappa_{1}-\kappa_{2})<\frac{1}{\kappa_{1}}\;\Leftrightarrow\; E_{01}\textnormal{ is unstable.}\]
On the other hand, eliminating $\epsilon_{1}$ from (\ref{eq:E11SS1Case1}--\ref{eq:E11SS2Case1})
yields the equation\[
\frac{\alpha}{\kappa_{2}+\epsilon_{2}}=(\frac{1}{\alpha}+\alpha)\epsilon_{2}+\frac{\kappa_{2}}{\alpha}-\kappa_{1}\]
which has at most 1 positive root, and it exists if and only if \[
\frac{\kappa_{2}}{\alpha}-\kappa_{1}<\frac{\alpha}{\kappa_{2}}\;\Leftrightarrow\; E_{10}\textnormal{ is unstable.}\]
Thus, $E_{11}$ exists if and only if both $E_{10}$ and $E_{01}$
are unstable. Furthermore, it is unique whenever it exists, and is
given by\begin{align*}
\epsilon_{1} & =\frac{(\alpha\kappa_{2}-\kappa_{1})-2\alpha^{2}\kappa_{1}+\sqrt{(\alpha\kappa_{2}-\kappa_{1})^{2}+4\left[(1+\alpha^{2})+\alpha\kappa_{2}\right]}}{2(1+\alpha^{2})},\\
\epsilon_{2} & =\frac{-\alpha(\alpha\kappa_{2}-\kappa_{1})-2\kappa_{2}+\sqrt{(\alpha\kappa_{2}-\kappa_{1})^{2}+4\left[(1+\alpha^{2})+\alpha\kappa_{2}\right]}}{2(1+\alpha^{2})}.\end{align*}
It turns out that $E_{11}$ is stable whenever it exists since (\ref{eq:E11SS1Case1}--\ref{eq:E11SS2Case1})
imply that \[
J_{11}=\frac{\kappa_{1}}{(\kappa_{1}+\epsilon_{1})^{2}}-2\epsilon_{1}-\alpha\epsilon_{2}=-\frac{\epsilon_{1}}{\kappa_{1}+\epsilon_{1}}\left(\kappa_{1}+2\epsilon_{1}+\alpha\epsilon_{2}\right)<0\]
and \[
J_{22}=\frac{\alpha\kappa_{2}}{(\kappa_{2}+\epsilon_{2})^{2}}-\epsilon_{1}-2\alpha\epsilon_{2}=-\frac{\epsilon_{2}}{\kappa_{2}+\epsilon_{2}}\left(\alpha\kappa_{2}+\epsilon_{1}+2\alpha\epsilon_{2}\right)<0\]
so that $\textnormal{tr}\, J(E_{11})<0$ and\[
\det J(E_{11})=\left(\kappa_{1}+\alpha\kappa_{2}\right)\epsilon_{1}+\alpha\left(\kappa_{1}+\alpha\kappa_{2}\right)\epsilon_{2}+4\alpha\epsilon_{1}\epsilon_{2}+2\alpha^{2}\epsilon_{2}^{2}>0.\]
Hence, the eigenvalues of $J(E_{11})$ have negative real parts.

We conclude that $E_{11}$ exists if and only if\begin{equation}
\textnormal{Both }E_{10}\textnormal{ and }E_{01}\textnormal{ are unstable}\Leftrightarrow\frac{1}{\kappa_{1}\left.\epsilon_{2}\right|_{E_{01}}}<\alpha<\kappa_{2}\left.\epsilon_{1}\right|_{E_{10}}.\label{eq:E11StabilityCase1App}\end{equation}
Furthermore, it is unique and stable whenever it exists.

\subsection{Disposition of the surfaces of $\alpha_{*}$, $\alpha^{*}$, and
$\alpha_{g}$}

The surface of $\alpha_{g}$ lies between the surfaces of $\alpha_{*}(\kappa_{1},\kappa_{2})$
and $\alpha^{*}(\kappa_{1},\kappa_{2})$, i.e., \begin{equation}
\alpha_{*}(\kappa_{1},\kappa_{2})<\alpha_{g}(\kappa_{1},\kappa_{2})<\alpha^{*}(\kappa_{1},\kappa_{2})\label{eq:alphaRelationsA}\end{equation}
 for all $\kappa_{1},\kappa_{2}>0$. To see this, observe that \[
\left(\frac{1}{x}+\frac{x}{2}\right)^{2}>1+\frac{x^{2}}{4}\Rightarrow\frac{-x+\sqrt{x^{2}+4}}{2}<\frac{1}{x}\]
 for all $x>0$. Hence\[
\frac{-\kappa_{1}+\sqrt{\kappa_{1}^{2}+4}}{2/\kappa_{2}}<\frac{-\kappa_{1}+\sqrt{\kappa_{1}^{2}+4}}{-\kappa_{2}+\sqrt{\kappa_{2}^{2}+4}}<\frac{2/\kappa_{1}}{-\kappa_{2}+\sqrt{\kappa_{2}^{2}+4}}\]
for all $\kappa_{1},\kappa_{2}>0$, and (\ref{eq:alphaRelationsA})
follows immediately from the definitions of $\alpha_{*}$, $\alpha^{*}$,
and $\alpha_{g}$.

\section{Stability analysis of case 2 ($n_{1}=2,\; n_{2}=1$)\label{a:StabilityAnalysis2}}

In this case, the steady states satisfy the equations\begin{align*}
0 & =\left(\frac{\epsilon_{1}}{\kappa_{1}^{2}+\epsilon_{1}^{2}}-\epsilon_{1}-\alpha\epsilon_{2}\right)\epsilon_{1},\\
0 & =\left(\alpha\frac{1}{\kappa_{2}+\epsilon_{2}}-\epsilon_{1}-\alpha\epsilon_{2}\right)\epsilon_{2}.\end{align*}
and the Jacobian at any $(\epsilon_{1},\epsilon_{2})$ is\[
J(\epsilon_{1},\epsilon_{2})=\left[\begin{array}{cc}
\frac{2\kappa_{1}^{2}\epsilon_{1}}{(\kappa_{1}^{2}+\epsilon_{1}^{2})^{2}}-2\epsilon_{1}-\alpha\epsilon_{2} & -\alpha\epsilon_{1}\\
-\epsilon_{2} & \frac{\alpha\kappa_{2}}{(\kappa_{2}+\epsilon_{2})^{2}}-\epsilon_{1}-2\alpha\epsilon_{2}\end{array}\right].\]
In what follows, we study the conditions on the parameter values for
the existence and stability of all four types of steady states.

\subsection{Trivial steady state}

The trivial steady, $E_{00}$, always exists, regardless of the parameter
values. The Jacobian is singular at this steady state, but we can
infer its stability from the dynamics on the invariant lines, $\epsilon_{1}=0$
and $\epsilon_{2}=0$. Indeed, in the neighborhood of $E_{00}$, \[
\left.\frac{d\epsilon_{1}}{dt}\right|_{\epsilon_{2}=0}\approx\epsilon_{1}^{2}\left(\frac{1}{\kappa_{1}^{2}}-1\right),\;\left.\frac{d\epsilon_{2}}{dt}\right|_{\epsilon_{1}=0}\approx\frac{\alpha}{\kappa_{2}}\epsilon_{2}>0.\]
Hence, $E_{00}$ is a nonhyperbolic saddle if $\kappa_{1}<1$ and
a nonhyperbolic unstable node if $\kappa_{1}>1$.

\subsection{Semitrivial steady states}

The semitrivial steady state, $E_{10}$, exists provided $\kappa_{1}<1$,
in which case it is unique, and given by\[
\epsilon_{1}=\sqrt{1-\kappa_{1}^{2}},\;\epsilon_{2}=0.\]
Since $\epsilon_{1}^{2}+\kappa_{e,1}^{2}=1$ at this steady state,
the Jacobian is\[
\left[\begin{array}{cc}
2\left(\kappa_{1}^{2}-1\right)\epsilon_{1} & -\alpha\epsilon_{1}\\
0 & \frac{\alpha}{\kappa_{2}}-\epsilon_{1}\end{array}\right],\]
and the eigenvalues are\[
\lambda_{1}=2\left(\kappa_{1}^{2}-1\right)\epsilon_{1}<0,\;\lambda_{2}=\frac{\alpha}{\kappa_{2}}-\epsilon_{1}.\]
Hence, $E_{10}$ is stable (as a node) if and only if

\begin{equation}
\lambda_{2}=\frac{\alpha}{\kappa_{2}}-\left.\epsilon_{1}\right|_{E_{10}}<0\;\Leftrightarrow\alpha<\kappa_{2}\sqrt{1-\kappa_{1}^{2}}.\label{eq:E10StabilityCase2App}\end{equation}

Analysis of the steady state, $E_{01}$, shows that this steady state
always exists, and is given by \[
\epsilon_{1}=0,\;\epsilon_{2}=\frac{-\kappa_{2}+\sqrt{\kappa_{2}^{2}+4}}{2}.\]
The Jacobian at this steady state is\[
\left[\begin{array}{cc}
-\alpha\epsilon_{2} & 0\\
-\epsilon_{2} & \frac{\alpha\kappa_{2}\epsilon_{2}}{\kappa_{2}+\epsilon_{2}}-2\alpha\epsilon_{2}\end{array}\right],\]
and the eigenvalues are\begin{equation}
\lambda_{1}=-\alpha\epsilon_{2}<0,\;\lambda_{2}=-\alpha\epsilon_{2}\frac{\kappa_{2}+2\epsilon_{2}}{\kappa_{2}+\epsilon_{2}}<0.\label{eq:E01StabilityCase2App}\end{equation}
 We conclude that $E_{01}$ always exists and is stable (as a node).

\subsection{Nontrivial steady state(s)}

\subsubsection{Existence}

The nontrivial steady states satisfy the equations\begin{align}
0 & =\frac{\epsilon_{1}}{\kappa_{1}^{2}+\epsilon_{1}^{2}}-\epsilon_{1}-\alpha\epsilon_{2},\label{eq:E11Case2SS1}\\
0 & =\alpha\frac{1}{\kappa_{2}+\epsilon_{2}}-\epsilon_{1}-\alpha\epsilon_{2}.\label{eq:E11Case2SS2}\end{align}
If $\kappa_{1}>0$, there are no nontrivial steady states, since (\ref{eq:E11Case2SS1})
cannot be satisfied for any $\epsilon_{1},\epsilon_{2}>0$. Indeed,\[
\kappa_{1}>1\Rightarrow\frac{\epsilon_{1}}{\kappa_{1}^{2}+\epsilon_{1}^{2}}-\epsilon_{1}-\alpha\epsilon_{2}=\epsilon_{1}\left(\frac{1}{\kappa_{1}^{2}+\epsilon_{1}^{2}}-1\right)-\alpha\epsilon_{2}<0\]
for all $\epsilon_{1},\epsilon_{2}>0$. Henceforth, we shall assume
that $0<\kappa_{1}<1$ and $\kappa_{2}>0$.

We begin by introducing a change of coordinates that reduces the problem
to the existence of roots on a finite interval. Letting $\epsilon_{2}=\nu\epsilon_{1}$,
we rewrite the above system as \begin{align*}
\frac{1}{\kappa_{1}^{2}+\epsilon_{1}^{2}} & =1+\alpha\nu,\\
\frac{\alpha}{\kappa_{2}+\nu\epsilon_{1}} & =\epsilon_{1}(1+\alpha\nu).\end{align*}
 Solving the first equation for $\epsilon_{1}$, and rewriting the
second equation, we obtain \begin{align*}
\epsilon_{1} & =\sqrt{\frac{1}{1+\alpha\nu}-\kappa_{1}^{2}},\\
\frac{\alpha}{1+\nu\alpha} & ==\nu\epsilon_{1}^{2}+\kappa_{2}\epsilon_{1}.\end{align*}
 Hence, we obtain the following equation for $\nu$: \[
\frac{\alpha}{1+\nu\alpha}=\nu\left(\frac{1}{1+\alpha\nu}-\kappa_{1}^{2}\right)+\kappa_{2}\sqrt{\frac{1}{1+\alpha\nu}-\kappa_{1}^{2}}.\]
 Multiplying through by $(1+\alpha\nu)$, we obtain \[
\alpha=\nu\left[1-\kappa_{1}^{2}(1+\alpha\nu)\right]+\kappa_{2}\sqrt{(1+\alpha\nu)\left[1-\kappa_{1}^{2}(1+\alpha\nu)\right]}.\]
 Finally, we let \[
z=\kappa_{1}^{2}(1+\alpha\nu),\quad\nu=\frac{1}{\alpha}\left(\frac{z}{\kappa_{1}^{2}}-1\right),\]
 and rewrite the above equation as \begin{equation}
1=\frac{1}{\alpha^{2}\kappa_{1}^{2}}(z-\kappa_{1}^{2})(1-z)+\frac{\kappa_{2}}{\alpha\kappa_{1}}\sqrt{z(1-z)}=f(z,\alpha).\label{eq:zEqn}\end{equation}
 Observe that $z$ uniquely determines $\nu$, and hence both $\epsilon_{1}$
and $\epsilon_{2}$. To ensure that both $\epsilon_{i}>0$, $z$ must
belong to the interval $(\kappa_{1}^{2},1)$. Thus, the problem is
reduced to the existence of roots of (\ref{eq:zEqn}) on the finite
interval $(\kappa_{1}^{2},1)$.

Several properties of $f$ are immediate: \[
f(1,\alpha)=0<1,\quad f(\kappa_{1}^{2},\alpha)=\frac{\kappa_{2}\sqrt{1-\kappa_{1}^{2}}}{\alpha},\]
 and \[
f_{zz}(z,\alpha)=-\frac{2}{\alpha^{2}\kappa_{1}^{2}}-\frac{\kappa_{2}\left[4z(1-z)+(1-2z)^{2}\right]}{4\alpha\kappa_{1}\left[z(1-z)\right]^{3/2}}<0,\quad\forall\ 0<z<1,\]
 so that $f(z,\alpha)$ is strictly concave down in $z$ for each
$\alpha>0$ and for all $\kappa_{1}^{2}<z<1$. In particular, this
implies that (\ref{eq:zEqn}) has a unique root $z\in[\kappa_{1}^{2},1]$
if $f(\kappa_{1}^{2},\alpha)>1$, and at most two roots if $f(\kappa_{1}^{2},\alpha)\leq1$.
Furthermore, we observe that $f_{z}(z,\alpha)\to-\infty$ as $z\to1$,\[
f_{z}(\kappa_{1}^{2},\alpha)=\frac{2(1-\kappa_{1}^{2})^{3/2}+\alpha\kappa_{2}(1-2\kappa_{1}^{2})}{2\alpha^{2}\kappa_{1}^{2}\sqrt{1-\kappa_{1}^{2}}},\]
 and \[
f_{z}(\kappa_{1}^{2},\alpha_{*})=\frac{2(1-\kappa_{1}^{2})+\kappa_{2}^{2}(1-2\kappa_{1}^{2})}{2\kappa_{1}^{2}\kappa_{2}^{2}(1-\kappa_{1}^{2})}.\]
 Hence, $f_{z}(\kappa_{1}^{2},\alpha_{*})>0$ if and only if $2(1-\kappa_{1}^{2})+\kappa_{2}^{2}(1-2\kappa_{1}^{2})>0$,
i.e., \[
\kappa_{1}<\beta\equiv\sqrt{\frac{1+\kappa_{2}^{2}}{2(1+\kappa_{2}^{2})}}.\]
Finally, we observe that $f(z,\alpha)$ decreases in $\alpha$ for
each fixed $z$, and $f(z,\alpha)\to0$ as $\alpha\to+\infty$.

We conclude that there are two scenarios

\begin{enumerate}
\item If $\kappa_{1}<\beta$, then there exists $\alpha^{*}>\alpha_{*}$
such that (\ref{eq:zEqn}) has only one root in $(\kappa_{1}^{2},1)$
for all $\alpha<\alpha_{*}$, two distinct roots in $(\kappa_{1}^{2},1)$
for all $\alpha_{*}<\alpha<\alpha^{*}$, and no roots for $\alpha>\alpha^{*}$.\\
When $\alpha=\alpha_{*}$, (\ref{eq:zEqn}) has two roots: $z=\kappa_{1}^{2}$
and another root in $(\kappa_{1}^{2},1)$. When $\alpha=\alpha^{*}$,
(\ref{eq:zEqn}) admits a double root in $(\kappa_{1}^{2},1)$.
\item If $\kappa_{1}\ge\beta$, then (\ref{eq:zEqn}) admits only one root
in $(\kappa_{1}^{2},1)$ for all $\alpha<\alpha_{*}$, and no roots
for $\alpha>\alpha_{*}$.\\
A single root $z=\kappa_{1}^{2}$ occurs if and only if $\alpha=\alpha_{*}$.
\end{enumerate}
Thus, we obtain the bifurcation diagram shown in Figure~\ref{f:BDcase2}.

\subsubsection{Computation of $\alpha^{*}(\kappa_{1},\kappa_{2})$}

We have shown above that $\alpha^{*}$ exists for all $\kappa_{1}<\beta$.
Here, we present an algorithm for computing $\alpha^{*}$, which rests
upon the fact that the nontrivial nullclines for $\epsilon_{1}$ and
$\epsilon_{2}$ touch at $\alpha=\alpha^{*}$.

Eliminating $\epsilon_{2}$ from (\ref{eq:E11Case2SS1}--\ref{eq:E11Case2SS2})
yields the quartic polynomial\begin{multline}
\epsilon_{1}^{4}\cdot(1+\alpha^{2})-\epsilon_{2}^{3}\cdot\alpha\kappa_{2}+\epsilon_{1}^{2}\cdot\left[2\alpha^{2}\kappa_{1}^{2}-(1-\kappa_{1}^{2})\right]\\
+\epsilon_{1}\cdot\alpha\kappa_{1}^{2}+\alpha^{2}\kappa_{1}^{4}=0.\label{eq:E11Case2E1}\end{multline}
This equation has two equal real roots if and only if the discriminant
is zero~\citep[p.~41]{Dickson}, i.e.,\begin{equation}
\triangle\equiv\frac{\alpha^{2}}{\left(1+\alpha^{2}\right)^{6}}\left[c_{0}\left(\alpha^{2}\right)^{3}+c_{1}\left(\alpha^{2}\right)^{2}+c_{2}\left(\alpha^{2}\right)+c_{3}\right]=0\label{eq:discrim4}\end{equation}
where\begin{align*}
c_{0} & =\kappa_{1}^{8}\left(4+\kappa_{2}^{2}\right)^{2}>0,\\
c_{1} & =\frac{\kappa_{1}^{6}}{4}\left[8\kappa_{1}^{4}\left(\kappa_{2}^{2}-4\right)-\left(\kappa_{2}^{2}+2\right)\left(\kappa_{2}^{2}+4\right)^{2}-4\kappa_{1}^{2}\left(\kappa_{2}^{2}+4\right)\left(\kappa_{2}^{2}+8\right)\right],\\
c_{2} & =\frac{\kappa_{1}^{4}}{16}\left[16\kappa_{1}^{8}+32\kappa_{1}^{6}\left(\kappa_{2}^{2}-6\right)+\left(\kappa_{2}^{2}+4\right)^{2}-4\kappa_{1}^{2}\left(\kappa_{2}^{2}+4\right)\left(5\kappa_{2}^{2}+12\right)\right.\\
 & \quad\quad\left.8\kappa_{1}^{4}\left(\kappa_{2}^{4}-11\kappa_{2}^{2}-44\right)\right]\\
c_{3} & =\frac{\kappa_{1}^{4}}{4}\left(1-\kappa_{1}^{2}\right)^{3}\left[4\left(1-\kappa_{1}^{2}\right)+\kappa_{2}^{2}\right]>0.\end{align*}
For every $\kappa_{1},\kappa_{2}>0$, equation~(\ref{eq:discrim4})
has three nonzero roots. One of these roots is negative since $c_{0}$
and $c_{3}$ are positive. Computations show that the the remaining
two roots are also positive. However, the nullclines touch in the
first quadrant, $\epsilon_{1},\epsilon_{2}>0$, only if $\kappa_{1}<\beta$,
and $\alpha$ is the largest positive root. Thus, $\alpha^{*}(\kappa_{1},\kappa_{2})$
is the largest of the three roots of~(\ref{eq:discrim4}).

\subsubsection{Stability}

The stability of the steady states follows from the geometry of the
nontrivial nullclines for $\epsilon_{1}$ and $\epsilon_{2}$. Indeed,
it is known from the theory of the generalized Lotka-Volterra model
for competing species that a nontrivial steady state is stable if
and only if in the neighborhood of the nontrivial steady state, the
nontrivial nullclines for both $\epsilon_{1}$ and $\epsilon_{2}$
have negative slopes, but the slope of the nontrivial nullcline for
$\epsilon_{2}$ is more negative (i.e., higher in absolute value)
than the slope of the nontrivial nullcline for $\epsilon_{1}$~\citep[Chap.~12]{hirschEcol}.

\bibliographystyle{C:/texmf/bibtex/bst/elsevier/elsart-harv}

\begin{thebibliography}{62}
\expandafter\ifx\csname
natexlab\endcsname\relax\def\natexlab#1{#1}\fi
\expandafter\ifx\csname url\endcsname\relax
  \def\url#1{\texttt{#1}}\fi
\expandafter\ifx\csname urlprefix\endcsname\relax\def\urlprefix{URL
}\fi

\bibitem[{Asensio et~al.(1963)Asensio, Avigad, and Horecker}]{ASENSIO1963}
Asensio, C., Avigad, G., Horecker, B.~L., Dec 1963. Preferential
galactose
  utilization in a mutant strain of \textit{E. coli}. Arch Biochem Biophys 103,
  299--309.

\bibitem[{Brandt et~al.(2004)Brandt, Kelpin, van Leuwen, and
  Kooijman}]{Brandt2003}
Brandt, B.~W., Kelpin, F. D.~L., van Leuwen, I. M.~M., Kooijman, S.
A. L.~M.,
  2004. Modelling microbial adaptation to changing availability of substrates.
  Water Research 38, 1004--1013.

\bibitem[{Chung and Stephanopoulos(1996)}]{chung96}
Chung, J.~D., Stephanopoulos, G., 1996. On physiological
multiplicity and
  population heterogeneity of biological systems. Chem.\ Eng.\ Sc. 51,
  1509--1521.

\bibitem[{Clark and Holms(1976)}]{clark76}
Clark, B., Holms, W.~H., 1976. Control of the sequential utilization
of glucose
  and fructose by {{\em Escherichia coli\/}}. J.\ Gen.\ Microbiol. 95,
  191--201.

\bibitem[{Cohn and Horibata(1959)}]{cohn59a}
Cohn, M., Horibata, K., 1959. Inhibition by glucose of the induced
synthesis of
  the $\beta$-galactoside-enzyme system of \textit{{E}scherichia coli}.
  {A}nalysis of maintenance. J.\ Bacteriol. 78, 601--612.

\bibitem[{Crampin et~al.(1999)Crampin, Gaffney, and Maini}]{Crampin1999}
Crampin, E.~J., Gaffney, E.~A., Maini, P.~K., 1999. Reaction and
diffusion on
  growing domains: Scenarios for robust pattern formation. Bull. Math. Biol.
  61, 1093--1120.

\bibitem[{Crampin et~al.(2002)Crampin, Hackborn, and Maini}]{Crampin2002}
Crampin, E.~J., Hackborn, W.~W., Maini, P.~K., Jul 2002. {P}attern
formation in
  reaction-diffusion models with nonuniform domain growth. Bull Math Biol
  64~(4), 747--769.

\bibitem[{Daughton et~al.(1979)Daughton, Cook, and Alexander}]{Daughton1979}
Daughton, C.~G., Cook, A.~M., Alexander, M., Mar 1979. Phosphate and
soil
  binding: factors limiting bacterial degradation of ionic
  phosphorus-containing pesticide metabolites. Appl Environ Microbiol 37~(3),
  605--609.

\bibitem[{Day and Lawrence(2000)}]{Day2000}
Day, S.~J., Lawrence, P.~A., Jul 2000. {M}easuring dimensions: the
regulation
  of size and shape. Development 127~(14), 2977--2987.

\bibitem[{Dickson(1914)}]{Dickson}
Dickson, L.~E., 1914. Elementary Theory of Equations, first edition
Edition.
  Elementary Theory of Equations.

\bibitem[{Egli(1995)}]{egli95}
Egli, T., 1995. The ecological and physiological significance of the
growth of
  heterotrophic microorganisms with mixtures of substrates. Adv.\ Microbiol.\
  Ecol. 14, 305--386.

\bibitem[{Egli et~al.(1982)Egli, K\"appeli, and Fiechter}]{egli82a}
Egli, T., K\"appeli, O., Fiechter, A., 1982. Regulatory flexibility
of
  methylotrophic yeasts in chemostat culture: {S}imultaneous assimilation of
  glucose and methanol at a fixed dilution rate. Arch.\ Microbiol. 131, 1--7.

\bibitem[{Gilbert(2002)}]{Gilbert2002}
Gilbert, S.~F., 2002. Enzymatic adaptation and the entrance of
molecular
  biology into embryology. In: Sarkar, S. (Ed.), The Biology and History of
  Molecular Biology: New Perspectives. Boston Studies in the Philosophy of
  Science. Springer.

\bibitem[{Hafen and Stocker(2003)}]{Hafen2003}
Hafen, E., Stocker, H., Dec 2003. {H}ow are the sizes of cells,
organs, and
  bodies controlled? PLoS Biol 1~(3), E86.

\bibitem[{Hamilton and Dawes(1959)}]{hamilton59}
Hamilton, W.~A., Dawes, E.~A., 1959. A diauxic effect with {{\em
{P}seudomonas
  aeruginosa}}. Biochem.\ J. 71, 25P--26P.

\bibitem[{Hamilton and Dawes(1960)}]{hamilton60}
Hamilton, W.~A., Dawes, E.~A., 1960. The nature of the diauxic
effect with
  glucose and organic acids in {{\em {P}seudomonas aeruginosa}}. Biochem.\ J.
  76, 70P.

\bibitem[{Hamilton and Dawes(1961)}]{hamilton61}
Hamilton, W.~A., Dawes, E.~A., 1961. Further observations on the
nature of the
  diauxic effect with {{\em {P}seudomonas aeruginosa}}. Biochem.\ J. 79, 25P.

\bibitem[{Harder and Dijkhuizen(1976)}]{harder76}
Harder, W., Dijkhuizen, L., 1976. Mixed substrate utilization. In:
Dean, A.
  C.~R., Ellwood, D.~C., Evans, C. G.~T., Melling, J. (Eds.), Continuous
  Culture 6: Applications and New Fields. Ellis Horwood, Chichester, Ch.~23,
  pp. 297--314.

\bibitem[{Harder and Dijkhuizen(1982)}]{harder82}
Harder, W., Dijkhuizen, L., 1982. Strategies of mixed substrate
utilization in
  microorganisms. Philos.\ Trans.\ R.\ Soc.\ London~B 297, 459--480.

\bibitem[{Hirsch and Smale(1974)}]{hirschEcol}
Hirsch, M.~W., Smale, S., 1974. Differential Equations, Dynamical
Systems, and
  Linear Algebra. Academic Press, New York, Ch.~12.

\bibitem[{Inada et~al.(1996)Inada, Kimata, and Aiba}]{inada96}
Inada, T., Kimata, K., Aiba, H., 1996. Mechanism responsible for the
  glucose-lactose diauxie in \textit{Escherichia coli}: {C}hallenge to the
  {cAMP} model. Genes Cells 1, 293--301.

\bibitem[{Ishihara and Kaneko(2006)}]{Ishihara2006}
Ishihara, S., Kaneko, K., Feb 2006. {T}uring pattern with proportion
  preservation. J Theor Biol 238~(3), 683--693.

\bibitem[{Jacob and Monod(1961)}]{jacob61}
Jacob, F., Monod, J., 1961. Genetic regulatory mechanisms in the
synthesis of
  proteins. J.\ Mol.\ Biol. 3, 318--356.

\bibitem[{Joseph et~al.(1981)Joseph, Danchin, and Ullmann}]{Joseph1981}
Joseph, E., Danchin, A., Ullmann, A., Apr 1981. {R}egulation of
galactose
  operon expression: {g}lucose effects and role of cyclic adenosine
  3',5'-monophosphate. J Bacteriol 146~(1), 149--154.

\bibitem[{Kamogawa and Kurahashi(1967)}]{Kamogawa1967}
Kamogawa, A., Kurahashi, K., Feb 1967. {I}nhibitory effect of
glucose on the
  growth of a mutant strain of {E}scherichia coli defective in glucose
  transport system. J Biochem (Tokyo) 61~(2), 220--230.

\bibitem[{Kimata et~al.(1997)Kimata, Takahashi, Inada, Postma, and
  Aiba}]{kimata97}
Kimata, K., Takahashi, H., Inada, T., Postma, P., Aiba, H., 1997.
{cAMP}
  receptor protein-{cAMP} plays a crucial role in glucose-lactose diauxie by
  activating the major glucose transporter gene in \textit{{E}scherichi coli}.
  Proc.\ Nat.\ Acad.\ Sci.\ USA 94, 12914--12919.

\bibitem[{Kompala et~al.(1986)Kompala, Ramkrishna, Jansen, and
  Tsao}]{kompala86}
Kompala, D.~S., Ramkrishna, D., Jansen, N.~B., Tsao, G.~T., 1986.
Investigation
  of bacterial growth on mixed substrates: Experimental evaluation of
  cybernetic models. Biotechnol.\ Bioeng. 28, 1044--1055.

\bibitem[{Kovarova-Kovar and Egli(1998)}]{kovarova98}
Kovarova-Kovar, K., Egli, T., 1998. Growth kinetics of suspended
microbial
  cells: {F}rom single-substrate-controlled growth to mixed-substrate kinetics.
  Microbiol. Mol. Biol. Rev. 62, 646--666.

\bibitem[{Kremling et~al.(2001)Kremling, Bettenbrock, Laube, Jahreis, Lengeler,
  and Gilles}]{Kremling2001}
Kremling, A., Bettenbrock, K., Laube, B., Jahreis, K., Lengeler,
J.~W., Gilles,
  E.~D., Oct 2001. The organization of metabolic reaction networks. {III}.
  {A}pplication for diauxic growth on glucose and lactose. Metab Eng 3~(4),
  362--379.

\bibitem[{Kuroda et~al.(1992)Kuroda, Osaki, Tsuda, and Tsuchiya}]{Kuroda1992}
Kuroda, M., Osaki, N., Tsuda, M., Tsuchiya, T., Jun 1992.
Preferential
  utilization of glucose over melibiose, and \textit{vice versa}, in a {PTS}
  mutant of \textit{{S}almonella typhimurium}. Chem Pharm Bull (Tokyo) 40~(6),
  1637--1640.

\bibitem[{Lee et~al.(1974)Lee, Fredrickson, and Tsuchiya}]{lee74}
Lee, I.~H., Fredrickson, A.~G., Tsuchiya, H.~M., 1974. Diauxic
growth of {{\em
  Propionibacterium shermanii}}. Appl.\ Microbiol. 28, 831--835.

\bibitem[{Lengeler and Lin(1972)}]{Lengeler1972}
Lengeler, J., Lin, E.~C., Nov 1972. {R}eversal of the
mannitol-sorbitol diauxie
  in \textit{{E}scherichia coli}. J Bacteriol 112~(2), 840--848.

\bibitem[{Liberman and Bleiweis(1984)}]{Liberman1984b}
Liberman, E.~S., Bleiweis, A.~S., Feb 1984. {R}ole of the
  phosphoenolpyruvate-dependent glucose phosphotransferase system of
  \textit{{S}treptococcus mutans} {GS}5 in the regulation of lactose uptake.
  Infect Immun 43~(2), 536--542.

\bibitem[{Liu et~al.(1998)Liu, Svoronos, and Koopman}]{Liu1998}
Liu, P.~H., Svoronos, S.~A., Koopman, B., Dec 1998. Experimental and
modeling
  study of diauxic lag of \textit{{P}seudomonas denitrificans} switching from
  oxic to anoxic conditions. Biotechnol Bioeng 60~(6), 649--655.

\bibitem[{Mahadevan et~al.(2002)Mahadevan, Edwards, and Doyle}]{Mahadevan2002}
Mahadevan, R., Edwards, J.~S., Doyle, F.~J., Sep 2002. Dynamic flux
balance
  analysis of diauxic growth in \textit{{E}scherichia coli}. Biophys J 83~(3),
  1331--1340.

\bibitem[{Monod(1942)}]{monod1}
Monod, J., 1942. Recherches sur la croissance des cultures
bact\'eriennes
  [{S}tudies on the growth of bacterial cultures]. Actualit\'es Scientifique et
  Industrielles 911, 1--215.

\bibitem[{Monod(1947)}]{monod47}
Monod, J., 1947. The phenomenon of enzymatic adaptation and its
bearings on
  problems of genetics and cellular differentiation. Growth 11, 223--289.

\bibitem[{Murray(1989)}]{murray}
Murray, J.~D., 1989. Mathematical Biology. Biomathematics Texts.
  Springer-Verlag, New York.

\bibitem[{Narang(1998{\natexlab{a}})}]{narang98b}
Narang, A., 1998{\natexlab{a}}. The dynamical analogy between
microbial growth
  on mixtures of substrates and population growth of competing species.
  Biotechnol.\ Bioeng. 59, 116--121.

\bibitem[{Narang(1998{\natexlab{b}})}]{narang98a}
Narang, A., 1998{\natexlab{b}}. The steady states of microbial
growth on
  mixtures of substitutable substrates in a chemostat. J.\ Theor.\ Biol. 190,
  241--261.

\bibitem[{Narang(2006)}]{Narang06b}
Narang, A., 2006. Comparitive analysis of some models of gene
regulation in
  mixed-substrate microbial growth, {J.} Theor. Biol., in press,
  doi:10.1016.jtbi.2006.03.017.

\bibitem[{Narang et~al.(1997{\natexlab{a}})Narang, Konopka, and
  Ramkrishna}]{narang97c}
Narang, A., Konopka, A., Ramkrishna, D., 1997{\natexlab{a}}. The
dynamics of
  microbial growth on mixtures of substrates in batch reactors. J.\ Theor.\
  Biol. 184, 301--317.

\bibitem[{Narang et~al.(1997{\natexlab{b}})Narang, Konopka, and
  Ramkrishna}]{narang97a}
Narang, A., Konopka, A., Ramkrishna, D., 1997{\natexlab{b}}. New
patterns of
  mixed substrate growth in batch cultures of \textit{{E}scherichia
  coli}~{K12}. Biotechnol.\ Bioeng. 55, 747--757.

\bibitem[{Neidhardt and Magasanik(1957)}]{Neidhardt1957}
Neidhardt, F.~C., Magasanik, B., Feb 1957. Reversal of the glucose
inhibition
  of histidase biosynthesis in \textit{{A}erobacter aerogenes}. J Bacteriol
  73~(2), 253--259.

\bibitem[{Notley-McRobb et~al.(1997)Notley-McRobb, Death, and
  Ferenci}]{McRobb1997}
Notley-McRobb, L., Death, A., Ferenci, T., Jun 1997. {T}he
relationship between
  external glucose concentration and c{AMP} levels inside {E}scherichia coli:
  implications for models of phosphotransferase-mediated regulation of
  adenylate cyclase. Microbiology 143 ( Pt 6), 1909--1918.

\bibitem[{Oehler et~al.(2006)Oehler, Alberti, and Müller-Hill}]{Oehler2006}
Oehler, S., Alberti, S., Müller-Hill, B., 2006. {I}nduction of the
lac promoter
  in the absence of {DNA} loops and the stoichiometry of induction. Nucleic
  Acids Res 34~(2), 606--612.

\bibitem[{Okada et~al.(1981)Okada, Ueyama, Niiya, Kanazawa, Futai, and
  Tsuchiya}]{Okada1981}
Okada, T., Ueyama, K., Niiya, S., Kanazawa, H., Futai, M., Tsuchiya,
T., Jun
  1981. Role of inducer exclusion in preferential utilization of glucose over
  melibiose in diauxic growth of \textit{{E}scherichia coli}. J Bacteriol
  146~(3), 1030--1037.

\bibitem[{Othmer and Pate(1980)}]{Othmer1980}
Othmer, H.~G., Pate, E., Jul 1980. {S}cale-invariance in
reaction-diffusion
  models of spatial pattern formation. Proc Natl Acad Sci U S A 77~(7),
  4180--4184.

\bibitem[{Panikov(1995)}]{panikov}
Panikov, N.~S., 1995. Microbial Growth Kinetics. Chapman \& Hall,
London,
  Ch.~3, p. 181.

\bibitem[{Plumbridge(2003)}]{plumbridge03}
Plumbridge, J., 2003. Regulation of gene expression in the {PTS} in
  \textit{{E}scherichia coli}: {T}he role and interactions of {M}lc. Curr.\
  Opin.\ Microbiol. 5, 187--193.

\bibitem[{Postma et~al.(1993)Postma, Lengeler, and Jacobson}]{Postma1993}
Postma, P.~W., Lengeler, J.~W., Jacobson, G.~R., Sep 1993.
  Phosphoenolpyruvate:carbohydrate phosphotransferase systems of bacteria.
  Microbiol Rev 57~(3), 543--594.

\bibitem[{Ptashne and Gann(2002)}]{Ptashne2}
Ptashne, M., Gann, A., 2002. Genes \& Signals. Cold Spring Harbor
Laboratory
  Press, Cold Spring Harbor, New York.

\bibitem[{Ramakrishna et~al.(1996)Ramakrishna, Ramkrishna, and
  Konopka}]{ramakrishna96}
Ramakrishna, R., Ramkrishna, D., Konopka, A., 1996. Cybernetic
modeling of
  growth in mixed, substitutable substrate environments: {P}referential and
  simultaneous utilization. Biotechnol.\ Bioeng. 52, 141--151.

\bibitem[{Roseman and Meadow(1990)}]{Roseman1990}
Roseman, S., Meadow, N.~D., Feb 1990. {S}ignal transduction by the
bacterial
  phosphotransferase system. {D}iauxie and the \textit{crr} gene ({J}. {M}onod
  revisited). J Biol Chem 265~(6), 2993--2996.

\bibitem[{Santillán and Mackey(2004)}]{Santillan2004}
Santillán, M., Mackey, M.~C., Mar 2004. Influence of catabolite
repression and
  inducer exclusion on the bistable behavior of the \textit{lac} operon.
  Biophys J 86~(3), 1282--1292.

\bibitem[{Serrano and O'Farrell(1997)}]{Serrano1997}
Serrano, N., O'Farrell, P.~H., Mar 1997. {L}imb morphogenesis:
connections
  between patterning and growth. Curr Biol 7~(3), R186--R195.

\bibitem[{Spiegelman(1948)}]{Spiegelman1948}
Spiegelman, S., 1948. Differentiation as the controlled production
of unique
  enzymatic patterns. In: Danielli, J.~F., Brown, R. (Eds.), Growth in relation
  to differentiation and morphogenesis. No.~II in Symposium of the Society for
  Experimental Biology. Academic Press, pp. 286--325.

\bibitem[{Tanaka et~al.(1967)Tanaka, Fraenkel, and Lin}]{Tanaka1967}
Tanaka, S., Fraenkel, D.~G., Lin, E.~C., Apr 1967. {T}he enzymatic
lesion of
  strain {MM}-6, a pleiotropic carbohydrate-negative mutant of
  \textit{{E}scherichia coli}. Biochem Biophys Res Commun 27~(1), 63--67.

\bibitem[{Thattai and Shraiman(2003)}]{Thattai2003}
Thattai, M., Shraiman, B.~I., Aug 2003. Metabolic switching in the
sugar
  phosphotransferase system of \textit{{E}scherichia coli}. Biophys J 85~(2),
  744--754.

\bibitem[{van Dedem and Moo-Young(1973)}]{vandedem73}
van Dedem, G., Moo-Young, M., 1973. Cell growth and extracellular
enzyme
  synthesis in fermentations. Biotechnol.\ Bioeng. 15, 419--439.

\bibitem[{Wong et~al.(1997)Wong, Gladney, and Keasling}]{Wong1997}
Wong, P., Gladney, S., Keasling, J.~D., 1997. Mathematical model of
the
  \textit{lac} operon: {i}nducer exclusion, catabolite repression, and diauxic
  growth on glucose and lactose. Biotechnol Prog 13~(2), 132--143.

\bibitem[{Yagil and Yagil(1971)}]{yagil71}
Yagil, G., Yagil, E., 1971. On the relation between effector
concentration and
  the rate of induced enzyme synthesis. Biophys. J. 11, 11--17.

\end{thebibliography}

\end{document}